\documentclass[amsmath,amssymb,preprint,longbibliography,aps,pra]{revtex4-1}
\usepackage{geometry}
\usepackage[utf8]{inputenc}
\usepackage[T1]{fontenc}
\usepackage{mathptmx}
\usepackage{fancyhdr}
\usepackage{multirow}
\usepackage{booktabs}
\usepackage{xcolor}
\usepackage{braket}
\usepackage[version=4]{mhchem}
\usepackage{listings}
\usepackage{caption}
\usepackage{graphicx}
\usepackage{float}
\usepackage{makecell}
\usepackage{array}
\usepackage{adjustbox}

\makeatletter
\newcommand{\smalloplus}{\mathbin{\mathpalette\make@small\oplus}}
\newcommand{\smallotimes}{\mathbin{\mathpalette\make@small\otimes}}

\newcommand{\make@small}[2]{%
  \vcenter{\hbox{%
    $\m@th\ifx#1\displaystyle\scriptstyle\else\ifx#1\textstyle\scriptstyle
     \else\scriptscriptstyle\fi\fi#2$%
  }}%
}
\makeatother

\renewcommand{\tensor}[1]{\mathrm{\mathbf{#1}}}

\begin{document}
\pagestyle{fancy}
\fancyhf{}
\cfoot{\thepage}
\title{Quantum-Classical Auxiliary Field Quantum Monte Carlo with Matchgate Shadows on Trapped Ion Quantum Computers}

% IonQ contributors

% Chemistry team
\author{Luning Zhao}
\email[Corresponding author: ]{zhao@ionq.co}
\author{Joshua J. Goings}
% All other IonQ contributors, alphabetically
\author{Willie Aboumrad}
\author{Andrew Arrasmith}
\author{Lazaro Calderin}
\author{Spencer Churchill}
\author{Dor Gabay}
\author{Thea Harvey-Brown}
\author{Melanie Hiles}
\author{Magda Kaja}
\author{Matthew Keesan}
\author{Karolina Kulesz}
\author{Andrii Maksymov}
\author{Mei Maruo}
\author{Mauricio Muñoz}
\author{Bas Nijholt}
\author{Rebekah Schiller}
\author{Yvette de Sereville}
\author{Amy Smidutz}
\author{Felix Tripier}
\author{Grace Yao}
\author{Trishal Zaveri}
% Project leadership
\author{Coleman Collins}
\author{Martin Roetteler}
\author{Evgeny Epifanovsky}

\affiliation{
IonQ Inc, College Park, MD, 20740, USA
}

% AstraZeneca contributors

\author{Arseny Kovyrshin}
\author{Lars Tornberg}
\author{Anders Broo}

\affiliation{
Data Science and Modelling, Pharmaceutical Sciences, R\&D, AstraZeneca Gothenburg, Pepparedsleden 1, Molndal SE-431 83, Sweden
}

% NVIDIA contributors

\author{Jeff R. Hammond}

\affiliation{
NVIDIA Helsinki Oy, Porkkalankatu 1, Helsinki, 00180 Finland
}

\author{Zohim Chandani}
\author{Pradnya Khalate}
\author{Elica Kyoseva}

\affiliation{
NVIDIA Corporation, 2788 San Tomas Expressway, Santa Clara,
95051, CA, USA
}

% AWS contributors (moved to Acknowledgments until the final version of the manuscript on AWS request)

\author{Yi-Ting Chen}
\author{Eric M. Kessler}
\author{Cedric Yen-Yu Lin}
\author{Gandhi Ramu}
\author{Ryan Shaffer}

\affiliation{
AWS Quantum Technologies, Seattle,
98121, WA, USA
}

\author{Michael Brett}
\author{Benchen Huang}
\author{Maxime R. Hugues}
\author{Tyler Y. Takeshita}

\affiliation{
AWS Worldwide Specialist Organization, Seattle,
98170, WA, USA
}

\date{\today}

\begin{abstract}
We demonstrate an end-to-end workflow to model chemical reaction barriers with the quantum-classical auxiliary field quantum
Monte Carlo (QC-AFQMC) algorithm with quantum tomography using matchgate shadows. The workflow operates within an accelerated quantum supercomputing environment with the IonQ Forte quantum computer and NVIDIA GPUs on Amazon Web Services.
We present several algorithmic innovations and an efficient GPU-accelerated execution, which achieves a several orders of magnitude
speedup over the state-of-the-art implementation of QC-AFQMC.
We apply the algorithm to simulate the oxidative addition step of the nickel-catalyzed Suzuki--Miyaura reaction using 24 qubits of IonQ Forte with 16 qubits used to represent the trial state, plus 8 additional ancilla qubits for error mitigation, resulting in the largest QC-AFQMC with matchgate shadow experiments ever performed on quantum hardware. We achieve a $9\times$ speedup in collecting matchgate circuit measurements, and our distributed-parallel post-processing implementation attains a $656\times$ time-to-solution improvement over the prior state-of-the-art. Chemical reaction barriers for the model reaction evaluated with active-space QC-AFQMC are within the uncertainty interval of $\pm4$~kcal/mol from the reference CCSD(T) result when matchgates are sampled on the ideal simulator and within 10~kcal/mol from reference when measured on QPU. This work marks a step towards practical quantum chemistry simulations on quantum devices while identifying several opportunities for further development.
\end{abstract}

\maketitle

\section{Introduction}

Computationally intensive tasks have become integral to drug discovery and development (DDD) driving advancement across the entire pharmaceutical R\&D value chain. 
There is a strong motivation to find ways to accomplish traditionally experimental tasks faster and cheaper with the aid of computer simulations. Quantum computing is an emerging technology that enables the study of a certain class of problems thought to be intractable with classical computation. The possibility of realizing these theoretical speedups in practice has sparked research in a range of application areas, including pharmaceutical R\&D. Therefore, it is 
unsurprising to see the recent development of quantum algorithms for many problems within 
DDD intended for near-term\cite{mull2019,robe2021,fing2018,smal2024,mens2023} and future 
fault-tolerant\cite{allc2022,weid2023} quantum computing devices. While the early drug 
discovery phase has been the primary target for computational methods, in this work 
we focus on scenarios arising in the drug development phase that require accurate electronic structure modeling. Specifically, we consider finding a synthetic pathway of desired purity and yield using transition metal (TM) catalysis during the drug development phase\cite{kovy2025}.

Some of the most significant chemical reactions in medicinal chemistry include amide bond formation, nucleophilic aromatic substitution, and Suzuki--Miyaura cross-coupling\cite{bost2018}. The first two reactions typically involve only organic compounds, and density functional theory (DFT)\cite{Yang89_book} calculations have sufficient resolution to give useful mechanistic insights. However, the Suzuki--Miyaura cross-coupling reactions include catalytic TM 
complexes that exhibit complicated electronic structure. In particular, unpaired {\it d}- or {\it f}-electrons 
lead to dense low-lying electronic states resulting in strong electron correlation. Additionally, the open-shell character of many transition metal complexes leads to states with different spin symmetries, shaping their reactivity, stability, and underlying reaction mechanisms. As a result,  multiple reaction spin channels emerge, allowing reactions to proceed through distinct spin states with different activation barriers. Moreover, because high-spin and low-spin states in transition metal complexes often lie close in energy, the spin state can shift during a reaction\cite{harv2014}, a phenomenon known as spin crossover.

The intricate electronic structure of TM complexes powers diverse catalytic mechanisms. At the same time, it makes it difficult to model their chemistry reliably using approximate single-reference methods since the wave function often exhibits a multi-reference character, and strong electron correlation plays a central role\cite{reih2009}. Single-reference methods such as DFT and even the coupled cluster singles and doubles with perturbative 
triples [CCSD(T)]\cite{headgordon89_479}, which is regarded as the gold standard in terms of accuracy, are not able to reliably model these systems.
In fact, CCSD(T) often exhibits well-known breakdowns for strongly correlated systems, including complexes with polynuclear transition metal cores, systems undergoing metal to insulator phase transitions, and multi-electron excited states. (It may still yield 
reasonable accuracy for mononuclear TM complexes\cite{rado2024}.)
Methods such as the complete active space self-consistent field method (CASSCF)\cite{Siegbahn80_157}, are capable of treating strong correlation, but their computational cost grows exponentially with the size of the active space. The density matrix 
renormalization group method (DMRG)\cite{whit1992,Zhai2023-yi}, originally developed for one-dimensional and quasi-one-dimensional systems, offers polynomial scaling in computational cost for such systems and has demonstrated strong performance on chemically relevant systems, including TM complexes\cite{mart2008,reih2009,chan2011}. While DMRG enables significantly larger active space computations compared to conventional approaches and is amenable to GPU acceleration\cite{Menczer:DMRG-DGXH100,Legeza:DMRG-CASSCF,chun2024}, its performance at large scales increasingly depends on available physical memory and memory bandwidth\cite{Menczer:DMRG-DGXH100,liu2022}.

This gives rise to one of the most promising applications of quantum computers pertaining to the DDD pipeline --- modeling strongly correlated catalytic systems. In recent years, a variety of different algorithms have been developed to solve the electronic structure problem using quantum computers. Within these algorithms, quantum phase estimation (QPE)\cite{Head-Gordon05_1704, white10_106} could in principle provide exact solutions to quantum chemistry problems in polynomial time, but even with recent innovations such as qubitization \cite{Chuang19_163,Neven18_041015} and low-rank tensor decomposition\cite{Babbush21_030305}, it demands error rates far beyond the capability of today's noisy quantum computers, and likely unachievable without robust error correction. The variational quantum eigensolver (VQE)\cite{Brien14_5213, Martinis16_031007, Gambetta17_23879,Google20_1084,Kim20_33} has been considered one of the most promising near-term algorithms. However, it requires an often prohibitive number of measurements on quantum computers\cite{Gonthier2022-lw}, and the predicted chemical properties are highly sensitive to device noise. Both factors limit its ability to efficiently deliver usable results on current-generation devices.

In this work, we focus on an alternative algorithm, the quantum-classical auxiliary field quantum Monte Carlo (QC-AFQMC)\cite{Huggins2022-lh} algorithm, which builds upon the classical AFQMC method\cite{Gubernatis97_7464, Krakauer03_136401, Motta2018-ft, Lee2022-rl, Shee2023-gs}. AFQMC is a projector Monte Carlo approach that solves for the ground state of the electronic Hamiltonian with imaginary time propagation. Although AFQMC is formally exact, it relies on an approximate trial state to control the fermionic phase problem through the phaseless constraint introducing a bias proportional to the trial state's deviation from the true ground state. Classical methods can generate multi-determinant trial wavefunctions (CASSCF and selected CI) or high-quality single-reference states from coupled cluster theory, but these approaches face intrinsic barriers: multi-determinant trials scale exponentially in cost, while single-reference methods fail in strongly correlated regimes. QC-AFQMC overcomes both by using quantum computers to prepare correlated trial states that capture multi-reference character without explicit enumeration. Early results\cite{Huggins2022-lh,Huang2024-hk} show QC-AFQMC reaching chemical accuracy with polynomial measurement cost, demonstrating notable noise resilience compared to VQE and sidestepping the steep overhead of QPE.

The core innovation in QC-AFQMC is not simply its hybrid structure, but the way it divides tasks: the quantum computer prepares the correlated trial state and performs shadow tomography measurements, after which all imaginary time propagation and observable estimation are carried out on the classical computer. Unlike VQE, which requires ongoing quantum-classical feedback for each parameter update, QC-AFQMC isolates quantum measurements to the start, avoiding the iterative quantum-classical loop. Despite this conceptual elegance, several studies\cite{Mazzola2022-rv, Amsler2023-hr, Kiser2023-rf, Huang2024-hk} have argued that the classical post-processing of QC-AFQMC is prohibitively expensive making the method impractical for systems of real scientific interest\cite{Mazzola2022-rv, Amsler2023-hr, Kiser2023-rf, Huang2024-hk}.
This work serves as a demonstration to the contrary. We significantly enhance the practicality of QC-AFQMC by reducing its time to solution through several key advances: tuning the performance of quantum hardware by increasing the throughput of job submission and execution, algorithmic improvements (inspired by some recent developements of AFQMC\cite{Jiang2024-cp}) leading to a computational cost reduction in post-processing, and GPU-accelerated implementations of routines required for post-processing via the NVIDIA CUDA Toolkit leveraging cuBLAS, cuSOLVER, and cuTENSOR. Overall, with these improvements we have been able to reduce the computational cost of QC-AFQMC energy evaluation and imaginary time propagation from $\mathcal{O}(N^{8.5})$ to $\mathcal{O}(N^{5.5})$, and the cost of imaginary time propagation from $\mathcal{O}(N^{7.5})$ to $\mathcal{O}(N^{4.5})$, where $N$ is molecule size. Executing the steps of QC-AFQMC via CUDA-Q on IonQ hardware hosted on Amazon Braket~\cite{braket} combined with post-processing on Amazon Web Services (AWS), we establish an efficient, end-to-end workflow for applying QC-AFQMC to complex chemical problems.

\begin{figure}[bt!]
\includegraphics[width=\textwidth]{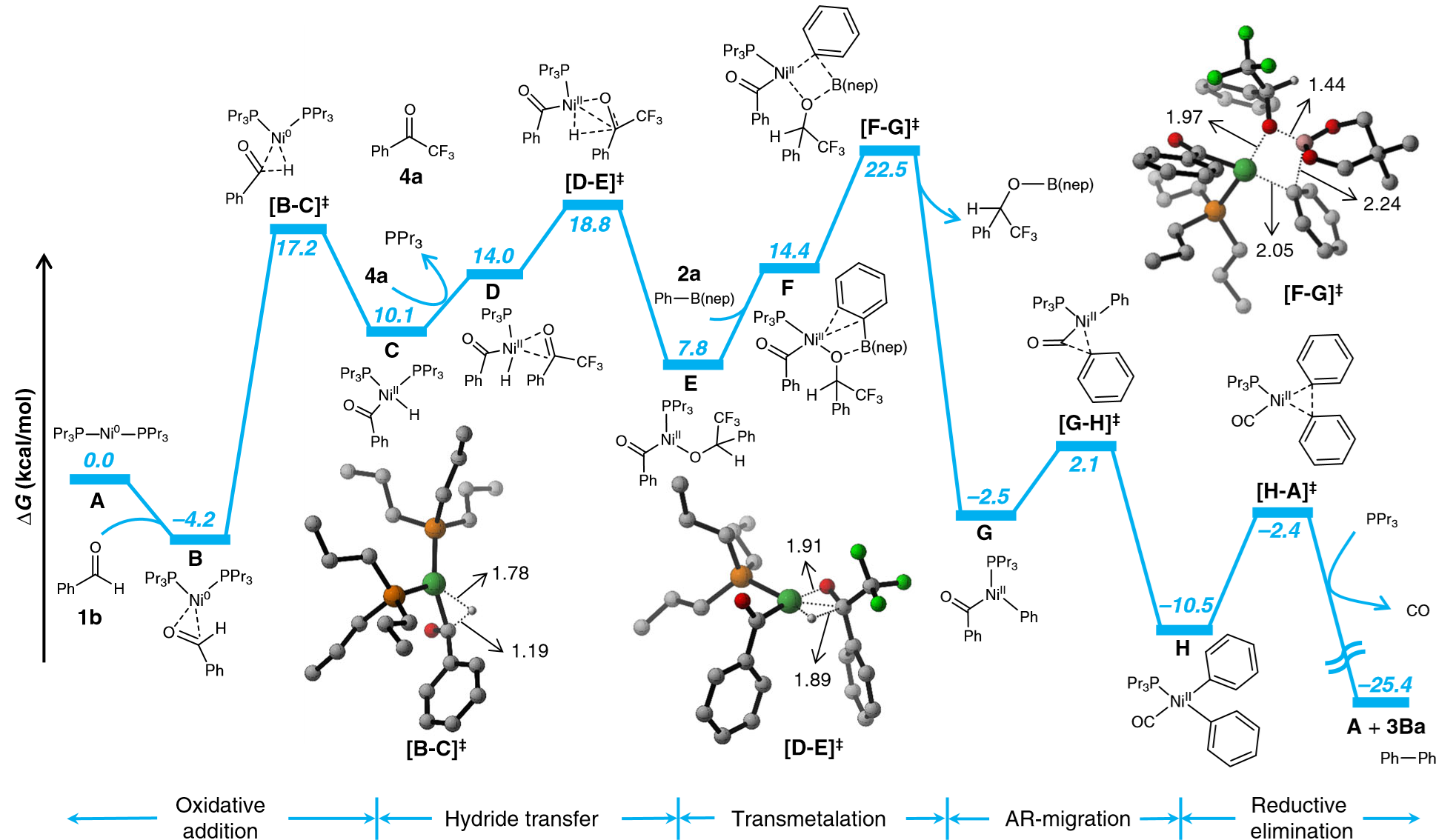}
\caption{Reaction mechanism and energy profile of the nickel-catalyzed deformylative Suzuki--Miyaura cross-coupling reaction. Our work focuses on modeling the oxidative addition step denoted {\bf B} $\rightarrow$ {\bf [B-C]$^\ddagger$} $\rightarrow$ {\bf C}. (Figure reproduced from Ref.~\cite{Guo2019-zc} under the terms of the Creative Commons CC BY license.)}
\label{fig:guo}
\end{figure}

This paper demonstrates the QC-AFQMC workflow by modeling a key step in the Suzuki--Miyaura cross-coupling reaction (Fig.~\ref{fig:guo}), which uses a nickel-based catalytic complex as a cost-effective alternative to palladium\cite{Guo2019-zc}.  Specifically, we focus on the prediction of the reaction barrier for the first step: oxidative addition of aryl halide to nickel bis(tripropylphosphine). This step begins with the complex denoted {\bf B} in Fig.~\ref{fig:guo}, proceeds through a transition state {\bf [B-C]$^\ddagger$}, and ends with the intermediate {\bf C}.

The paper is organized as follows. Section~\ref{sec:methods} introduces the classical AFQMC algorithm, outlining its strengths and limitations, then presents QC-AFQMC and the classical shadow sampling technique at its core. Once the theoretical framework is established, Section~\ref{sec:comp-details} describes the computational details: the chosen chemical reaction, active space selection, algorithmic improvements, and key aspects of implementation and execution. Section~\ref{sec:results} reports QC-AFQMC energies for the species in the reaction and compares them to results from other methods. We then discuss time to solution, separating quantum execution from classical post-processing. The paper concludes with a discussion of open challenges and future directions. 

\section{Methods}
\label{sec:methods}
\subsection{The electronic structure problem}
The chemical reactivity of a molecular system can be modeled by connecting reactants, products, and relevant transition states on a potential energy surface (PES), which is the energy of the molecule as a function of its atomic geometric configuration. Each point of the PES represents a solution to the ground-state energy problem, which can be determined computationally from first principles by solving the electronic Schr\"odinger equation:
\begin{equation}
    \label{eqn:schrodinger_eqn}
    H\ket{\Psi_g}=E_g\ket{\Psi_g},
\end{equation}
in which $H=H_1+H_2$ is the electronic Hamiltonian consisting of one- and two-body contributions, $\ket{\Psi_g}$ is the 
ground state wave function, and $E_g$ is the energy.

Solving the Schr\"odinger equation exactly on a classical computer has exponential computational and memory complexity with respect to system size, which is prohibitive for practical applications. This motivated the development of many physics-informed approximations and heuristics that trade approximation error (usually not directly controllable) for time to solution. 
We have discussed some of the most widely used approximate 
methods such as CCSD(T), CASSCF, and DMRG in the 
Introduction. A notable class of classical electronic structure methods
relies on projector quantum Monte Carlo (QMC). These 
techniques\cite{Rajagopal01_33, Zhao20_174105} employ stochastic walkers to represent the system's wavefunction, evolving them through imaginary time propagation to generate ground-state samples. The precise form of projector QMC is highly adaptable, determined by factors such as the walker's Hilbert space, the propagator's mathematical structure, and the nature of walker interactions. This flexibility gives rise to a range of QMC variants\cite{Umrigar143_164105, Booth17_176403, Malone2022-dc}, including the AFQMC algorithm, which we will now explore in detail.

\subsection{Classical AFQMC}
The AFQMC method\cite{Zhang2018-dn,Motta2018-ft,Lee2022-rl,Shee2023-gs} is a powerful tool for modeling the ground state of 
molecules and materials. In AFQMC, the wave function is expressed as a linear combination
of walkers 
\begin{equation}
    \label{eqn:afqmc_wfn}
    \left|\Psi\right>=\sum_l{w_l\left|\phi_l\right>},
\end{equation}
where each walker $\left|\phi_l\right>$ is a Slater determinant with its corresponding weight $w_l$. 
One can obtain the ground state by evolving the wave function in imaginary time $\tau$
\begin{equation}
    \label{eqn:afqmc_imag_time_prop}
    \left|\Psi(\tau)\right>=e^{-\tau H}\sum_l{w_l\left|\phi_l\right>}.
\end{equation}
As the system evolves, contributions from excited states decay exponentially, and eventually only the ground state survives.
This evolution can be expressed as a product of short-time propagators
\begin{equation}
    \label{eqn:afqmc_imag_time_prop_trotterized}
     \left|\Psi(\tau)\right>=\left[e^{-\Delta\tau H}\right]^N\sum_l{w_l\left|\phi_l\right>}, \quad \Delta\tau=\frac{\tau}{N}.
\end{equation}
To carry this out in practice, each propagator must be efficiently approximated and applied to the set of walkers.
The electronic Hamiltonian that defines the propagator is given by
\begin{equation}
    \label{eqn:hamiltonian}
    H=H_1+H_2=\sum_{pq}{t_{pq}a^\dagger_pa_q}+\frac{1}{2}\sum_{pqrs}{g_{pqrs}a^\dagger_pa^\dagger_qa_sa_r},
\end{equation}
where $t$ and $g$ are one- and two-electron integrals, and $a^\dagger_p (a_p)$ are the creation (annihilation)
operators for the $p^{th}$ orbital. Using Cholesky decomposition, the Hamiltonian can be rewritten as 
\begin{equation}
    \label{eqn:ham_cholesky_decom}
    H=v_0-\frac{1}{2}\sum_{\gamma}{v^2_\gamma},
\end{equation}
where $v_0$ is the modified one-body term and $v_\gamma=i\mathcal{L}_\gamma$, with $\mathcal{L}_\gamma$
denoting the Cholesky vector associated with the two-body term
\begin{equation}
    \label{eqn:cholesky_vec}
    \mathcal{L}_\gamma=\sum_{pq}{L^\gamma_{pq}a^\dagger_pa_q}, \quad g_{pqrs}=\sum_{\gamma}{L^\gamma_{pr}L^{\gamma\ast}_{sq}}.
\end{equation}
In QC-AFQMC, the Cholesky decomposition is done with a modified algorithm\cite{Malone2022-dc}, and 
its cost scaling is $\mathcal{O}(N^4)$. Now using the second-order Trotter-Suzuki decomposition, the imaginary time propagator becomes 
\begin{equation}
    \label{eqn:2nd_trotterized_imag_time_prop}
    e^{-\Delta\tau H}\approx e^{-\frac{\Delta\tau}{2}v_0}\prod_{\gamma}{e^{\frac{\Delta\tau}{2}v^2_\gamma}}e^{-\frac{\Delta\tau}{2}v_0}.
\end{equation}
The Hubbard-Stratonovich transformation\cite{stratonovich57_416, hubbard59_77} converts the term involving $v^2_\gamma$ into an integral over an auxiliary field, resulting in the exponentials of one-body operators coupled to this auxiliary field
\begin{equation}
    \label{eqn:hs_transform}
    e^{\frac{\Delta\tau}{2}v^2_\gamma}=\int{\frac{dx_\gamma}{\sqrt{2\pi}}e^{-\frac{x^2_\gamma}{2}}e^{\sqrt{\Delta\tau}x_\gamma v_\gamma}}.
\end{equation}
Substituting into Eq.~\ref{eqn:2nd_trotterized_imag_time_prop}, we obtain
\begin{equation}
    \label{eqn:imag_time_prop_integral}
    e^{-\Delta\tau H}\approx\int{d\textbf{x}p(\textbf{x})B(\textbf{x})},
\end{equation}
in which the operator $B(\textbf{x})$ is defined as 
\begin{equation}
    \label{eqn:operator_b}
    B(\textbf{x})=e^{-\frac{\Delta\tau}{2}v_0}\prod_{\gamma}{e^{\sqrt{\Delta\tau}x_\gamma v_\gamma}}e^{-\frac{\Delta\tau}{2}v_0},
\end{equation}
and $p(\textbf{x})$ is a multi-dimensional Gaussian distribution over the auxiliary field variable $\textbf{x}$.

According to Eq.~\ref{eqn:imag_time_prop_integral} and~\ref{eqn:operator_b}, to propagate the walkers, one
could first sample $\textbf{x}$ from $p(\textbf{x})$, then construct and apply the operator $B(\textbf{x})$ to each walker.
Since $B(\textbf{x})$ is the exponential of a one-body operator, its action on a Slater determinant
simply yields another determinant: 
\begin{equation}
    \label{eqn:walker_prop}
    \left|\phi^\prime_l(\textbf{x})\right>=B(\textbf{x})\left|\phi_l\right>.
\end{equation}

Finally, with the propagated walkers that sample from the 
ground state, AFQMC energy is a normalized weighted sum of the local energies, 
\begin{equation}
    \label{eqn:afqmc_energy}
    E=\frac{\sum_l{w_lE_L(\phi_l)}}{\sum_l{w_l}},
\end{equation}
in which the local energy is computed as 
\begin{equation}
    \label{eqn:local_energy}
    E_L(\phi)=\frac{\braket{\Psi_T|H|\phi}}{\braket{\Psi_T|\phi}}.
\end{equation}

\subsection{The phase problem}
The approach described above is usually referred to as the ``free-projection'' version of AFQMC. Although it is 
formally exact, in reality it suffers from the phase problem\cite{Gubernatis97_7464, Krakauer03_136401,zhang13_15}, which causes instabilities in the energy evaluation.
In order to mitigate the phase problem, a trial wave function $\left|\Psi_T\right>$ is introduced to guide 
the propagation. The trial wave function should be a close approximation to the exact ground state, and the
propagator in Eq.~\ref{eqn:imag_time_prop_integral} is modified with\cite{Krakauer03_136401}
\begin{equation}
    \label{eqn:imag_time_prop_integral_modified}
    \int{\left<\Psi_T|\phi^{'}(\textbf{x}-\bar{\textbf{x}})\right>p(\textbf{x}-\bar{\textbf{x}})B(\textbf{x}-\bar{\textbf{x}})\frac{1}{\left<\Psi_T|\phi\right>}d\textbf{x}},
\end{equation}
in which a constant shift $\bar{\textbf{x}}$ is also introduced, which does not affect the equality of 
Eq.~\ref{eqn:imag_time_prop_integral}. The optimal choice of $\bar{\textbf{x}}$ will be determined later. The above equation could be written as 
\begin{equation}
    \label{eqn:imag_time_prop_integral_modified_simple}
    \int{p(\textbf{x})w(\textbf{x},\phi)B(\textbf{x}-\bar{\textbf{x}})d\textbf{x}},
\end{equation}
where
\begin{equation}
    \label{eqn:weight_factor}
    w(\textbf{x},\phi)=\frac{\left<\Psi_T|\phi^{'}(\textbf{x}-\bar{\textbf{x}})\right>}{\left<\Psi_T|\phi\right>}e^{\textbf{x}\cdot\bar{\textbf{x}}-(\bar{\textbf{x}}\cdot\bar{\textbf{x}}/2)}.
\end{equation}
In each step, an $\bar{\textbf{x}}$ is first decided for each 
walker, and the walker is propagated to a new walker with $B(\textbf{x}-\bar{\textbf{x}})$. Here, $w(\textbf{x},\phi)$ is the importance sampling function, which could be taken into account by allowing every walker to have a weight factor and update
it during the propagation\cite{Zhang17_1364-use-Motta2018-ft-instead, Huang2024-hk}. 

The optimal choice of $\bar{\textbf{x}}$ is determined by minimizing the fluctuation of $w(\textbf{x},\phi)$ 
with respect to $\textbf{x}$, and one obtains\cite{Krakauer03_136401}
\begin{equation}
    \label{eqn:force_bias}
    \bar{x}_\gamma=-\sqrt{\Delta\tau}\frac{\braket{\Psi_T|v_\gamma|\phi}}{\braket{\Psi_T|\phi}},
\end{equation}
which is usually referred to as force bias in AFQMC. With this choice of $\bar{\textbf{x}}$, 
one could further approximate the weight factor as $w(\textbf{x},\phi)\approx\mathrm{exp}[-\Delta\tau E_L(\phi)]$\cite{Krakauer03_136401}, in which $E_L(\phi)$ is the local energy defined in Eq.~\ref{eqn:local_energy}. 

If the trial wave function is exact, then one can show that the 
local energy is real valued and equal to the exact energy. In this case 
all the weights also become real valued and positive, which mitigates the phase problem. However, for approximate trial wave functions, the local
energy is a complex number. Therefore a common approximation is to replace it
with its real part. $w(\textbf{x},\phi)\approx\mathrm{exp}[-\Delta\tau \mathrm{Re}\{E_L(\phi)\}]$, and the overall weight factor is
\begin{equation}
    \label{eqn:weight_factor_final}
    w(\textbf{x},\phi)\approx\mathrm{exp}[-\Delta\tau \mathrm{Re}\{E_L(\phi)\}]\mathrm{max}\left(0, \mathrm{cos}\left[\mathrm{arg\left(\frac{\braket{\Psi_T|B(\textbf{x}-\bar{\textbf{x}})|\phi}}{\braket{\Psi_T|\phi}}\right)}\right]\right),
\end{equation}
in which an additional factor is introduced to prevent abrupt phase changes during the random walk\cite{Zhang17_1364-use-Motta2018-ft-instead}.

\subsection{QC-AFQMC}
The introduction of the weight factor in Eq.~\ref{eqn:weight_factor_final} is
an approximation to the imaginary time propagation, and it leads to a bias in energy
predictions. This is similar to the fixed node approximation\cite{Rajagopal01_33} in diffusion Monte
Carlo. The quality of the approximation depends on the quality of the trial state, and, more
specifically, on the quality of the nodal structure of the trial state. One could show
that if the trial state is the exact ground state, then the weight factor is no longer
an approximation, and the predicted energy is also exact. On the other hand, if the trial
state is a poor approximation to the ground state, it will introduce a significant bias
to the predicted energy\cite{Amsler2023-hr}. 

Unfortunately, only a few trial state options are efficient to prepare on classical computers. The 
most widely used trial state is the single determinant state. However, single determinant 
wave functions are poor approximations to the ground state for strongly correlated systems,  
and using them as trial states leads to large errors in energy. In such 
cases, one needs to use multi-determinant trial states\cite{Mahajan2022-aa, Sharma21_4786}, but the size of the multi-determinant
expansion grows exponentially with system size, and in practice further approximations need to
be introduced to make the approach efficient, such as truncating the multi-determinant expansion
to a predefined, limited number of terms, which then leads to errors in the predicted energy. 

The development of the quantum computer offers a different prospective for AFQMC. Efficient quantum state preparation techniques\cite{Duncan20_10515, Gambetta17_23879,Yelin24_1422} for the ground state of molecules and materials open up a possibility of using them as the trial state for AFQMC.
Huggins {\it et al.}\cite{Huggins2022-lh}
introduced the quantum-classical AFQMC (QC-AFQMC) method, in which one prepares the trial state on a quantum computer and uses it to
guide the imaginary time propagation in AFQMC. Furthermore, Huggins {\it et al.} demonstrated that even with a simple perfect pairing wave function prepared on the quantum computer, the QC-AFQMC method yields more accurate energy estimates in both molecules and materials compared with
classical AFQMC using single-determinant trial states.
The QC-AFQMC approach is compatible with any quantum state preparation technique, including
ansatzes used in the variational quantum 
eigensolver (VQE) approach such as those based on unitary coupled cluster (UCC) 
theory\cite{Tavernelli20_124107,Singh22_14834}, hardware-efficient circuits\cite{Gambetta17_23879}, and Clifford+T ansatz\cite{Yelin24_1422}, as well as other states one
could prepare non-variationally.

\subsection{Classical shadows and the evaluation of overlaps}

In order for a quantum state to be used as the trial state in AFQMC, one needs to efficiently 
evaluate its overlap with the walker states: $\braket{\Psi_T|\phi}$. Since both the trial state
and the walker state could be prepared on the quantum computer, the most straightforward
way for evaluating the overlap is to use the Hadamard test. However, this requires a large 
number of measurements and back and forth communication between quantum and classical computers. 
To tackle this issue, the initial study by Google\cite{Huggins2022-lh} proposed to use classical shadow techniques. The classical shadow method was first proposed by Huang {\it et al.}\cite{Huang2020-jc, Preskill22_1397}. This method tries to store a ``classical copy'' of the quantum state by measuring the quantum state in
a number of random bases. These measurement results are
classical shadows of the quantum state, and by inverting the measurement channel, one could reconstruct the quantum state's density matrix, and use it to compute reduced quantities such as expectation values, fidelity, and non-linear functions. This method is especially advantageous if one wants to predict many properties of a quantum state
simultaneously, as it has been shown that one could achieve so with very 
few measurements that only scales logarithmically with the number of properties.  

QC-AFQMC is a scenario well suited for classical shadows, as one needs to evaluate the overlap between the trial state $|\Psi\rangle$ and many different walkers simultaneously. However, although classical shadow is efficient in terms of the number of measurements one needs, to compute properties one needs to invert the measurement channel classically, and whether such an inversion can be done efficiently depends on both the randomized basis and the 
properties of interest. The initial study of QC-AFQMC used randomized Clifford basis,
which turned out to be very inefficient in the classical post processing to compute overlaps in AFQMC, with a cost
that scales exponentially with system size\cite{Mazzola2022-rv}. However, a follow up study shows that such an 
exponential bottleneck could be reduced to polynomial by switching to a basis generated 
by randomized matchgate matrices\cite{Wan2023-ns, Kiser2023-rf, Huang2024-hk}. 
The first demonstration of the matchgate shadow based QC-AFQMC on quantum hardware was demonstrated in 2024 by a team from AWS and the University of Chicago\cite{Huang2024-hk} on a system with 4 qubits. 

In order to compute the overalp $\braket{\Psi_T|\phi}$ with matchgate shadows, we follow the same procedure described in Wan {\it et al.}\cite{Wan2023-ns}. The protocol consists of two phases: we first prepare a trial state $|\Psi_T\rangle$ via VQE, and construct a new state $\ket{\Psi}=\frac{1}{\sqrt{2}}(\ket{0}^{\otimes N}+\ket{\Psi_T})$. The reason for preparing such a state is because overlaps $\braket{\Psi_T|\phi}$ could 
be formulated as its expectation values with respect to certain
observables defined by $\ket{\phi}$. Second, we perform shadow measurements by composing the state $\ket{\Psi}$ circuit with Gaussian circuits constructed from orthogonal matrices $\tensor{Q}_p \in \mathbb{R}^{2N \times 2N}$. These matrices can be derived from either Haar-random orthogonal matrices or random signed permutation matrices, with the latter ensuring unit determinant by construction. In our study, signed permutation matrices are used. For each measurement $\ket{b}$, we obtain a covariance matrix $\tensor{C}_{\ket{b}}$ that contributes to the classical shadow representation.
The overlap computation exploits the mathematical structure of matchgate circuits through efficient Pfaffian evaluation\cite{Wimmer2012-ck}. For a fermionic state represented by determinant $p$, we construct an antisymmetric matrix:
\begin{equation}
\label{eqn:a_mat}
\tensor{A}_{p\ket{b}}(z) = \tensor{C}^{(s)}_{\ket{\textbf{0}}} + z \cdot \tensor{B}^{(s)}_{p\ket{b}}.
\end{equation}
Here, $\tensor{C}^{(s)}_{0}$ represents the selected submatrix of the vacuum state covariance matrix, and $\tensor{B}^{(s)}_{p\ket{b}}$ combines the shadow measurement with the walker structure through the transformation:
\begin{equation}
\label{eqn:b_mat}
\tensor{B}_{p\ket{b}} = \tensor{W}^\ast \mathbf{M}_\phi \tensor{Q}_p^T\mathbf{C}_{\ket{b}} \tensor{Q}_p\mathbf{M}^T_\phi  \tensor{W}^\dagger,
\end{equation}
where $\tensor{W}$ encodes the fermionic structure of the system through carefully constructed block rotations, and $\mathbf{M}_\phi$ is the matrix that defines the walker $\ket{\phi}$. The definitions of $\tensor{W}$, $\tensor{C}$ and $\mathbf{M}_\phi$ could be found in the Supplementary Information. 
The overlap between trial state and walker $\ket{\phi}$ is computed by first finding the coefficients of the polynomial
defined by the Pfaffian of $\tensor{A}_{p\ket{b}}(z)=\sum_x^l{c_{p\ket{b}x}z^x}$, which could be
calculated by evaluating the Pfaffian at Chebyshev nodes \cite{Kiser2023-rf}
\begin{equation}
    z_k = \cos\left(\frac{2k+1}{2N}\pi\right),
\end{equation}
and performing polynomial interpolation to obtain the coefficients. 
The final contribution to the overlap for this shadow takes the form:
\begin{equation}
\braket{\Psi_T|\phi}=\frac{1}{N_{\mathrm{shadow}}}\sum_p{o_p}, \quad \mathrm{and} \quad o_p = \frac{i^{\eta/2}}{2^{N-\eta/2}} \sum_{x=0}^{\ell} c_{p\ket{b}x} \binom{2N}{2x} \bigg/ \binom{N}{x},
\end{equation}
where $\ell = N - \lfloor \eta/2 \rfloor$ defines the polynomial degree, and $\eta$ is the number of Fermions in the 
system. The coefficients $c_{p\ket{b}x}$ are determined through polynomial interpolation at the Chebyshev nodes.
As been shown in Wan {\it et al.}\cite{Wan2023-ns}, to bound the error of the computed overlap by $\epsilon$, the required number of measurements scales as $\mathcal{O}(\sqrt{N}\mathrm{ln}N/\epsilon^2)$ .  The overall cost for the above algorithm for computing overlaps scales as $\mathcal{O}(N^{4.5})$ per walker, as 
computing matrix Pfaffian scales as $\mathcal{O}(N^{3})$, and one needs to do it 
for every grid point and shadow. The number of grid points (Chebyshev nodes) and shadows scale as 
$\mathcal{O}(N)$ and $\mathcal{O}(N^{0.5})$, respectively, so the overall cost for 
overlaps is $\mathcal{O}(N^{4.5})$. 

\subsection{Force bias and local energy}
\label{subsec:improvement}
Despite achieving efficient overlap evaluation through quantum measurements, QC-AFQMC still faces another computational bottleneck: without algorithmic improvements, computing the local energy and force bias terms would scale as $\mathcal{O}(N^{8.5})$ and $\mathcal{O}(N^{7.5})$ per walker. Each step requires evaluating the local energy:
\begin{equation}
E_L(\phi)=\frac{\langle\Psi_T|H|\phi\rangle}{\langle\Psi_T|\phi\rangle},
\end{equation}
and the force bias terms:
\begin{equation}
\bar{x}_\gamma = -\sqrt{\Delta\tau}\frac{\langle\Psi_T|v_\gamma|\phi\rangle}{\langle\Psi_T|\phi\rangle}.
\end{equation}

The most straightforward way of evaluating the local energy and force bias is
to enumerate every term in the Hamiltonian and the $v_\gamma$ operator, apply
each term to $\phi$ to obtain a new walker, and evaluate the overlap of the new
walker with the trial state. Since the cost for computing one overlap scales as 
$\mathcal{O}(N^{4.5})$, computing local energies scales as $\mathcal{O}(N^{8.5})$ as
there are $\mathcal{O}(N^{4})$ terms in the Hamiltonian. For force bias, although
there are only $\mathcal{O}(N^{2})$ terms in each $v_\gamma$, one needs to compute it
for every Cholesky vector, and since there are $\mathcal{O}(N)$ Cholesky vectors, 
the overall cost for force bias scales as $\mathcal{O}(N^{7.5})$. 

However, a more efficient algorithm has been developed. As noted by Jiang et al.\cite{Jiang2024-cp}, since these quantities represent 
variations of the overlap under state transformations, they can be computed through 
(algorithmic) differentiation of the overlap expressions. For example, the force
bias could be written as 
\begin{equation}
    \label{eqn:force_bias_diff}
    \frac{\braket{\Psi_T|v_{\gamma}|\phi}}{\braket{\Psi_T|\phi}}=i\frac{\braket{\Psi_T|\sum_{pq}{L_{pq}^{\gamma}a^\dagger_pa_q}|\phi}}{\braket{\Psi_T|\phi}}=i\frac{\partial}{\partial\lambda}\mathrm{ln}\braket{\Psi_T|e^{\lambda\sum_{pq}{L_{pq}^{\gamma}a^\dagger_pa_q}}|\phi}\big|_{\lambda=0}=i\frac{\partial}{\partial\lambda}\mathrm{ln}\braket{\Psi_T|\tilde{\phi}(\lambda)}\big|_{\lambda=0},
\end{equation}
and it becomes the derivative of the trial wave function overlap with a modified 
walker, with respect to a auxiliary variable $\lambda$. 

The one-body part of the energy shares the same form as the force bias, and it could
be computed in the same way. The two-body part of the energy is
\begin{equation}
    \label{eqn:energy_tb_diff}
    \begin{split}
           &\sum_\gamma\frac{\braket{\Psi_T|\sum_{pqrs}{L_{pq}^{\gamma}L_{rs}^{\gamma}a^\dagger_pa_qa^\dagger_ra_s}|\phi}}{\braket{\Psi_T|\phi}} \\
           &=\sum_{\gamma}\frac{\partial^2}{\partial\lambda_1\partial\lambda_2}\braket{\Psi_T|e^{\lambda_1\sum_{pq}{L_{pq}^{\gamma}a^\dagger_pa_q}}e^{\lambda_2\sum_{rs}{L_{rs}^{\gamma}a^\dagger_ra_s}}|\phi}\big|_{\lambda_1=0, \lambda_2=0}/\braket{\Psi_T|\phi} \\
           &=\sum_{\gamma}\frac{\partial^2}{\partial\lambda_1\partial\lambda_2}\braket{\Psi_T|\tilde{\phi}(\lambda_1,\lambda_2)}\big|_{\lambda_1=0, \lambda_2=0}/\braket{\Psi_T|\phi}, 
    \end{split}
\end{equation}
which becomes the second order partial derivative of the overlap. 

These derivatives could be computed using numerical differentiations, and it transforms the computational complexity from $\mathcal{O}(N^{8.5})$ per walker\cite{Kiser2023-rf, Huang2024-hk} to $\mathcal{O}(N^{5.5})$\cite{Jiang2024-cp} for local energies, and from $\mathcal{O}(N^{7.5})$ to 
$\mathcal{O}(N^{5.5})$\cite{Jiang2024-cp} for computing all force bias. 
The efficiency stems from the scaling properties: the overlap, and one-body energy terms maintain $\mathcal{O}(1)$ scaling relative to the base overlap computation, while the two-body energy and the number of force bias term scales as $\mathcal{O}
(N_{\gamma})$ relative to the overlap, where $N_{\gamma}$ is the number of Cholesky vectors.

We further notice that by using algorithmic differentiation instead of finite difference, one could further reduce the computational cost, especially for force bias.
Notice that the derivative of the matrix Pfaffian is 
\begin{equation}
    \label{eqn:Pfaffian_der}
    \frac{\partial\mathrm{Pf}(\tensor{A}(\lambda))}{\partial\lambda}=\frac{\mathrm{Pf}(\tensor{A}(\lambda))}{2}\mathrm{Tr}(\tensor{A}(\lambda)^{-1}\frac{\partial\tensor{A}}{\partial\lambda})\big|_{\lambda=0}, 
\end{equation}
since we only need the derivative at $\lambda=0$, the matrix Pfaffian and inverse 
only need to be computed once for each time step, and they could be reused later. For force bias, only
the trace needs to be recomputed for each Cholesky vector. The cost for computing 
the trace of product of inverse and matrix derivative scales as $\mathcal{O}(N^2)$ for
each Cholesky vector, so the overall cost for computing all force bias becomes 
$\mathcal{O}(N^{4.5})+\mathcal{O}(N_\gamma)\cdot\mathcal{O}(N^2)\cdot\mathcal{O}(N^{1.5})\sim\mathcal{O}(N^{4.5})$, one order of magnitude reduction from numerical differentiations. 

The local energy requires the second order derivatives of the matrix Pfaffian,
\begin{equation}
    \label{eqn:Pfaffian_2nd_der}
    \begin{split}
        &\frac{\partial^2 \mathrm{Pf}(\tensor{A}(\lambda_1,\lambda_2))}{\partial \lambda_1 \partial \lambda_2} \\
        &=\frac{\mathrm{Pf}(\tensor{A}(\lambda_1,\lambda_2))}{2}\left\{\mathrm{tr}\left[\tensor{A}^{-1}(\lambda_1,\lambda_2)\frac{\partial^2 \tensor{A}(\lambda_1,\lambda_2)}{\partial\lambda_1\partial\lambda_2}\right]\right. \\
        &\left.-\mathrm{tr}\left[\tensor{A}^{-1}(\lambda_1,\lambda_2)\frac{\partial \tensor{A}(\lambda_1,\lambda_2)}{\partial\lambda_1}\tensor{A}^{-1}(\lambda_1,\lambda_2)\frac{\partial \tensor{A}(\lambda_1,\lambda_2)}{\partial\lambda_2}\right]\right. \\
        &\left.+\frac{1}{2}\mathrm{tr}\left[\tensor{A}^{-1}(\lambda_1,\lambda_2)\frac{\partial \tensor{A}(\lambda_1,\lambda_2)}{\partial\lambda_1}\right]\mathrm{tr}\left[\tensor{A}^{-1}(\lambda_1,\lambda_2)\frac{\partial \tensor{A}(\lambda_1,\lambda_2)}{\partial\lambda_2}\right]\right\}\big|_{\lambda_1=0,\lambda_2=0} \\
    \end{split}    
\end{equation}
again since we only need the derivative at $\lambda_1=\lambda_2=0$, the matrix Pfaffian
and inverse only need to be computed once for each time step. Unlike force bias, the middle term
now requires explicit matrix-matrix multiplication, and algorithmic differentiation provides
the same big-$\mathcal{O}$ scaling as numerical differentiation. However, using
algorithmic differentiation converts the computational bottleneck from computing 
matrix Pfaffian to matrix products, which could be easily accelerated with GPUs. 

\subsection{Virtual correlation energy}
\label{sec:vce}
In the QC-AFQMC algorithm, the trial wave function is prepared on quantum computers, with a one-to-one mapping 
from orbitals to qubits. However, for large scale chemical
systems such as the TM complexes, even a small basis set
produces hundreds of orbitals. In contrast, most of today's quantum computers only have a few tens of qubits. 
To tackle this issue, the most common option is to reduce
the system to an active space. Since the active space only
contains a fraction of the orbitals in the full space, 
it becomes feasible to prepare its wavefunction on today's
quantum computers. 

However, accurate modeling of electron correlations inside
the active space is not enough to obtain quantitatively 
accurate predictions to chemical properties, and 
one also has to take into account of the electron correlations outside the active space. This presents a distinct challenge in QC-AFQMC. Huggins {\it et al.}\cite{Huggins2022-lh} introduced the virtual correlation energy (VCE) technique to address this through orbital space partitioning, enabling quantum resources to focus on the active space while treating remaining correlation effects with the trial state of a single Slater determinant. The original proposal had a significant overhead in evaluating the local energy using overlaps, resulting in costs $\mathcal{O}(N^4)$ more expensive than the overlap evaluation. Building on the improvements by Jiang {\it et al.}\cite{Jiang2024-cp}, our implementation eliminates this overhead, making the method significantly more efficient for large basis sets.

Suppose one has already prepared the trial wavefunction
in the active space $\ket{\Psi_{T,a}}$ (with $N$ orbitals), then we can write the trial wave function in the full space (with $N_B$ orbitals) as:
\begin{equation}
    \left|\Psi_{T}\right\rangle= \left|\Xi_{\mathrm{c}}\right\rangle\otimes\left|\Psi_{T,a}\right\rangle\otimes |0_v\rangle,
\end{equation}
where $\left|\Xi_{\mathrm{c}}\right\rangle$ represents frozen core orbitals as a Slater determinant, and $|0_v\rangle$ denotes the virtual orbital vacuum state. 
One could show that for a walker $\ket{\phi}$ that lives
in the full space, its overlap with $\ket{\Psi_T}$ could be computed as:
\begin{equation}
    \label{eqn:vce_overlap}
    \left\langle\Psi_{\textrm{T}}|\phi\right\rangle = \operatorname{det}(\Sigma_c R)\left\langle \Psi_{\textrm{T},a}|\tilde{\phi}_a\right\rangle/(\mathrm{det}(U^\dagger)\mathrm{det}(V)),
\end{equation}
where $\Sigma_c$, $\mathrm{det}(U^\dagger)$, and $\mathrm{det}(V)$ capture the contributions from the core space through matrix operations, and $R$ is the renormalization factor. This formulation enables QC-AFQMC
to run in the full space with the trial wavefunction defined
in an active space, and therefore capture electron correlations outside the active space. 
VCE allows for basis set convergence while concentrating quantum resources on the strongly correlated active space. A detailed derivation of Equation 
\ref{eqn:vce_overlap} could be found in the Supplementary Information. 

\section{Computational details}
\label{sec:comp-details}

\subsection{Workflow}

\begin{figure}
\includegraphics[trim={1.5in 0.25in 1.75in 1.25in},clip,width=\textwidth]{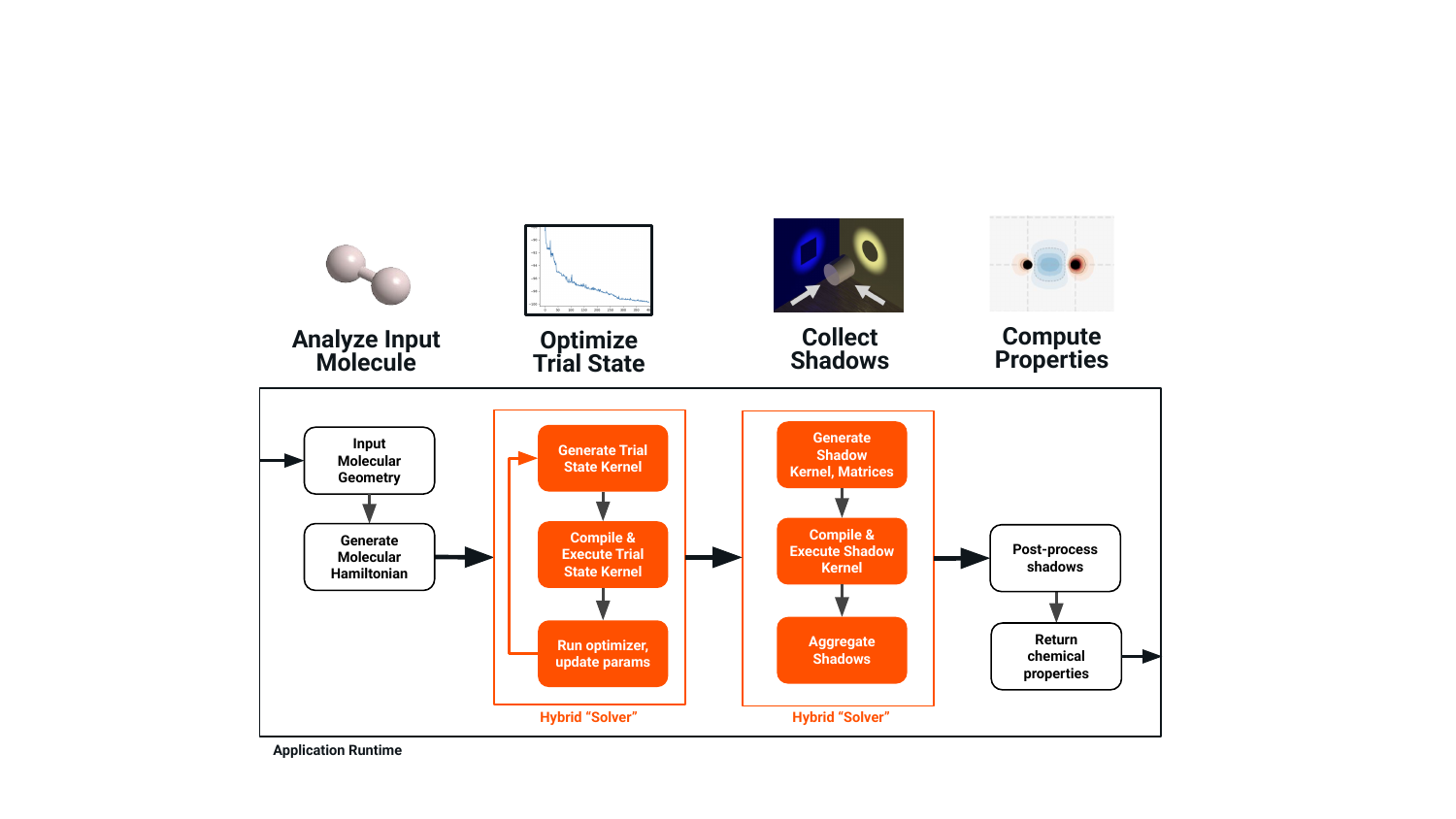}
\caption{Diagram of the computational workflow for QC-AFQMC. The hybrid steps indicated in orange are executed using a combination of QPU and CPUs. White boxes indicate classical computations using a combination of CPUs and GPUs.}
\label{fig:workflow}
\end{figure}

The QC-AFQMC computation proceeds in multiple steps executed on a variety of hardware as depicted in the chart in Fig.~\ref{fig:workflow}. In this section, we discuss each step in detail.

\paragraph{Analyze input molecule} The job begins with the specification of a molecular system and its geometry, which then undergoes a classical simulation at the self-consistent field (SCF) level to obtain a set of molecular orbitals and the molecular Hamiltonian. At this stage, we also determine the size of the active space. See Sections~\ref{sec:model_reaction} and~\ref{sec:active_space} for details.

\paragraph{Optimize trial state} With the Hamiltonian defined, the job enters the state preparation stage. In this work, we obtain the QC-AFQMC trial state using VQE with the unitary pair coupled-cluster doubles (upCCD) ansatz \cite{Zhao22_60}. Using CUDA-Q, we construct the kernel for the trial VQE state and variationally optimize the VQE parameters which is executed on an ideal simulator. 

\paragraph{Collect shadows} Next, the trial state is sampled using matchgate shadows on a quantum computer. Matchgate shadow circuits are constructed as CUDA-Q kernels and submitted for execution on the IonQ Forte quantum processing unit (QPU) from the Amazon Braket Hybrid Job environment. Upon collecting the measurements, the workflow aggregates them to prepare for post-processing.

\paragraph{Compute properties} Finally, the workflow finishes with performing AFQMC imaginary time propagation as a post-processing step to evaluate the energy of the molecule.

All electronic structure calculations were performed using the \texttt{PySCF}\cite{Sun2020-mh} framework, interfaced with \texttt{LibXC}\cite{Lehtola2018-lz} for density functional theory evaluations. Orbital entanglement analysis employed the \texttt{Block2} density matrix renormalization group software\cite{Zhai2023-yi}, accessed through the open-source \texttt{ActiveSpaceFinder} package\cite{HQS-Quantum-SimulationsUnknown-hl}. Quantum Monte Carlo calculations utilized \texttt{iPie}\cite{Jiang2024-rp,Malone2022-dc}, with our matchgate shadow implementation building upon routines from the open-source \texttt{symmetry\allowbreak-adjusted\allowbreak-classical\allowbreak-shadows} repository\cite{ZhaoUnknown-mf,Zhao2020-um,Zhao2023-ji}.
More implementation details on the QC-AFQMC workflow can be found in the Supplementary Information. 

\subsection{Model chemical reaction}
\label{sec:model_reaction}

For this demonstration we selected the oxidative addition step of the nickel-catalyzed deformylative
Suzuki--Miyaura cross-coupling reaction shown in Fig.~\ref{fig:guo}. We used the molecular geometries of the complexes B, [B-C]$^\ddagger$ and C from Ref.~\cite{Guo2019-zc}. To make the demonstration computationally tractable, we applied two approximations that reduce the computational effort while retaining relevant properties. First, we truncated the molecular structures from 77 atoms to 41 atoms, which reduces the computational cost of post-processing. On top of that, we selected an active space to reduce quantum computing requirements.

The truncation of molecular structures was carried out following established protocols in computational organometallic chemistry\cite{Macgregor2016-jv} beginning from the original, 77-atom reference system. Peripheral substituents that do not participate in the reaction mechanism were excised via selective cleavage of aliphatic $\ce{C-C}$ bonds at $sp^3$-hybridized centers, with $\ce{-CH2CH2CH3}$ groups replaced by $\ce{-CH3}$ substituents. This strategy preserves the integrity of the $\ce{P}$ coordination sphere while avoiding the disruption of critical $\ce{C-P}$ bonds.
The original and reduced molecules are shown
in Fig.~\ref{fig:molecular_structures}, and atomic cartesian coordinates could be found in the Supplementary
Information. In this study, we will be using the reduced 41-atom systems in QC-AFQMC calculations. 

\begin{figure}[tb!]
\centering
\scalebox{0.8}{
\begin{tabular}{@{} >{\centering\arraybackslash}m{3cm} @{\hspace{4mm}} ccc @{}}
& \multicolumn{3}{c}{\textbf{Reaction coordinate}} \\[2mm]
& B (reactant) & [B-C]$^\ddagger$ & C (product) \\[4mm]
\textbf{Original} \par (77 atoms) &
\adjustbox{valign=c}{\includegraphics[width=0.25\textwidth]{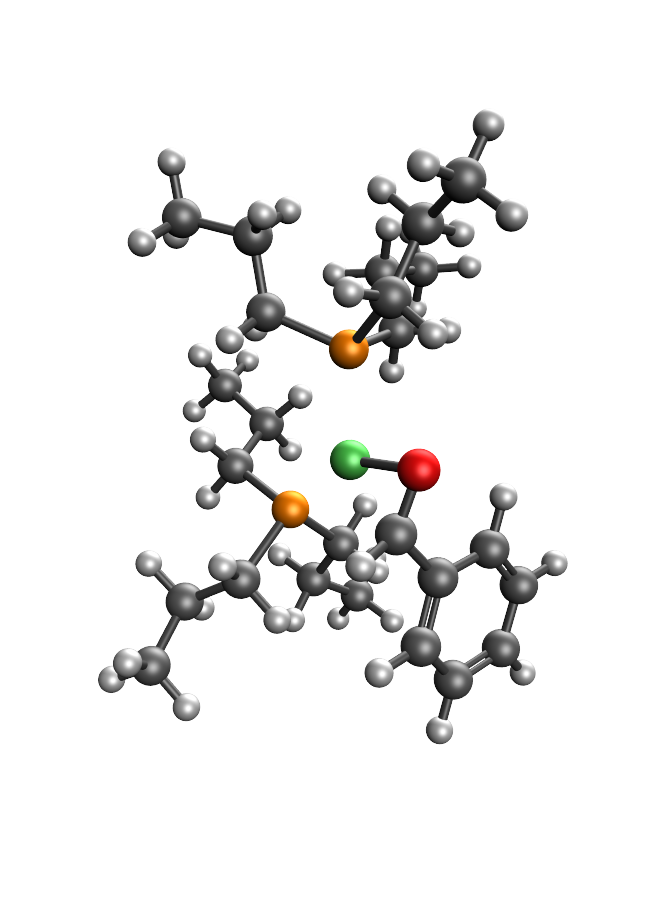}} &
\adjustbox{valign=c}{\includegraphics[width=0.25\textwidth]{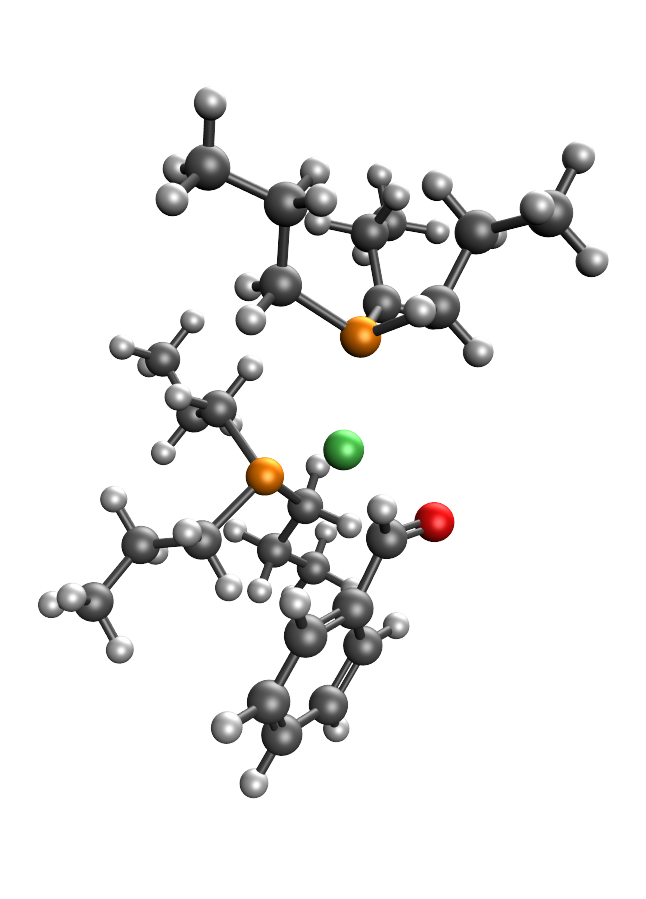}} &
\adjustbox{valign=c}{\includegraphics[width=0.25\textwidth]{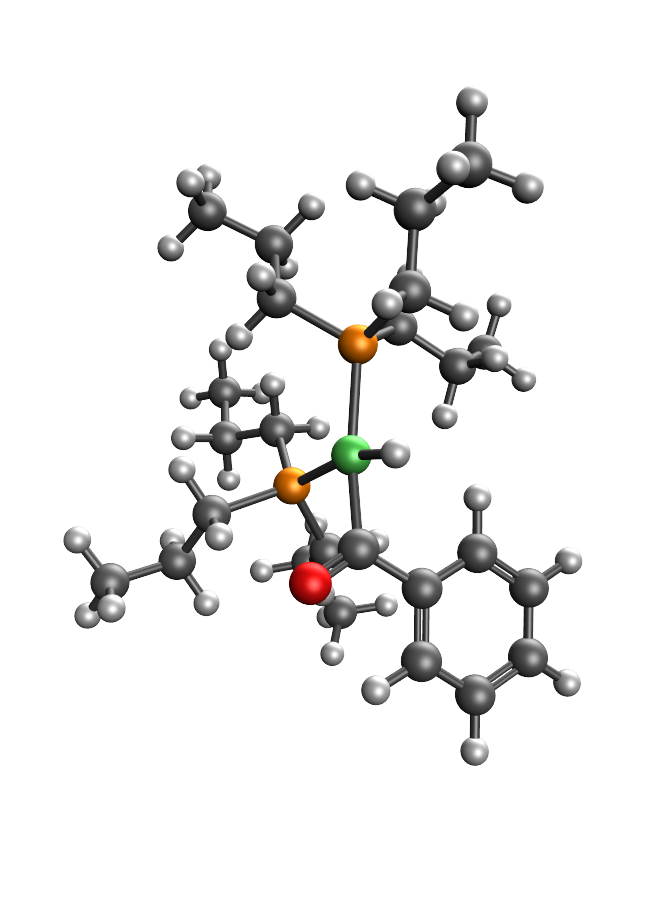}} \\[8pt]
\textbf{Reduced} \par (41 atoms) &
\adjustbox{valign=c}{\includegraphics[width=0.25\textwidth]{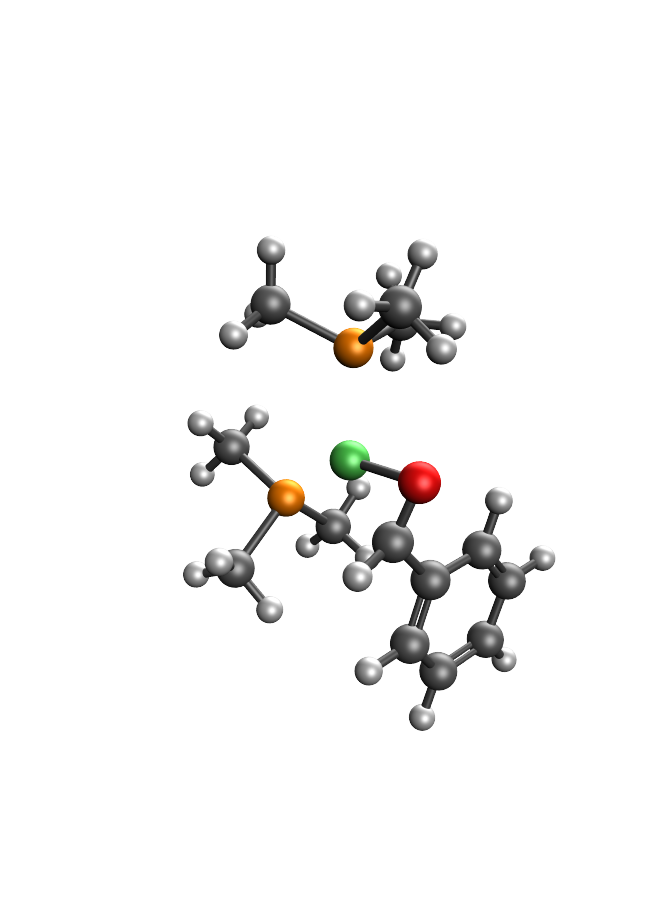}} &
\adjustbox{valign=c}{\includegraphics[width=0.25\textwidth]{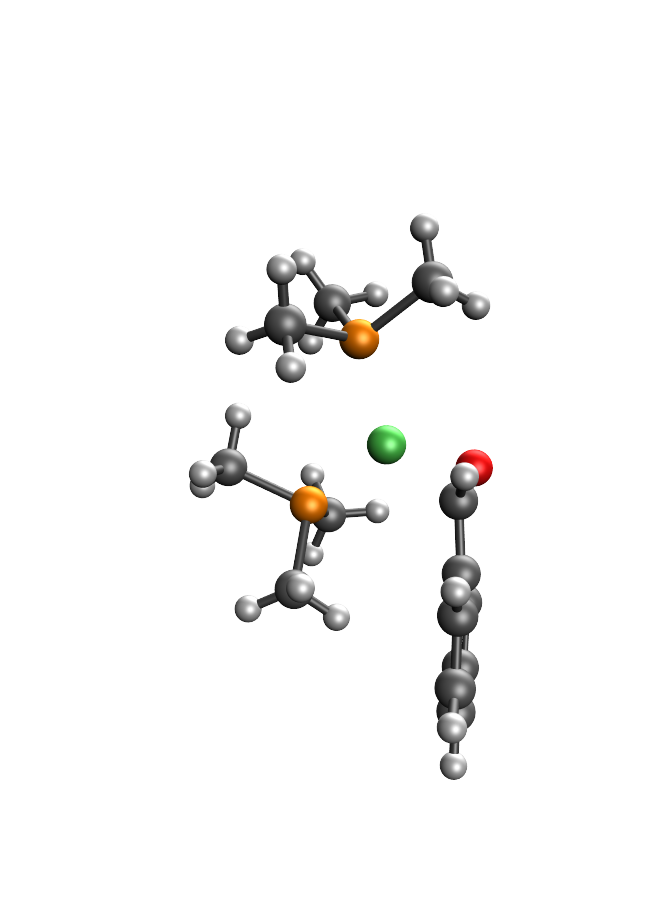}} &
\adjustbox{valign=c}{\includegraphics[width=0.25\textwidth]{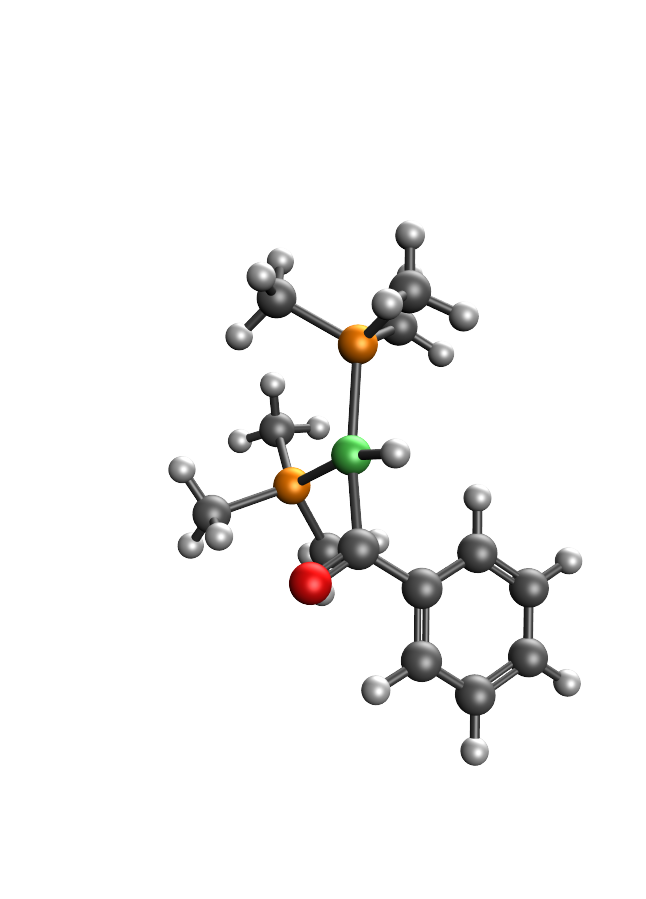}}
\end{tabular}
}
\caption{3D structures of the molecular system at each reaction coordinate (B, [B-C]$^\ddagger$, C) and truncation level. Atoms are colored as: Nickel (green), Phosphorus (orange), Oxygen (red), Carbon (gray), and Hydrogen (white).}
\label{fig:molecular_structures}
\end{figure}

\begin{figure}[h]
    \centering
    \includegraphics[width=0.8\textwidth]{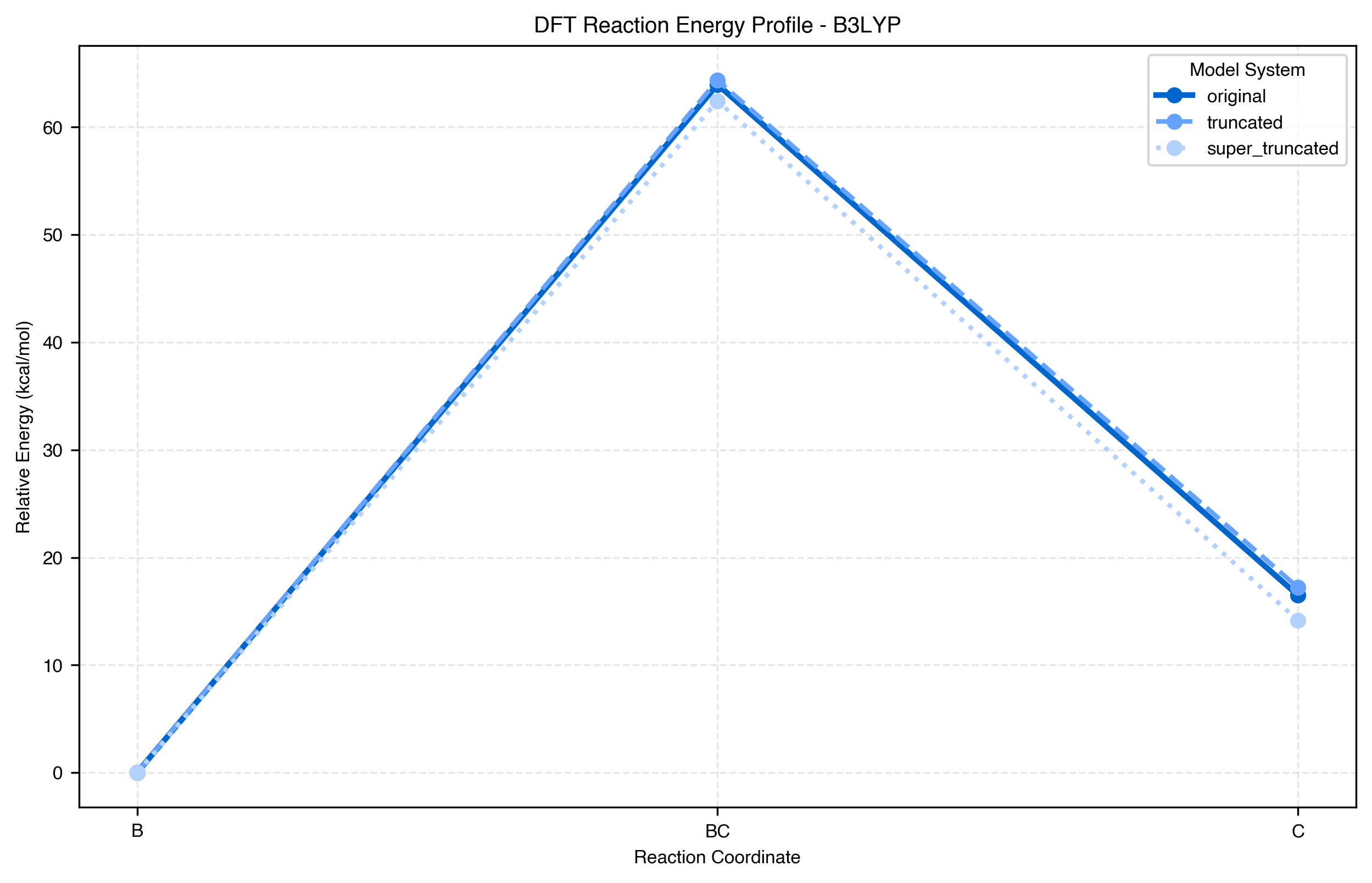}
    \caption{DFT reaction energy profile computed with B3LYP/STO-3G. The energy of complex B is set as the zero-point reference. The energy barrier and relative energies are largely preserved across truncation levels, indicating that the minimal model captures the essential reaction features.}
    \label{fig:dft_b3lyp}
\end{figure}

\begin{figure}[h]
    \centering
    \includegraphics[width=0.8\textwidth]{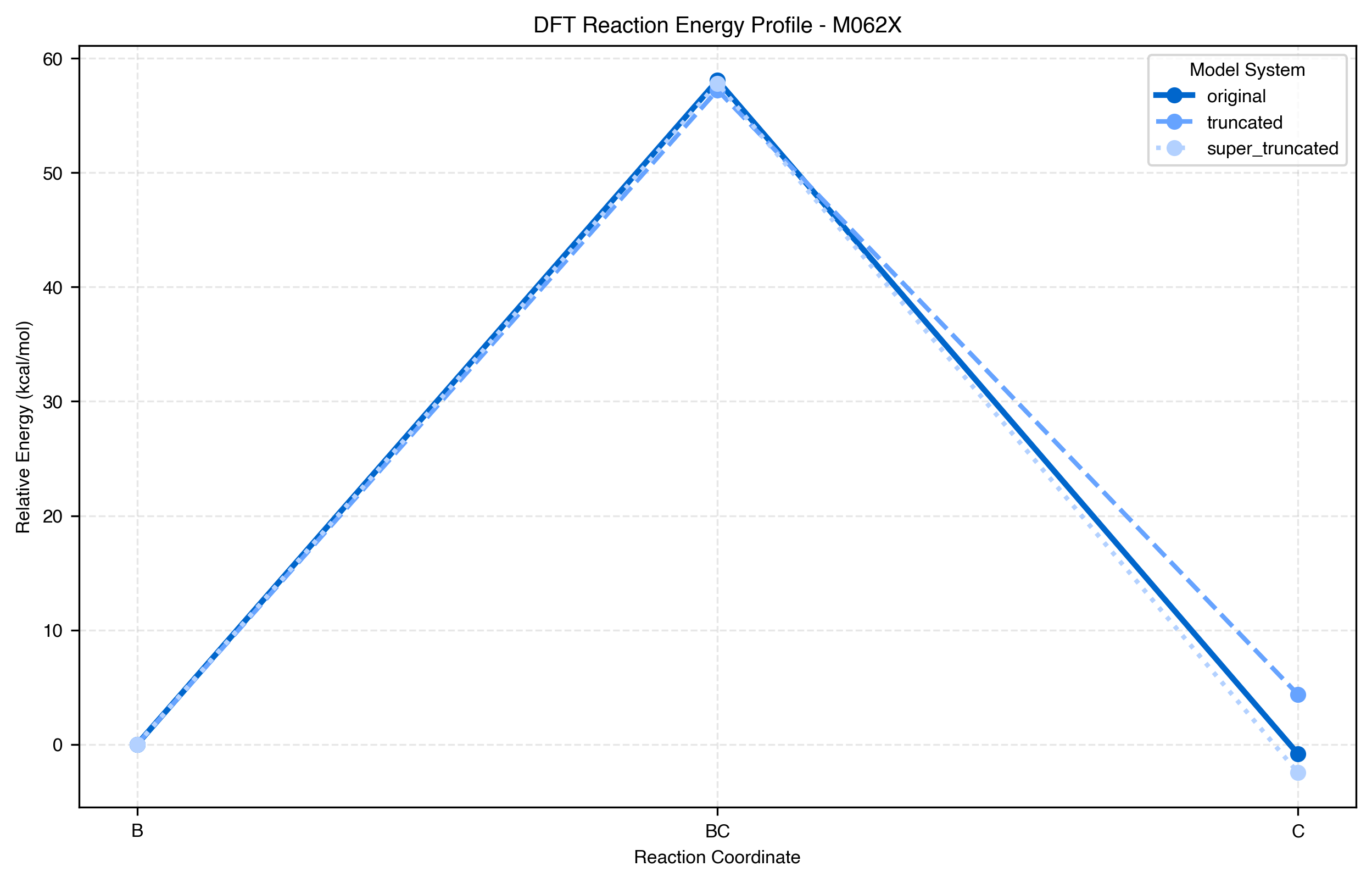}
    \caption{DFT reaction energy profile computed with M06-2X/STO-3G. The energy of complex B is set as the zero-point reference. The agreement between the full, reduced, and minimal models confirms that the reaction pathway is well-described even with substantial truncation.}
    \label{fig:dft_m062x}
\end{figure}

\begin{figure}[h]
    \centering
    \includegraphics[width=0.8\textwidth]{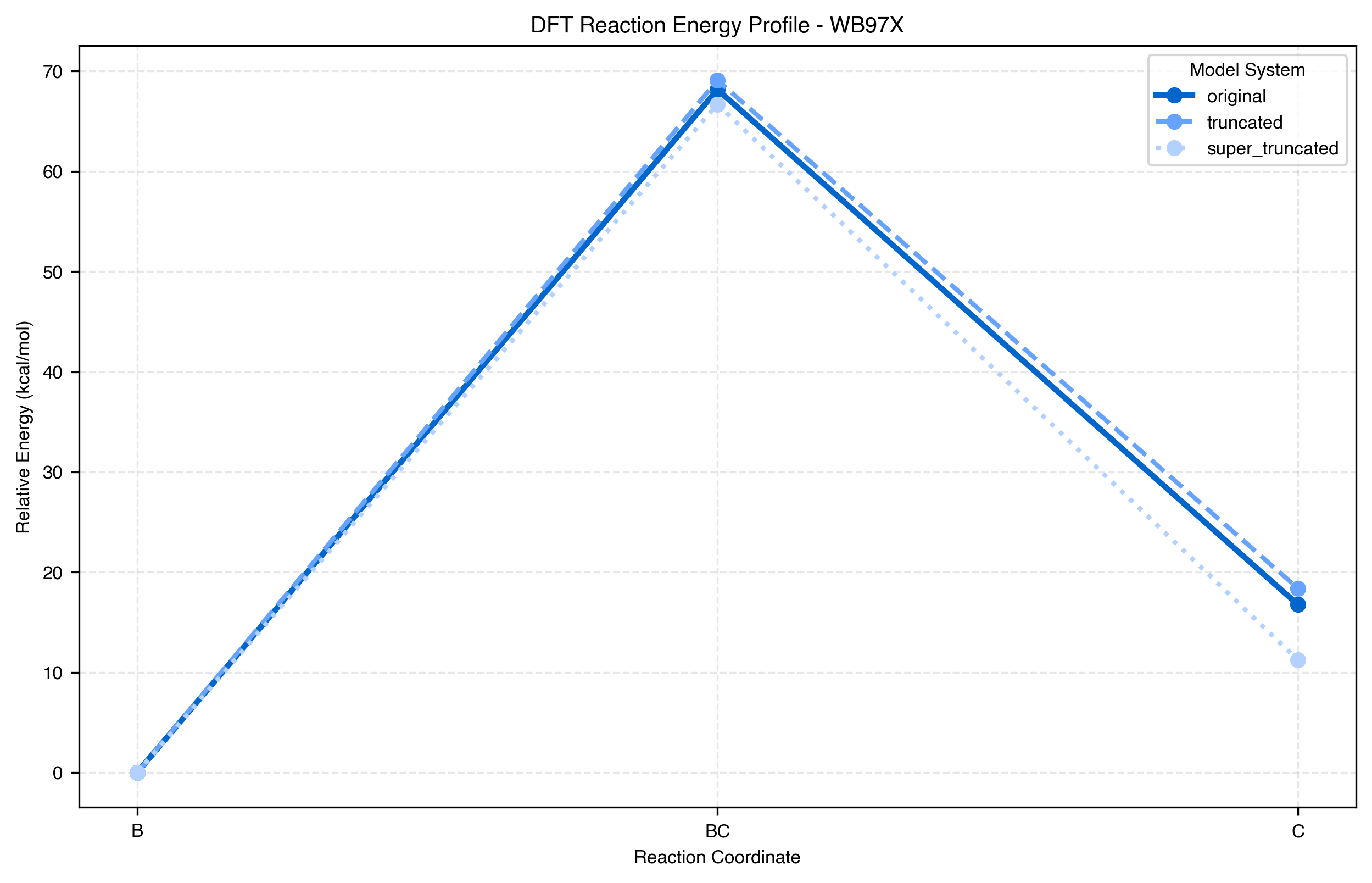}
    \caption{DFT reaction energy profile computed with $\omega$B97X/STO-3G. The energy of complex B is set as the zero-point reference. The energy barrier remains consistent across truncation levels, with minimal deviations in the transition state energy.}
    \label{fig:dft_wb97x}
\end{figure}

\begin{figure}[h]
    \centering
    \includegraphics[width=0.8\textwidth]{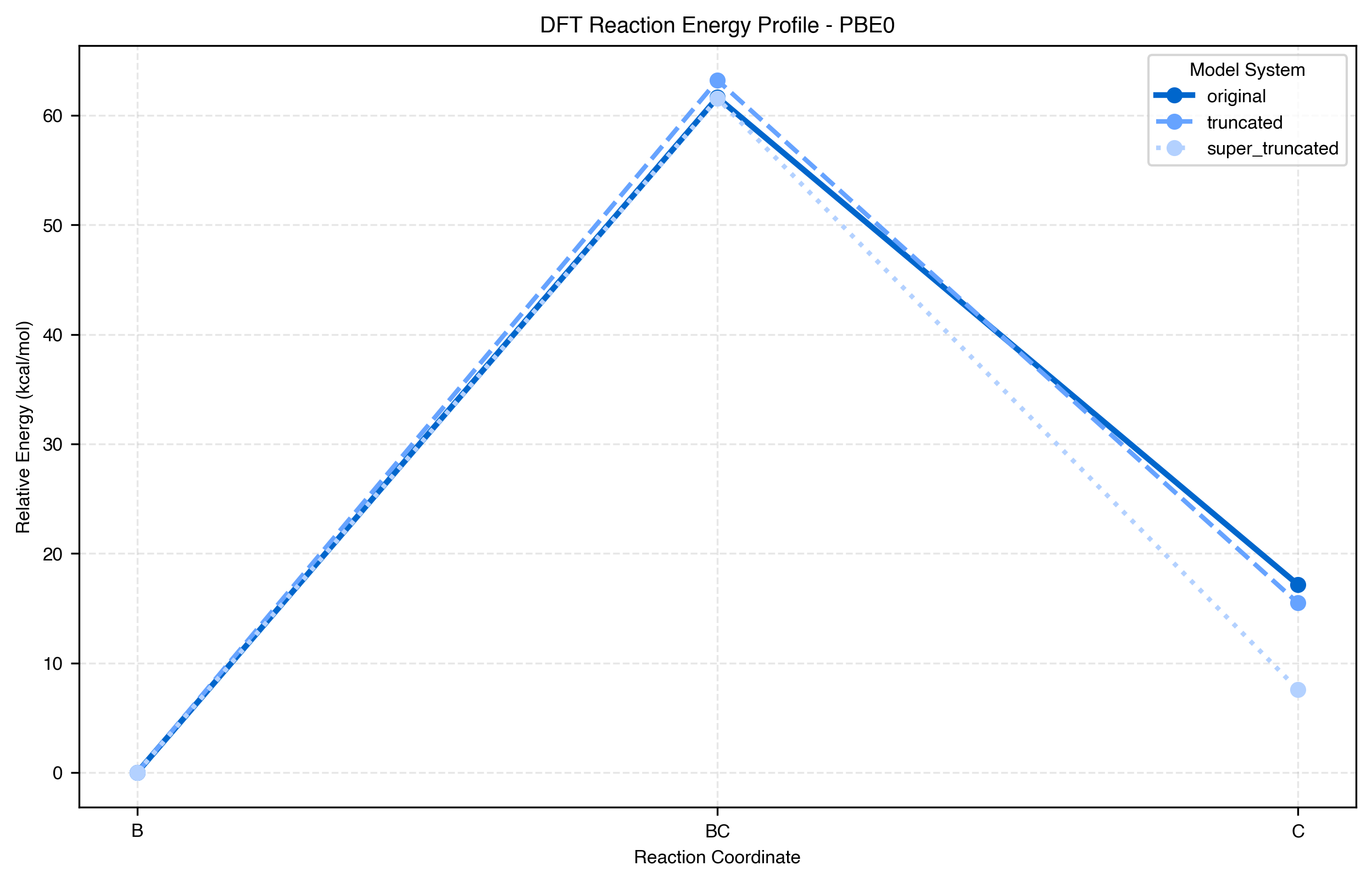}
    \caption{DFT reaction energy profile computed with PBE0/STO-3G. The energy of complex B is set as the zero-point reference. The overall reaction energy trend remains stable, validating the robustness of the truncation approach.}
    \label{fig:dft_pbe0}
\end{figure}

To validate this molecular reduction, we first conducted density functional theory calculations using four distinct functionals ($\omega$B97X,\cite{Chai2008-dr} B3LYP,\cite{Vosko1980-uw,Lee1988-pi,Becke1993-nv,Stephens1994-cm} PBE0,\cite{Perdew1996-hp,Ernzerhof1999-qw,Adamo1999-dx} and M06-2X\cite{Zhao2008-wi}) with the STO-3G basis set \cite{Pietro1983-wq,Pietro1981-tx,Pietro1980-gb,Hehre1969-cj,Hehre1972-cv} (which is sufficient for the purposes of this demonstration, we discuss the scalability of QC-AFQMC with the basis set in Section~\ref{sec:results}). The reaction energy profiles (Supplementary Fig.~\ref{fig:dft_b3lyp}--\ref{fig:dft_pbe0}) show remarkable consistency across all truncation levels we considered. Energetic differences (Supplementary Table~\ref{tab:dft_energies}) were on the order of the expected statistical error margins of AFQMC ($\sim$1--2 kcal/mol).

\subsection{Active space determination}
\label{sec:active_space}

Having verified that our reduced models preserve the energetics across different DFT functionals, we then used tools from quantum information theory to analyze the electronic structure, which provides quantitative indicators of orbital correlation. This framework employs two complementary measures derived from DMRG wave functions: single-orbital entropy and orbital-orbital mutual information. Single-orbital entropy quantifies the quantum entanglement between individual orbitals and their environment -- approaching a theoretical maximum of $\ln(4) \approx 1.4$ at complete entanglement. Previous studies\cite{Stein2016-yg, Stein2019-ih} have established that orbitals exhibiting entropy above 0.14 (10\% of maximum) reliably indicate strong static correlation effects.

\begin{figure*}[ht]
    \centering
    \begin{tabular}{@{}c@{\hspace{2mm}}ccc@{}}
        & \multicolumn{3}{c}{\textbf{Reaction Coordinate}} \\[2mm]
        & B (reactant) & BC (transition state) & C (product) \\[2mm]
        \rotatebox{90}{\parbox{4cm}{\centering \textbf{Original}\\ (77 atoms)}} &
        \includegraphics[width=0.28\textwidth]{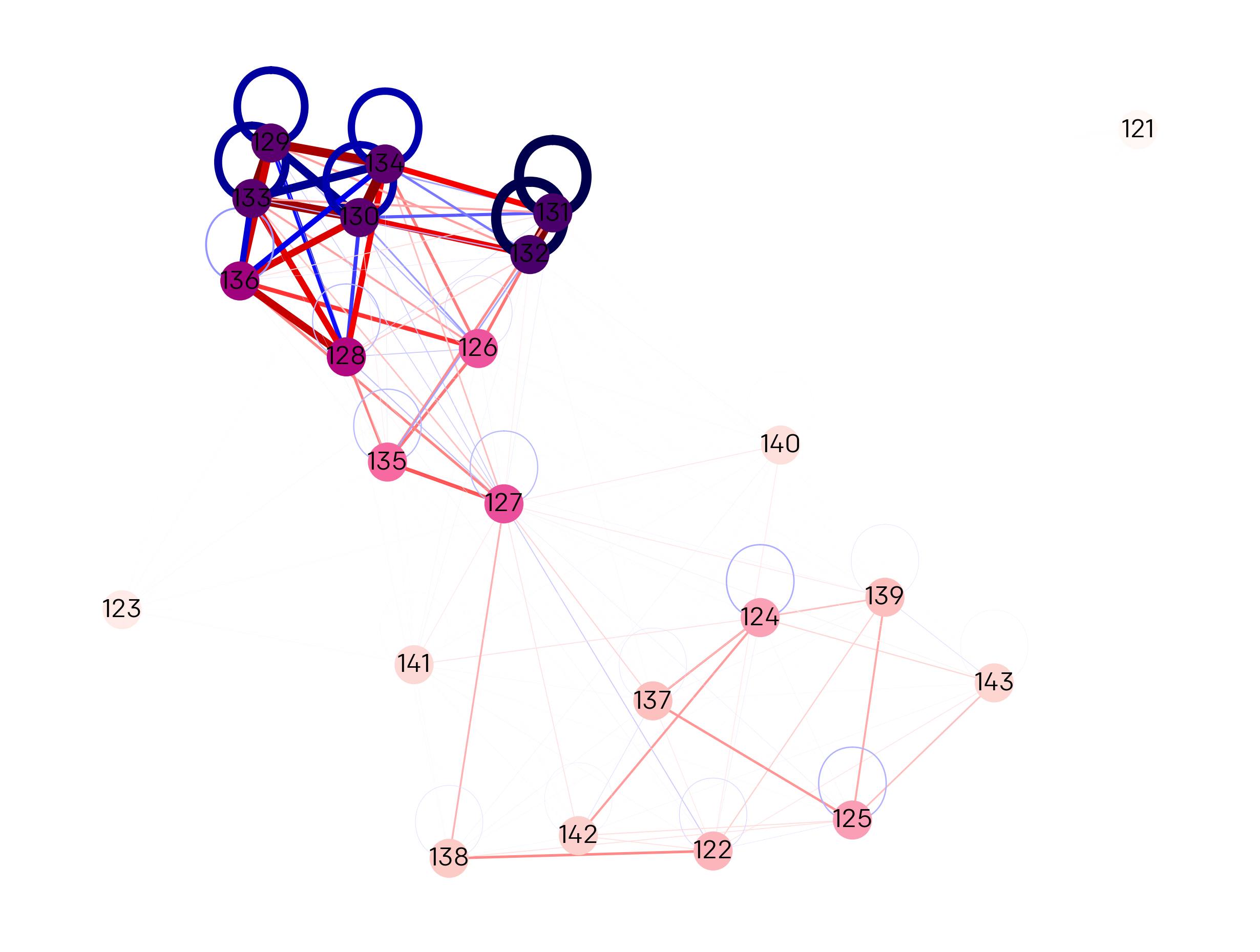} &
        \includegraphics[width=0.28\textwidth]{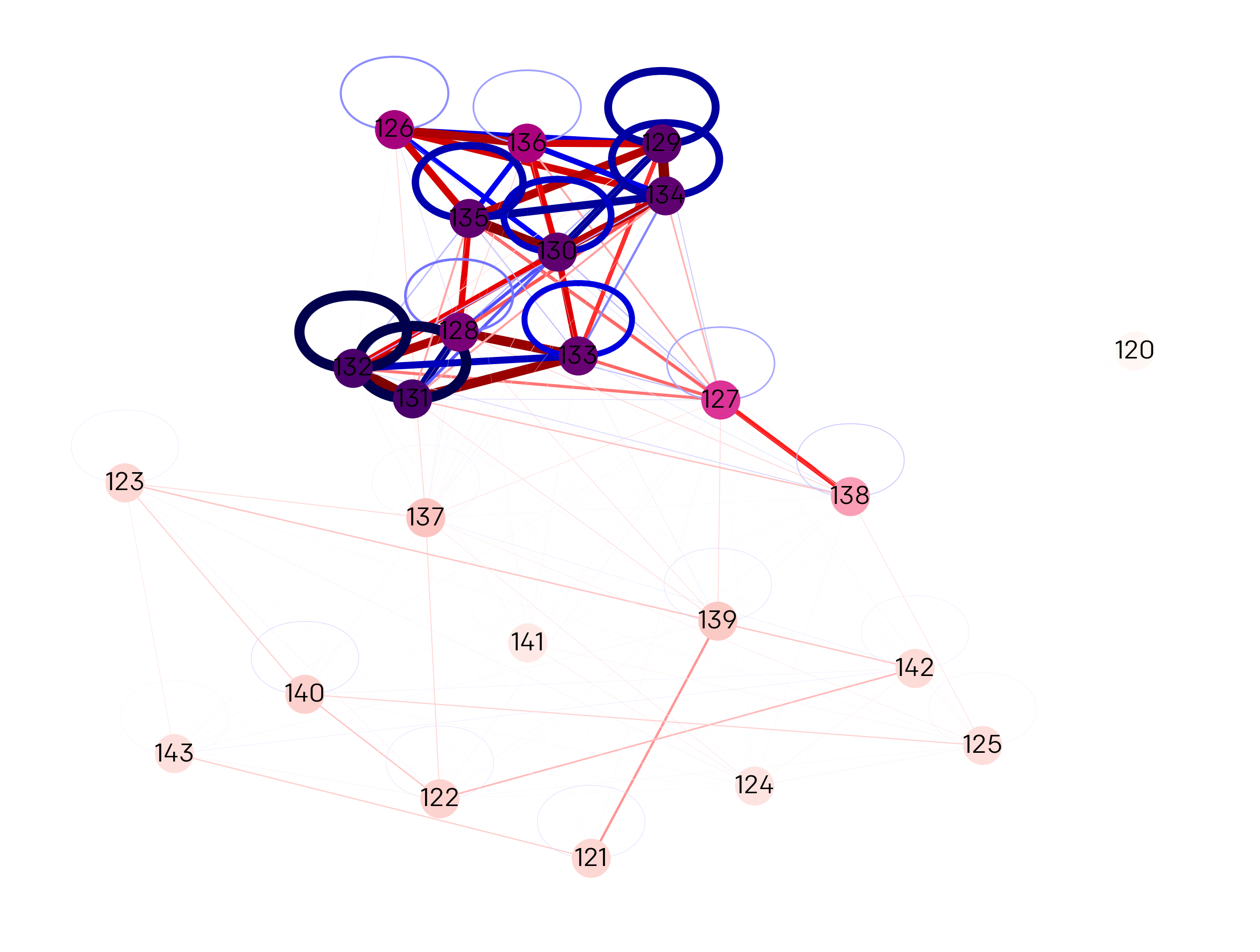} &
        \includegraphics[width=0.28\textwidth]{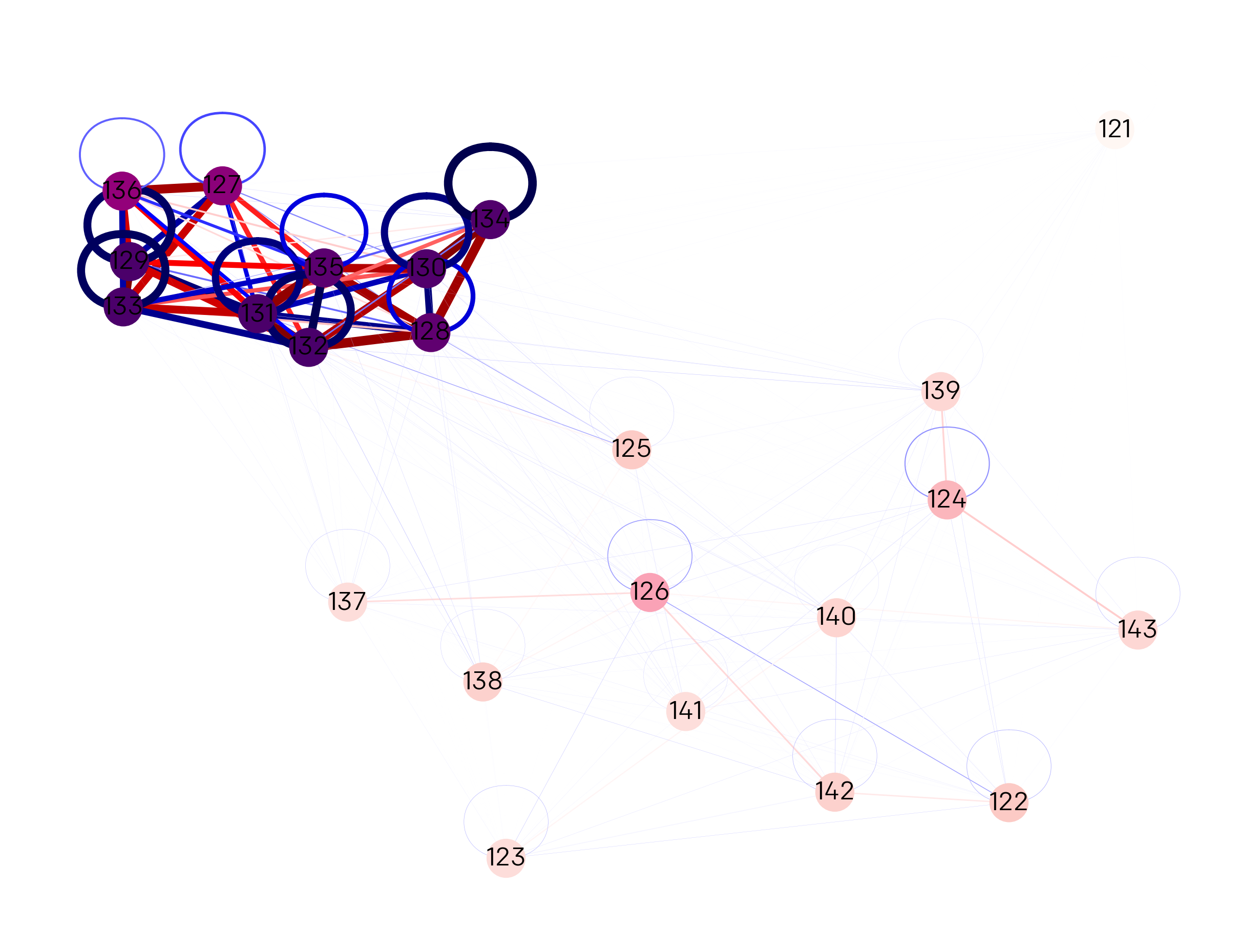} \\[2mm]
        \rotatebox{90}{\parbox{4cm}{\centering \textbf{Reduced}\\ (41 atoms)}} &
        \includegraphics[width=0.28\textwidth]{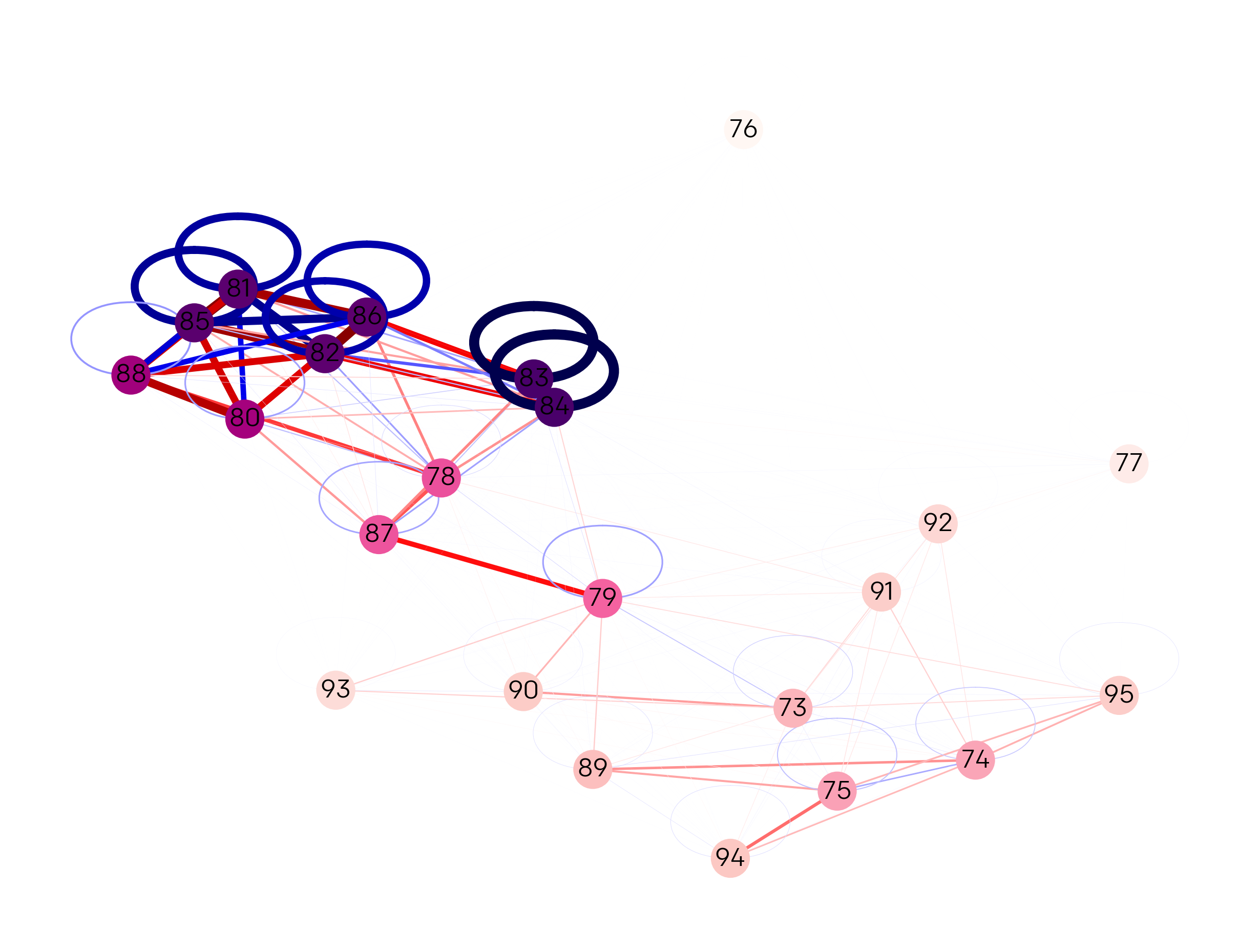} &
        \includegraphics[width=0.28\textwidth]{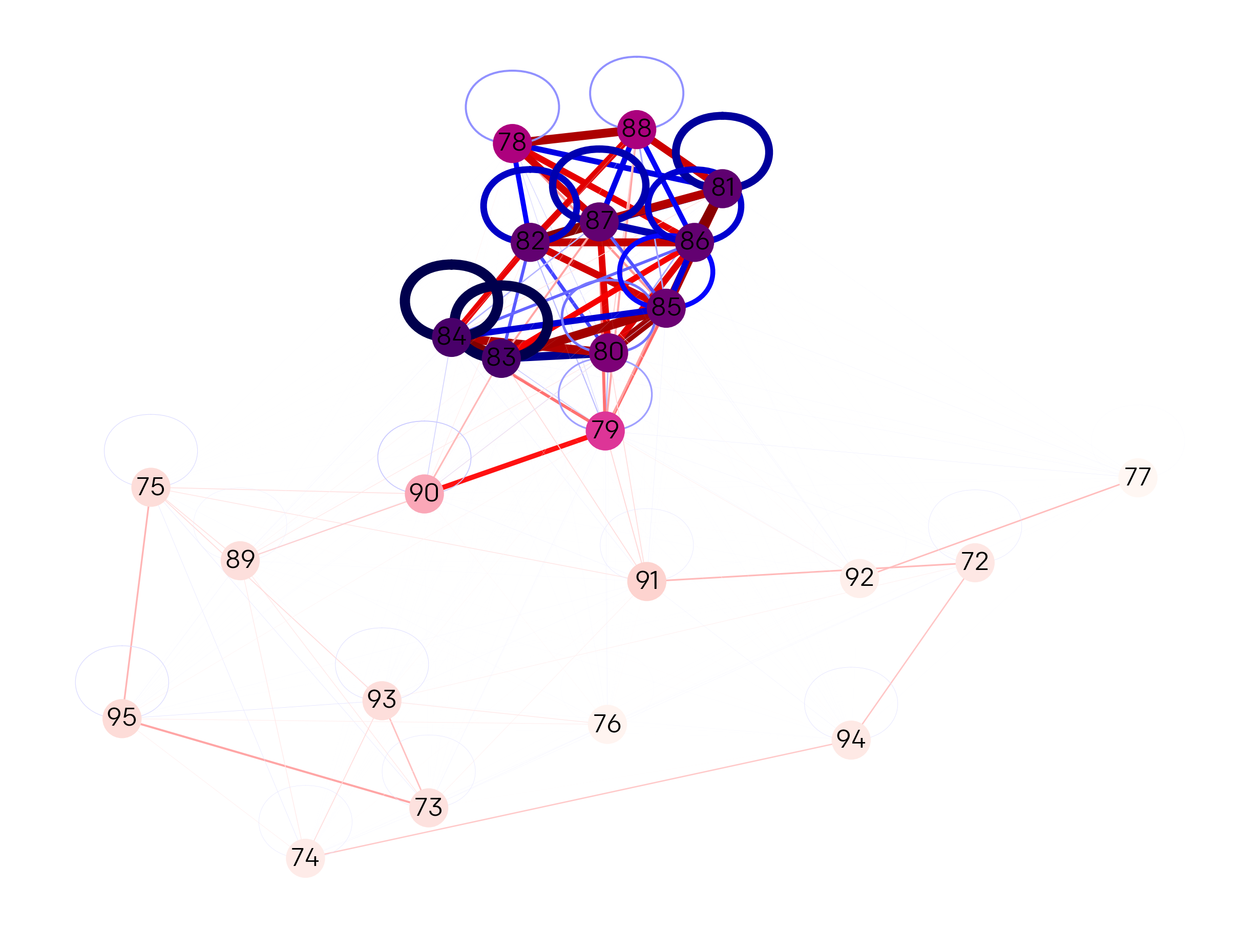} &
        \includegraphics[width=0.28\textwidth]{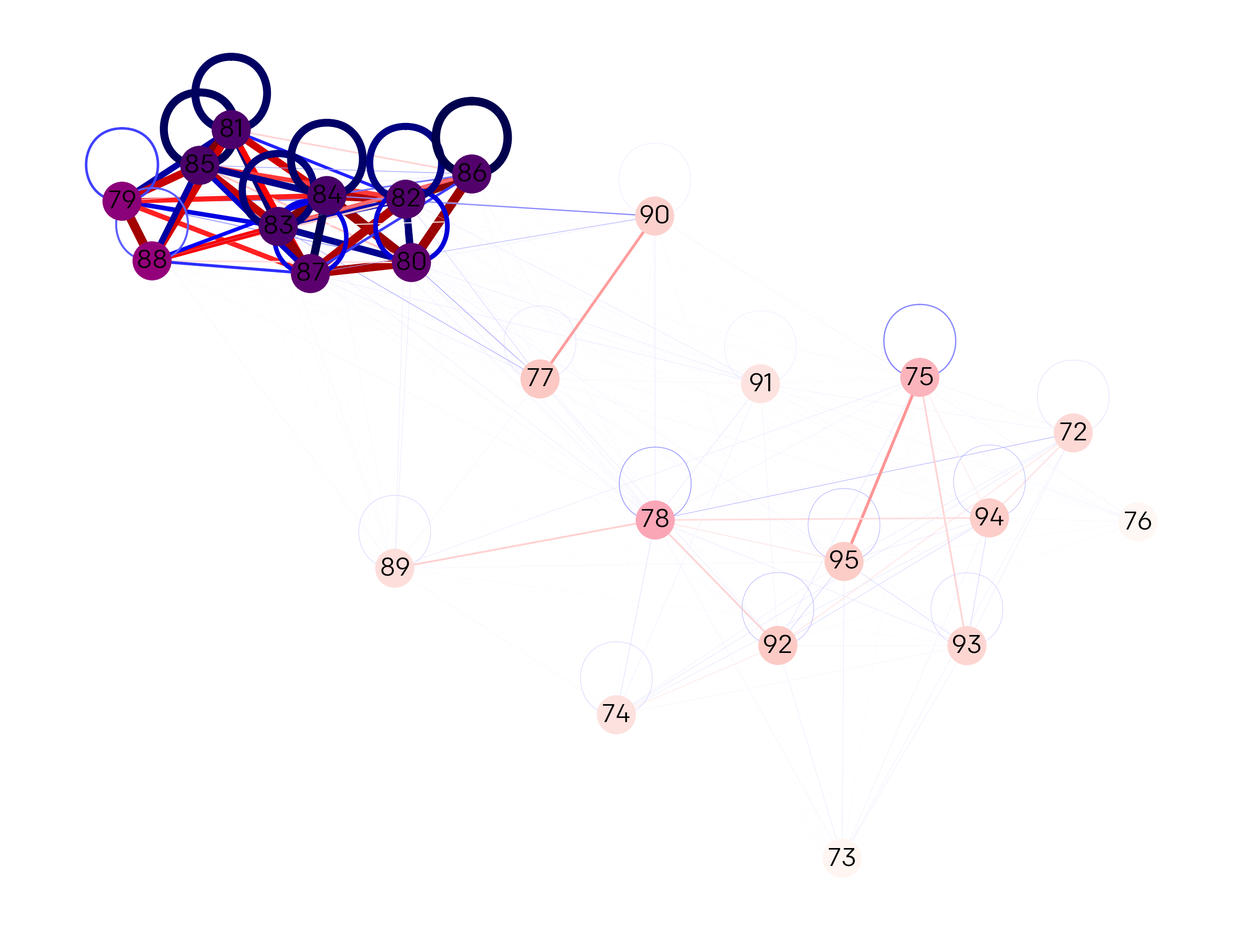} \\[2mm]
        \rotatebox{90}{\parbox{4cm}{\centering \textbf{Minimal}\\ (34 atoms)}} &
        \includegraphics[width=0.28\textwidth]{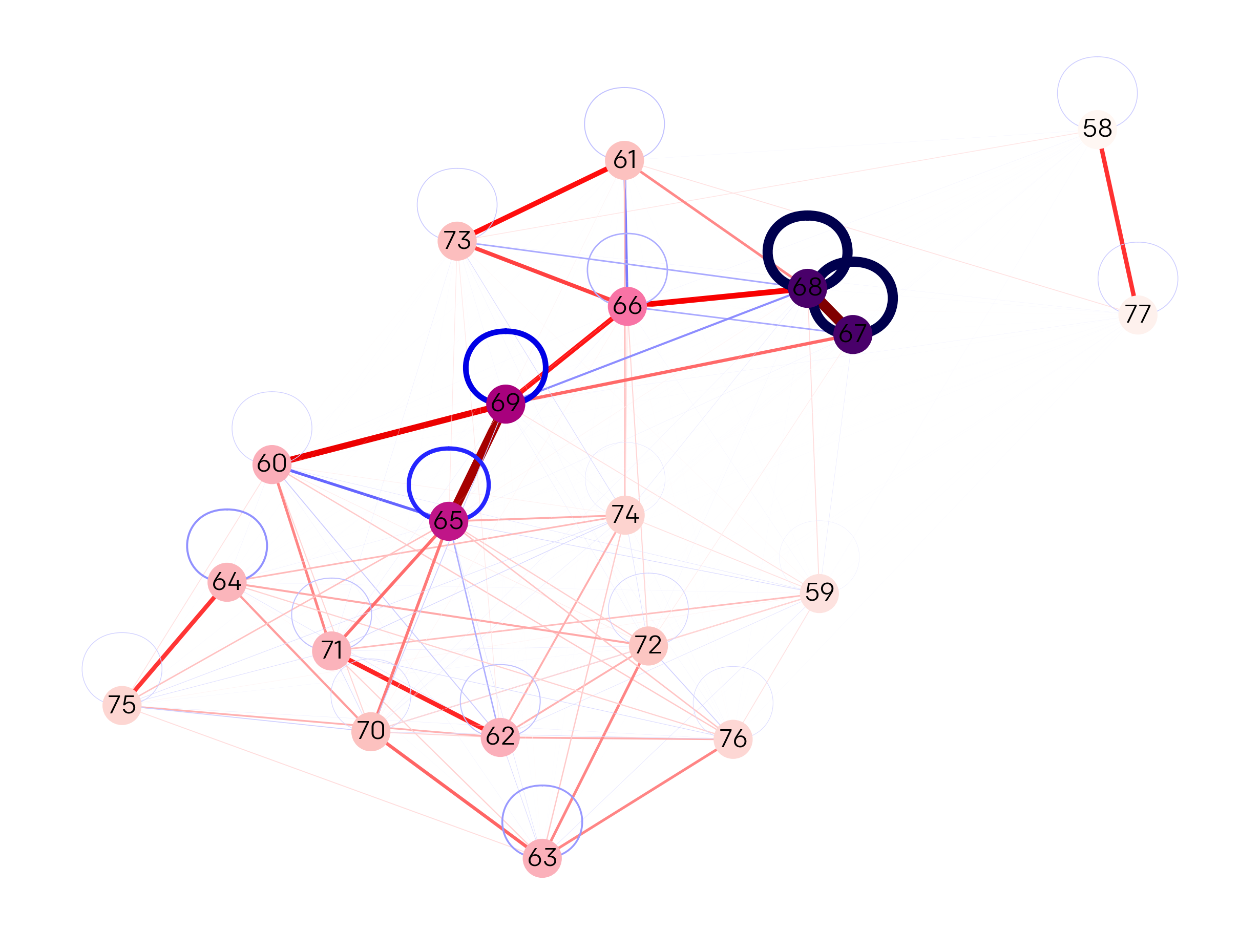} &
        \includegraphics[width=0.28\textwidth]{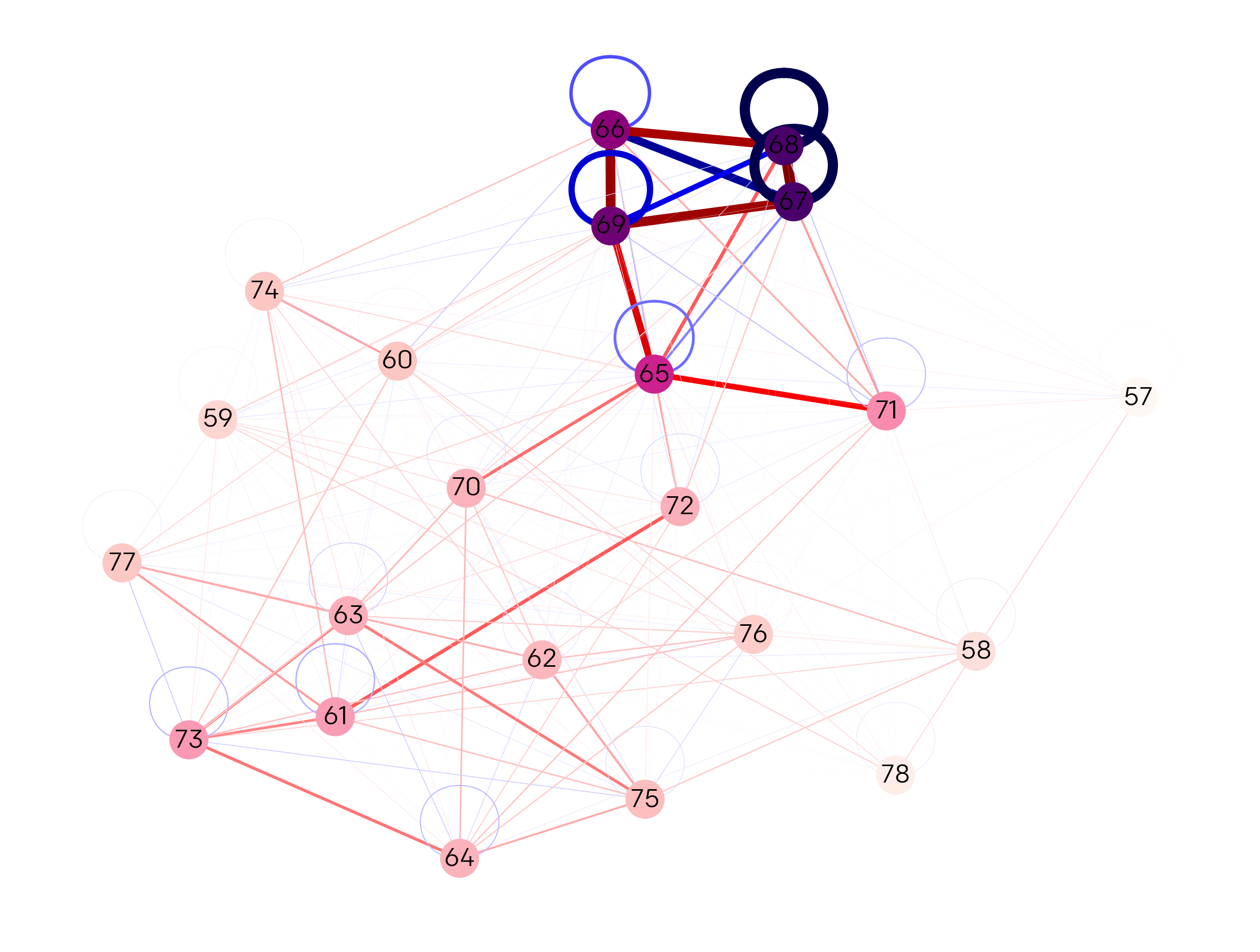} &
        \includegraphics[width=0.28\textwidth]{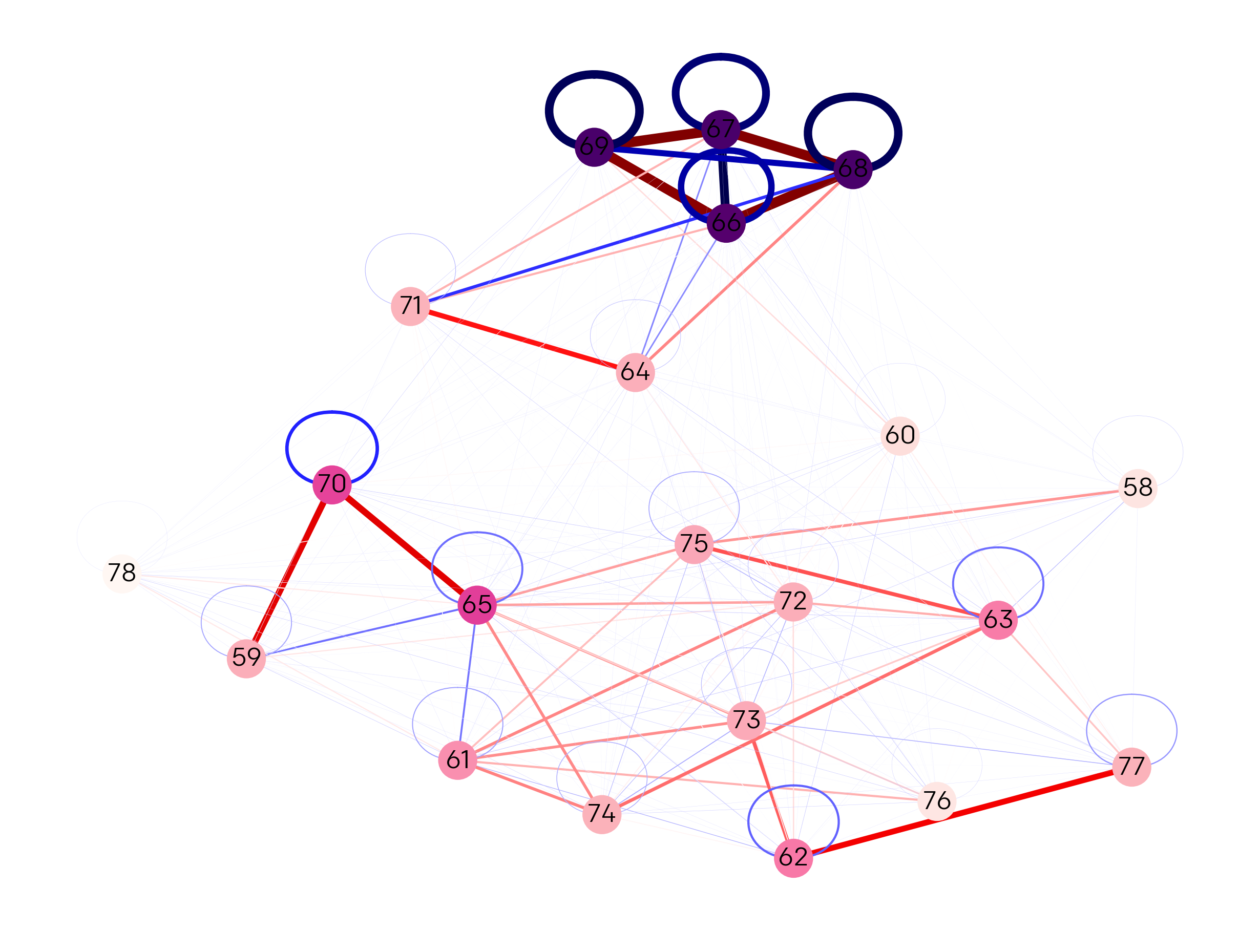}
    \end{tabular}
    \caption{Mutual information networks across truncation levels and reaction coordinates. Node intensity indicates single-orbital entropy (orbital entanglement), while connection width represents two-orbital mutual information from the density cumulant matrix. The largest active space cluster of approximately (12,12) in the original transition state (BC) confirms the necessity of strongly correlated multireference treatment. The natural clustering of these graphs provides complementary validation for single-orbital entropy-based active space selection. QC-AFQMC with virtual correlation energy maintains resilience to active space truncation effects when core entangled orbitals are captured by a multireference wavefunction, such as a VQE ansatz.}
    \label{fig:entanglement_networks}
\end{figure*}

Applying this analysis to the molecules in Fig.~\ref{fig:molecular_structures} revealed striking patterns. As one could see in Fig.~\ref{fig:truncated_entropy_reduced_main} (additional plots available in the Supplementary Information), for the
reduced systems, single-orbital entropy profiles across the reaction coordinate (initial complex B, transition state [B-C]$^\ddagger$, and product C) identified a pronounced cluster of frontier orbitals (orbital numbers 80--87, shown in Fig.~\ref{fig:truncated_entropy_reduced_main}) maintaining high entanglement throughout. These orbitals also exhibit strong mutual correlation patterns in their network structure (Supplementary Figure \ref{fig:entanglement_networks}), with the profiles remaining consistent across all truncation levels—providing independent validation of our reduction strategy.

\begin{figure}[tb!]
    \centering
    \includegraphics[width=0.8\textwidth]{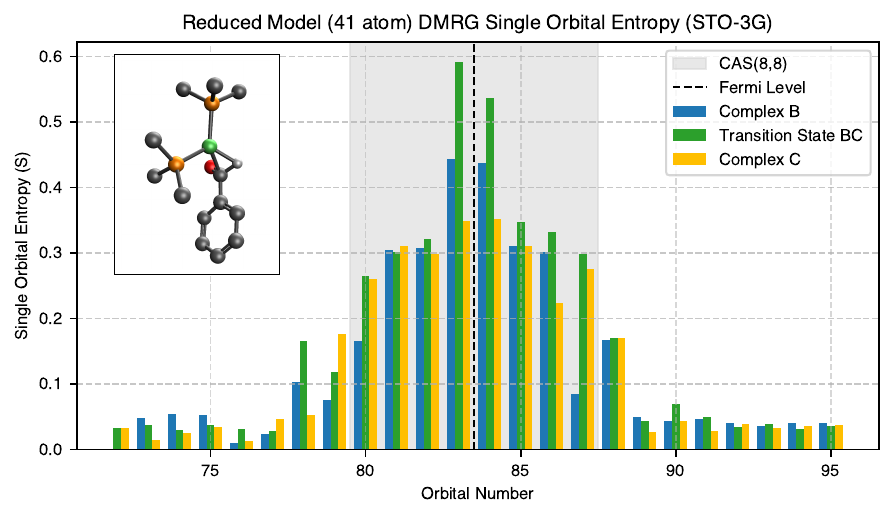}
    \caption{Single Orbital Entropy for the reduced model (41 atoms). The major entangled orbitals remain largely unchanged, validating the truncation approach.}
    \label{fig:truncated_entropy_reduced_main}
\end{figure}

The correlation network analysis, derived from two-electron reduced density matrix elements (Supplementary Information Fig.~\ref{fig:entanglement_networks}), quantifies these orbital-orbital interactions through edge weights representing correlation strengths. The resulting network topology naturally suggests a hierarchy of active spaces based on established metrics\cite{Stein2016-yg}: from a primary (12e,11o) space capturing orbitals with entropy $S > 0.1$, through an intermediate (8,8) space for orbitals exceeding $S > 0.2$, to a minimal (6,6) space for the most strongly correlated orbitals ($S > 0.3$). From this hierarchy, we selected the (8,8) active space, shown in Fig.~\ref{fig:truncated_entropy_reduced_main}, as it provides a balanced description of the key correlations for the demonstrations in this study. Additional correlations are captured by the virtual correlation energy procedure, as discussed in Sec.~\ref{sec:vce}.

These systematically constructed active spaces serve as trial states for QC-AFQMC calculations, with residual dynamical correlation handled efficiently through the VCE technique\cite{Huggins2022-lh}. This balanced treatment proves particularly valuable for nickel-catalyzed \ce{C-H} activation reactions, where accurate correlation treatment directly impacts mechanistic understanding. The integration of quantum information metrics with QC-AFQMC methodology thus establishes a robust framework for investigating electronic structure reorganization in transition metal catalysis, especially in cases where traditional single-reference methods prove inadequate.

\subsection{VQE trial state}
In our study, we use the VQE algorithm with the upCCD ansatz as the trial state for QC-AFQMC. The upCCD ansatz is 
\begin{equation}
    \label{eqn:upccd}
    \ket{\Psi}=e^{T-T^\dagger}\ket{\Phi_0}, \quad T=\sum_{ia}{t_i^aa^\dagger_{a\alpha}a^\dagger_{a\beta}a_{i\beta}a_{i\alpha}},
\end{equation}
in which $\ket{\Phi_0}$ is the HF state. $a^\dagger_{p\alpha}(a^\dagger_{p\beta})$ and 
$a_{p\alpha}(a_{p\beta})$ are the fermionic creation and annihilation operators in the $p^{th}$ 
spin up (down) orbital. 

As we have shown before\cite{Zhao22_60}, the upCCD ansatz is one of the most efficient 
VQE ansatz. It could be compiled with shallow circuits that contain only $O(N^2)$ entangling 
gates, and only a constant number of measurements are needed to compute energy. In this work, 
we obtain the optimal parameters of the upCCD ansatz by minimizing the energy with an ideal
quantum simulator using the COBYLA optimizer implemented in CUDA-Q. For such a small system, the
VQE simulation only takes a few minutes. However, we also note here that for a large scale 
system, simulating VQE with an ideal simulator becomes infeasible, and parameter optimizations on
quantum computers is also very challenging. Fortunately, there are methods\cite{Moreno24_05068, Hirsbrunner24_1538} to obtain
the optimal or close-to-optimal ansatz parameters classically without doing parameter optimizations 
on quantum computers. For upCCD, the most straightforward way is to use the optimal parameters of 
the classical pCCD ansatz. Solving optimal parameters of pCCD only scales cubically with the 
system size, which makes it a scalable way of preparing upCCD trial states for QC-AFQMC. 

\subsection{Execution of matchgate circuits on QPU via CUDA-Q}
\label{sec:execution}

This study used CUDA-Q which is an open source development platform built for accelerated quantum supercomputing developed by NVIDIA \cite{cudaq}. It streamlines the creation of hybrid applications and promotes both productivity and scalability. By offering a unified programming model in Python and C++ for GPUs, QPUs, and CPUs, CUDA-Q enables seamless integration of classical and quantum resources within a single application, ensuring optimal performance and efficiency.

CUDA-Q introduces the concept of a "quantum kernel" to distinguish between host and quantum device code, with each kernel specifying a target for compilation and execution. The platform also includes the NVQ++ compiler, which supports split compilation by lowering quantum kernels into multi-level intermediate representation (MLIR) and quantum intermediate representation (QIR). This approach ensures tight coupling between classical and quantum operations, facilitating accelerated execution of large-scale quantum workloads.

CUDA-Q is qubit-agnostic, allowing users to target QPUs of various modalities such as trapped ion, photonic, superconducting, neutral atoms and other architectures as their hardware matures. Its circuit simulation engine leverages NVIDIA's cuQuantum SDK, which supports statevector, density matrix, and tensor network simulations \cite{cuquantum}. These simulations of user-defined quantum kernels can be executed on an array of backends, including CPUs, GPUs, and multinode GPU clusters scaling to supercomputers. Users can seamlessly switch between simulation and execution on quantum hardware. All executables have parallelisation built into their functionality, hence, execution of quantum kernels can be parallelised amongst multi-GPU architectures today and multi-QPU architectures in the future. Additionally, GPU-accelerated quantum dynamics simulations provides accurate modeling of open quantum systems, essential for designing, characterizing, optimizing, and scaling the next-generation of QPUs.

CUDA-Q also offers specialised libraries for quantum error correction and algorithms \cite{cudaqx}, interoperability with the broader CUDA ecosystem and cloud-based hardware access via services such as Amazon Braket and IonQ. This allows developers to leverage GPU acceleration for hybrid quantum-classical workflows, utilise optimised kernels for algorithms like VQE and QAOA, and seamlessly integrate with a wide range of tools for AI, data science, and HPC ultimately simplifying the execution of complex quantum simulations on both cloud-based and on-premise quantum hardware.

Finally, complex algorithmic workflows, such as the QC-AFQMC example, rely on accelerated quantum supercomputing architectures for quantum circuit execution (through CUDA-Q on QPUs) and classical post-processing (via CUDA-based libraries on NVIDIA GPUs). The NVIDIA ecosystem of SDKs ensures these tightly coupled workflows run on their respective accelerated platforms with minimal overhead and seamless data communication, empowering researchers to focus on innovation rather than infrastructure.

\paragraph{Quantum hardware}
The experimental demonstration was performed on the IonQ Forte QPU executed via CUDA-Q on Amazon Braket \cite{Chen2024-co}. In these systems, trapped \ce{^{171}Yb+} ions serve as the qubits, with quantum information encoded in two hyperfine levels of the ground state. Ions are generated via laser ablation and selective ionization before being loaded into a surface linear Paul trap in a compact integrated vacuum package. Qubit states are manipulated by illuminating individual ions with pulses of 355~nm light that drive two-photon Raman transitions, thereby enabling the implementation of arbitrary single-qubit rotations and ZZ-type entangling two-qubit gates.

IonQ Forte integrates acousto-optic deflectors (AOD) that allow for independent steering of each laser beam to its respective ion, substantially reducing beam alignment errors across the ion chain. This optical architecture, combined with a robust control system and control software suite that automates calibration and optimizes gate execution, has enabled the realization of larger qubit registers with enhanced gate fidelities. Consequently, IonQ Forte establishes a scalable and high-fidelity platform for quantum information processing, marking a notable progression in trapped-ion QPU technology.

Several specific improvements were made to the control software and firmware suite to maximize the circuit execution rate (\textit{throughput}) of circuits for this workload. This performance tuning took advantage of shadow circuits' specific and somewhat uncommon structure. Notably, each shadow circuit is pregenerated, only run for a single shot, and consists of a common trial state, followed by classical shadows sourced from a common pool. This allowed for aggressive caching of waveforms and other real-time instructions that were common among the shadow circuits within the real-time control subsystem, as well enhanced pipelining and parallelization efforts up and down the control stack to ensure the now-much-faster circuit execution was never blocked by classical overhead. Specific results of this performance tuning are discussed in Section ~\ref{subsec:time_to_solution}. 

\paragraph{Error mitigation techniques}
\label{sec:error_mitigation}
Classical shadow techniques enjoy the benefit of responding predictably to Markovian, invertible, gate-independent quantum error channels, in the sense that expectation values are simply rescaled~\cite{chen2021robust,koh2022classical,zhao2024group}. For computing overlaps, as done in QC-AFQMC, the rescaling of expectation values cancels out, making the technique robust to these simple errors~\cite{Huggins2022-lh}. This robustness does not hold, however, for cases where the error channels are context dependent or non-invertible~\cite{brieger2025stability}. On our platform, one particularly important such channel is a spontaneous emission event that can happen during the Raman ZZ gates. This error channel results in the ion entering a state where it is insensitive to all further gates~\cite{ozeri2007errors,moore2023photon}, meaning that the placement of the gate where the error occurs drastically changes its impact. Further, given that the error represents a complete loss of information, the resulting shadow channel is non-invertible.

While this type of spontaneous emission error has the potential to inject non-trivial bias into a shadow evaluation of overlaps, it is fairly easy to detect and remove via post-selection. To that end, we use leakage error detection gadgets~\cite{stricker2020} attached at the end of the circuit. Each gadget requires two ZZ gates and an ancillary qubit. To reduce the overhead, we attach the gadgets on half of the qubits, alternating even and odd, thus requiring 8 additional qubits. If any of the ancilla is flagged, the corresponding measurement is discarded.

\subsection{GPU-accelerated post-processing}

We implement key linear algebra routines in the QC-AFQMC
post-processing with NVIDIA GPUs on AWS ParallelCluster. The construction 
of the $\textbf{A}$ and $\textbf{B}$ matrices defined in Equation \ref{eqn:a_mat} and \ref{eqn:b_mat}, along with 
the derivative of $\textbf{A}$ used in force bias 
and local energy evaluations, and matrix and vector products
needed to compute Pfaffian derivatives in Eq.~\ref{eqn:Pfaffian_der} and \ref{eqn:Pfaffian_2nd_der}, are 
implemented with NVIDIA's cuBLAS and cuTENSOR linear algebra packages. cuBLAS allows for highly efficient matrix multiplications 
on GPUs. cuTENSOR extends those capabilities to multidimensional arrays. The polynomial fitting needed to compute overlaps 
and their derivatives are implemented with NVIDIA's cuSOLVER
package, which allows for efficient linear system solutions 
on GPUs.

\section{Results}
\label{sec:results}

\subsection{Particle number and effect of hardware noise}

We first prepared the trial state of B, {[B-C]}$^\ddagger$, and C (Fig.~\ref{fig:molecular_structures}), which represent the reactant, transition state, and product, respectively, along the reaction coordinate for the oxidative addition of aldehyde to \ce{Ni(0)} in the nickel-catalyzed Suzuki--Miyaura cross-coupling. For each molecule, we obtained a restricted Hartree--Fock solution followed by VQE parameter optimization with the unitary pair coupled cluster double (upCCD) ansatz in an active space. We followed the procedure described in Section~\ref{sec:model_reaction} to define an (8,8) active space for all three systems, which corresponds to selecting orbitals with single orbital entropy $S>0.2$.

\begin{table*}[tbh!]
    \centering
    \caption{Particle number of the trial state evaluated from matchgate shadow measurements computed for the three molecular structures. There are 8 active space electrons in the trial state, which is the ground truth value.}
    \label{tab:particle_number}
    \renewcommand{\arraystretch}{1.2}
    \begin{tabular}{lcccc}
\toprule
Molecule & \multicolumn{2}{c}{Ideal simulator} & \multicolumn{2}{c}{Forte QPU} \\
 & Shadows & Particle number & Shadows & Particle number \\
\midrule
B & 58,482 & 7.998 & 60,365 & 10.663 \\
{[B-C]}$^\ddagger$ & 58,482 & 8.086 & 58,375 & 10.598 \\
C & 58,482 & 7.932 & 61,118 & 10.514 \\
\bottomrule
\end{tabular}
\end{table*}

To sample the trial state, we set up matchgate shadow tomography on the ideal simulator with 58,482 randomly generated shadows. On Forte QPU, we used 60,365, 58,375, and 61,118 matchgate shadows for system B, {[B-C]}$^\ddagger$, and C, respectively, following the error mitigation procedure described in Section~\ref{sec:error_mitigation}. The difference in the number of shadows is due to a slightly different rate of rejection at the error mitigation step.

We first assessed the quality of the collected matchgate shadow measurements by evaluating the total number of particles in the trial state by taking the trace of the one-particle reduced density matrix (RDM), which was computed from the measured matchgates. The details of how RDMs are computed could be found in the Supplementary Information. Because we used the (8,8) active space in our models, the ground truth number of electrons is known (8), making it a convenient metric to validate the measurements. The results from the ideal simulator and the Forte QPU for all three molecules are shown in Table~\ref{tab:particle_number}. The particle numbers measured from the ideal simulator of all three molecules are close to 8, as expected, with the remaining discrepancy with the integer value due to the finite sample of matchgates. The measured particle numbers on Forte QPU are around 10.5, which deviate from the ideal value due to hardware noise. The amount of difference between the particle number measured on QPU and the ideal simulator could also serve as an indicator of the noise level.

\subsection{Reaction energy}

\begin{figure}[tbh!]
    \centering
    \includegraphics[width=\textwidth]{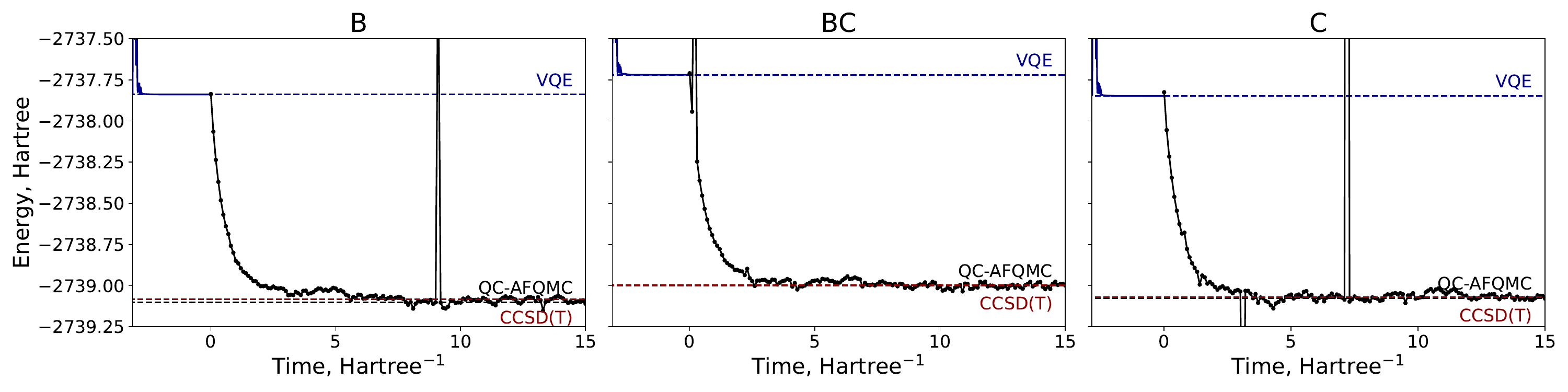}\\
    \includegraphics[width=\textwidth]{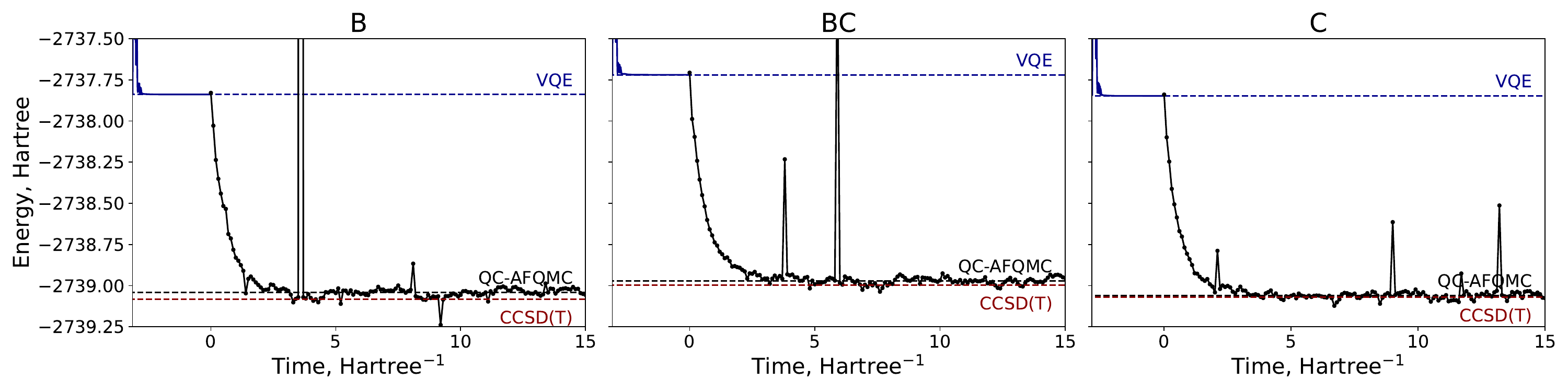}
     \caption{Convergence of the QC-AFQMC energy vs. imaginary time using matchgate shadow measurements collected using the ideal simulator (top) and Forte QPU (bottom). The dashed lines indicate the VQE/upCCD (blue) and CCSD(T) (red) reference energies, as well as the converged QC-AFQMC energy (black).}
    \label{fig:energy_vs_imag_time_2}
\end{figure}

We then applied QC-AFQMC to model the ground state of the three molecules using both simulated and QPU measurements. The imaginary time propagation of AFQMC, with force-bias and energies computed using the algorithm described in Section \ref{subsec:improvement}, ran with 1,280 walkers and 1,500 time steps of 0.01~a.u. for a total of 15~a.u. imaginary time. We plot the energy as a function of the imaginary time in Fig.~\ref{fig:energy_vs_imag_time_2}, and the comparison of the final converged QC-AFQMC energies, including their confidence intervals, for both ideally simulated matchgates and those collected from Forte QPU, with HF, VQE, and CCSD(T) is shown in Table~\ref{tab:energies_compare2}.

\begin{table*}[tbh!]
    \centering
    \caption{Total electronic energies and correlation energies computed for the three molecular structures. All energies are in Hartrees.}
    \label{tab:energies_compare2}
    \renewcommand{\arraystretch}{1.2}
    \begin{tabular}{lccc}
\toprule
Method & B & {[B-C]}$^\ddagger$ & C \\
\midrule
 RHF & $-2737.810\,245$ & $-2737.686\,838$ & $-2737.822\,695$ \\
 CCSD\footnotemark[1] & $-2739.045\,232$ & $-2738.952\,254$ & $-2739.040\,565$ \\
 CCSD(T)\footnotemark[1] & $-2739.081\,251$ & $-2738.996\,283$ & $-2739.068\,637$ \\
 VQE/upCCD & $-2737.839\,079$ & $-2737.719\,325$ & $-2737.847\,562$ \\
\midrule
 QC-AFQMC (Ideal simulator) \\
 \hspace{2em} Initial & $-2737.836$ & $ -2737.710$ & $-2737.825$ \\
 \hspace{2em} Converged & $-2739.085(4)$ & $-2738.994(4)$ & $-2739.064(4)$ \\
\midrule
 QC-AFQMC (Forte QPU) \\
 \hspace{2em} Initial & $-2737.829$ & $-2737.707$ & $-2737.839$ \\
 \hspace{2em} Converged & $-2739.040(4)$ & $-2738.972(3)$ & $-2739.059(3)$ \\
\bottomrule
\end{tabular}
    \footnotetext[1]{CCSD and CCSD(T) energies computed with 29 frozen orbitals (frozen-core approximation).}
\end{table*}

For all three molecules, the initial energies closely match the VQE results, which is expected since the initial step represents a sample of the trial state. As imaginary time evolution proceeds, QC-AFQMC recovers between 1.22 and 1.29 Hartree of correlation energy. This reflects the method's ability to capture correlation effects within the active space through the trial state, while also accounting for electron correlation outside the active space via VCE. As shown in Fig.~\ref{fig:energy_vs_imag_time_2}, we occasionally observe spikes in the measured energy with a magnitude of almost 2 Hartrees above the converged value. We attribute these anomalies to numerical errors in computing VCE. Fortunately, the spikes quickly subside and do not affect overall convergence. For our final energy estimates, we discard the first 50 blocks for equilibration, then remove any remaining outliers (outliers are defined to be points that are at least 200~mHartree higher or lower in energy than its adjacent points) from the next 100 blocks, and use the rest for reblocking analysis to obtain the mean energy and its error bars\cite{Petersen89_461}. 

Unlike the particle number via one-particle RDM, the energy is much more resilient to hardware noise. Because hardware noise causes an overestimate of the number of electrons by more than 2, one might intuitively expect that it could result in errors in the energy on the order of several Hartrees. However, the largest error we observed in energy between the ideal simulator and the QPU is only a few tens of mHartree. This disparity arises from fundamental differences in the mathematical formalisms used to evaluate energies and RDMs. Energy is computed as ratios of overlaps, a structure known to be robust to noise due to error cancellation, as pointed out in Huang~{\it et al.}\cite{Huang2024-hk}. Conversely, individual one-particle RDM elements share a similar formalism as single overlaps, which do not benefit from the same error cancellation mechanism.

\begin{table*}[tbh!]
    \centering
    \caption{Reaction barriers of the oxidative addition reaction estimated with RHF, CCSD(T), and QC-AFQMC. All energies are in kcal/mol.}
    \label{tab:reaction_diagram}
    \renewcommand{\arraystretch}{1.2}
    \begin{tabular}{lcc}
\toprule
Method & B $\rightarrow$ {[B-C]}$^\ddagger$ & C $\rightarrow$ {[B-C]}$^\ddagger$ \\
\midrule
 DFT\footnote{Free energies in solution [SMD (1,4-Dioxane)] evaluated with M06/Def2-TZVPP//$\omega$B97xD/Def2-TZVP(Ni)/Def2-SVP(non-metal).\cite{Guo2019-zc} } & 21.4 & 7.1 \\
 RHF & 77.4 & 85.3 \\
 CCSD(T) & 53.3 & 45.4 \\
 VQE/upCCD & 75.1 & 80.5 \\
 \midrule
 QC-AFQMC (Ideal simulator) & 57(4) & 44(4) \\
 QC-AFQMC (Forte QPU) & 43(3) & 55(3) \\
\bottomrule
\end{tabular}
\end{table*}

Table~\ref{tab:reaction_diagram} shows the reaction barrier computed with various methods. As a reference, we use CCSD(T), which is widely regarded as reliable for systems with moderate correlation and has recently demonstrated good accuracy for spin energetics in mononuclear iron complexes\cite{rado2024}. As can be seen in Table~\ref{tab:reaction_diagram}, among single-reference methods, RHF severely overestimates the reaction barrier, while DFT (which includes the treatment of solvent) exhibits its typical behavior of underestimating the reaction barrier. 

QC-AFQMC based on the ideal matchgate samples produces a reaction barrier profile that closely resembles that from CCSD(T) with reaction energies within 4~kcal/mol from the reference. In contrast, QC-AFQMC based on noisy Forte QPU samples, are 10~kcal/mol off from CCSD(T). Furthermore, in this case the noisy matchgate shadows cause the energy barriers to shift in the opposite directions changing the picture qualitatively: unlike CCSD(T) and ideal QC-AFQMC, which place molecule B lower in energy than C, noisy QC-AFQMC predicts C to be lower in energy. This result highlights the importance of developing novel error mitigation techniques\cite{Miyake24_57} for QC-AFQMC on noisy hardware.

The uncertainty interval in QC-AFQMC reaction barriers (up to $\pm4$~kcal/mol) is due to the statistical sampling of AFQMC. It can be improved by increasing the number of AFQMC walkers or the length of imaginary time propagation (with a proportional increase in computational effort in post-processing), as well as by enhancing the quality of the trial state.

Using larger basis sets, ideally approaching the complete basis set (CBS) limit, is expected to improve electron correlation treatment and may slightly increase barrier heights. Additionally, environmental effects must be also considered\cite{haya2023,rado2024}, as interactions with solvents and additives can significantly influence spin-state energetics. As the nuclear charge of TM increases, relativistic effects, such as kinematic effects and spin-orbit coupling, become more pronounced. These effects can be addressed with varying levels of accuracy and approximation using, for example, Zeroth-Order Regular Approximation (ZORA)\cite{lent1993}, Douglas-Kroll-Hess (DKH) method\cite{wolf2002}, and relativistic effective core potentials\cite{dolg2002}.

\subsection{Time to solution with Forte QPU}\label{subsec:time_to_solution}

\subsubsection{Quantum processing time}

In the course of collecting matchgate shadow measurements on Forte QPU, we executed 300,983 circuits in total, including a number of trial runs and results that were later discarded in error mitigation post-selection. Those circuits were executed over a period of several weeks, during which this workload was mixed with other shared workloads on the system. Additionally, during certain periods of time, the QPU was performing technical tasks such as calibration and characterization.

\begin{figure}[tb!]
    \centering
    \includegraphics[width=0.48\textwidth]{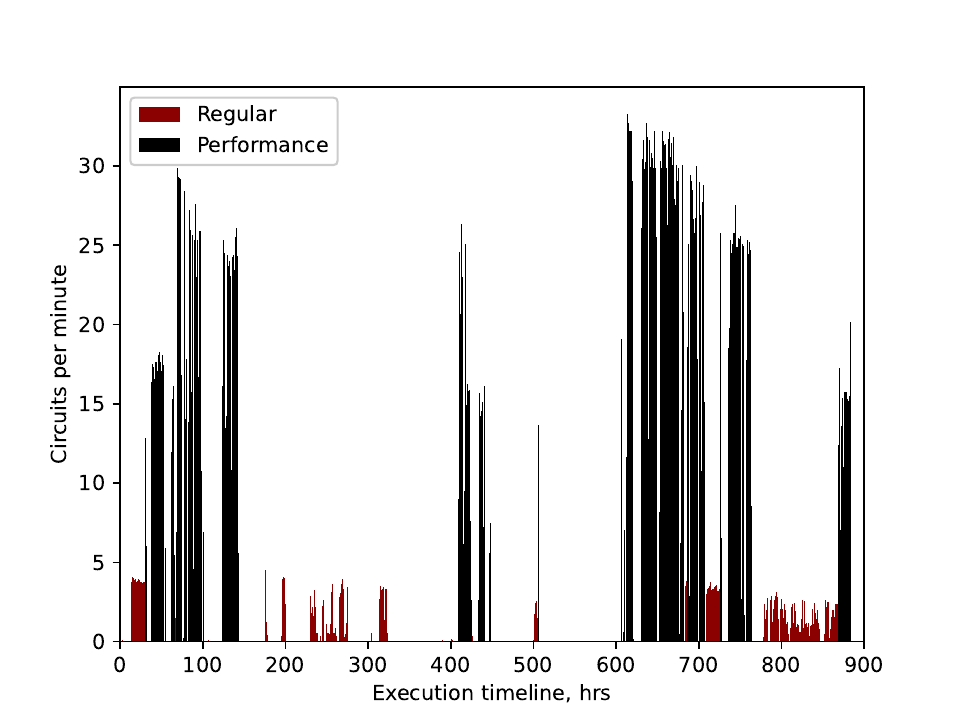}
    \includegraphics[width=0.48\textwidth]{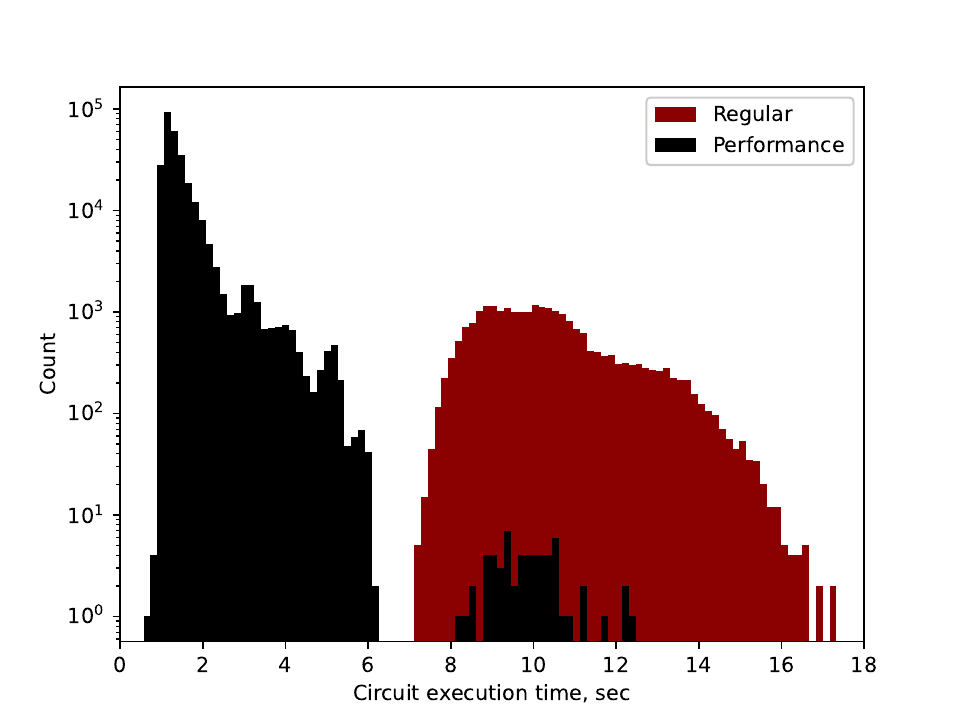}
    \caption{Rates (per minute) of processing matchgate shadow circuits on Forte QPU over the entire execution timeline (left panel).
    Distribution of matchgate shadow circuit execution times (seconds) before and after performance tuning of Forte QPU for the workload (right panel).}
    \label{fig:quantum_processing}
\end{figure}

Fig.~\ref{fig:quantum_processing} (left panel) shows the rate of executing the circuits on Forte over the entire execution timeline. Non-pertinent workloads are not shown in the chart. During the run, we switched between the regular and performance-tuned processing regimes, which now allows us to directly compare the rate of execution thus evaluating the effect of problem-specific enhancements implemented on Forte QPU's control software and firmware (see Section~\ref{sec:execution} for further details).
The right panel of Fig.~\ref{fig:quantum_processing} depicts the distribution of individual circuit execution times with and without performance tuning. While this execution time captures all the essential execution steps, it does not count overheads such as circuit submission and waiting in the queue.

\begin{table}[bth!]
    \centering
    \caption{Evaluation of the speedup factor due to tuning the performance of Forte QPU for the QC-AFQMC workload.}
    \label{tab:forte_speedup}
\begin{tabular}{lcccc}
\toprule
Regime & Median execution time, sec & Number of circuits & Projected total time, days\footnote{Assuming the workload ran uninterrupted in this regime.} & Speedup \\
\midrule
Regular & 9.9 & 24,101 & 34.5 \\
Performance & 1.1 & 276,882 & 3.8 & 9$\times$ \\
\bottomrule
\end{tabular}
\end{table}

It is now possible to summarize the timing results in Fig.~\ref{fig:quantum_processing} and distill them to an execution speedup achieved through the performance tuning of Forte QPU. Table~\ref{tab:forte_speedup} shows the median circuit execution time with and without performance tuning permitting direct comparison. While a matchgate shadow circuit executes, on average, in 9.9~sec in the regular regime, this time reduces to 1.1~sec with all the performance improvements activated, yielding a $9\times$ speedup. Had we run this entire workload without any interruption and without the performance improvements, it would have taken us almost 35~days to complete. In the performance regime, all circuit measurements would have completed in under 4~days. This demonstrates the importance of performance tuning (and, ultimately, co-design) of the QPU to specific workloads.

There are many avenues for further reducing execution time. Developing new, economical trial state ans{\"a}tze and optimizing matchgate circuit generation could make preparation more efficient. Improved error mitigation strategies would allow for faster execution rates. Advances in classical shadow techniques tailored to electronic structure might yield more sample-efficient shadows, lowering the total number of required circuits. Finally, faster QPU hardware and parallel measurement scheduling across multiple devices would further accelerate the overall workflow.

\subsubsection{HPC post-processing time}

The second key contributor to the time to solution of QC-AFQMC is the classical post-processing step, which converts matchgate measurements to QC-AFQMC energies. We executed the post-processing workload on AWS ParallelCluster equipped with NVIDIA H100 and H200 GPUs. Because of  difficulties obtaining a large allocation of GPU-enabled instances at once, post-processing took place in stages across various resources. Despite the fragmented execution, normalized timings exhibit remarkable consistency (Table~\ref{tab:gpu_timings}): using P5-type instances with NVIDIA H100, full processing takes 0.29 GPU-hours per shadow (179,858 shadows in total across the three molecules); on P5en-type instances with higher-performance NVIDIA H200s, processing time reduces to 0.22 GPU-hours per shadow. This translates to about 17,400 and 13,200 GPU-hours per molecule on H100 and H200, respectively.

\begin{table}[tbh!]
    \centering
    \caption{Details of computational resources used for GPU-accelerated QC-AFQMC post-processing on AWS for all three molecules and 1,280 walkers.}
    \label{tab:gpu_timings}
\begin{tabular}{lccccc}
\toprule
Molecule & AWS instance type & Instances & GPUs & \makecell[t]{Median time\\per block, min} & \makecell[t]{Projected total\footnote{For 150 blocks of QC-AFQMC.}\\ GPU-hours per shadow} \\
\midrule
B  & P5 (H100) & 10 & 80 & 86.6 & 0.29 \\
{[B-C]}$^\ddagger$ & P5 (H100) & 5 & 40 & 166.6 & 0.29 \\
{[B-C]}$^\ddagger$ & P5 (H100) & 20 & 160 & 44.1 & 0.30 \\
C  & P5en (H200) & 5 & 40 & 135.4 & 0.22 \\
C  & P5 (H100) & 20 & 160 & 44.4 & 0.29 \\
\bottomrule
\end{tabular}
\end{table}

We will now compare the timings attained by the improved post-processing algorithm with the performance of the state-of-the-art QC-AFQMC implementation by Huang~{\it et al.}\cite{Huang2024-hk} to quantify computational speedups. There are several major differences in the implementation and problem setup between the two demonstrations. These are shown in a side-by-side comparison in Table~\ref{tab:huang_comparison}. For example, size of the problem, number of qubits, number of shadows and walkers, accounting for the active space, and hardware used are all different between the two studies.

\begin{table}[p!]
    \centering
    \caption{Side-by-side comparison of the demonstration in this work with the current state-of-the-art result by Huang {\it et al.}\cite{Huang2024-hk}}
    \label{tab:huang_comparison}
\begin{tabular}{lcc}
\toprule
     & Ref.~\cite{Huang2024-hk} & This work \\
\midrule
Molecular system & \makecell[t]{Nitrogen vacancy center\\in diamond} & \makecell[t]{Ni catalytic complexes} \\
Number of basis functions & & 130 \\
Number of fields & & 1,169 \\
Active space & (4,3) & (8,8) \\
Number of qubits & 4 & 16 \\
Frozen orbitals & Included in core energy & Included via VCE \\
Trial state & VQE/UCCSD & VQE/upCCD \\
\midrule
QPU & IonQ Aria & IonQ Forte \\
Number of shadows & 4,000 & \makecell[t]
{60,365 (B)\\58,375 ([B-C]$^\ddagger$)\\61,118 (C)} \\
Number of measurements per shadow & 10 & 1 \\
Number of walkers & 4,800 & 1,280 \\
\midrule
Imaginary time step & 0.4 Ha$^{-1}$ & 0.01 Ha$^{-1}$ \\
Number of time steps & 400 & \makecell[t]{1,500\\(150 blocks, 10 steps each)} \\
Total propagation time & 160 Ha$^{-1}$ & 15 Ha$^{-1}$ \\
\midrule
Post-processing resources & 4,800 CPU cores & \makecell[t]{40--160 NVIDIA H100\\ and H200 GPUs} \\
\bottomrule
\end{tabular}
\end{table}

Huang~{\it et al.} provide runtime projections for a series 
of increasingly larger systems, including water, benzene, the chromium dimer, based on a simple performance model fitted to observed timings. For example, the projected runtime per step for the benzene molecule 
with 12 qubits on a million CPU cores is 6 hours. We will use these projections in order to compare the performance of the two algorithms making appropriate adjustments for the implementation differences. Because it is difficult to properly account for VCE, Table~\ref{tab:huang_comparison2} shows two comparisons, with and without VCE processing time. Ignoring the VCE correction puts this work at a disadvantage assuming it is free for Huang~{\it et al.}'s algorithm, which yields a lower-end estimate of the speedup at $656$. Naively correcting for VCE adds to the cost of Huang~{\it et al.}'s algorithm, but it is likely a significant overestimate, therefore we treat the resulting speedup of $4.58\times 10^7$ as a higher-end estimate.
The time per step of our 16-qubit imaginary time propagation is only 1.8 minutes, achieved with 320 NVIDIA H200 GPUs (benchmarked separately). Comparing this with a 6-hour projection for a 12-qubit system and one million CPU cores, our work achieves several orders of magnitude improvement, significantly enhancing the practicality of the QC-AFQMC algorithm. 

\begin{table}[t!]
    \centering
    \caption{Comparison of the projections of time to compute one step (force bias and local energy) with one walker and one shadow measurement between the state of the art algorithm by Huang {\it et al.}\cite{Huang2024-hk} (hydrogen molecule) and this work.
    All time units are in seconds.}
    \label{tab:huang_comparison2}
\begin{tabular}{lcc}
\toprule
     & Ref.~\cite{Huang2024-hk}  & This work \\
\midrule
Baseline: 4 qubits, (2,2) space, 160,000 shadows & 60 & \\
Baseline: time per shadow measurement & $3.75\times 10^{-4}$ &  \\
Baseline extrapolated to 16 qubits assuming $\mathcal{O}(N_q^8)$ scaling & 24.58\\
\midrule
\makecell[lt]{This work: 16 qubits, (8,8) space, 179,858 shadows,\\ 1 block, VCE, H200 GPU} & & 746.83 \\
This work: time per shadow & & $4.15\times 10^{-3}$ \\
This work: adjust for one force bias per block\footnote{In this study, energy was evaluated every 10 time steps. Energy computation alone takes $3.71\times 10^{-4}$ GPU-sec per shadow.} & & $7.49\times 10^{-4}$ \\
Assume that GPU provides a $50\times$ speedup over CPU\footnote{Typical observed speedup on a single NVIDIA H200 GPU over a single core of AMD Rome CPU in QC-AFQMC post-processing.} & & $3.74\times 10^{-2}$ \\
Speedup factor over baseline ignoring VCE (lower-end estimate) & & 656 \\
\midrule
Adjustment for VCE\footnote{VCE assumes the full system has 130 spatial orbitals.} assuming $\mathcal{O}(N_B^4N_q^4)$ scaling & $1.71 \times 10^6$ & \\
Speedup factor over baseline (higher-end estimate) & & $4.58\times 10^7$ \\
\bottomrule
\end{tabular}
\end{table}

\begin{figure}[htbp]
    \centering
    \includegraphics[width=0.8\linewidth]{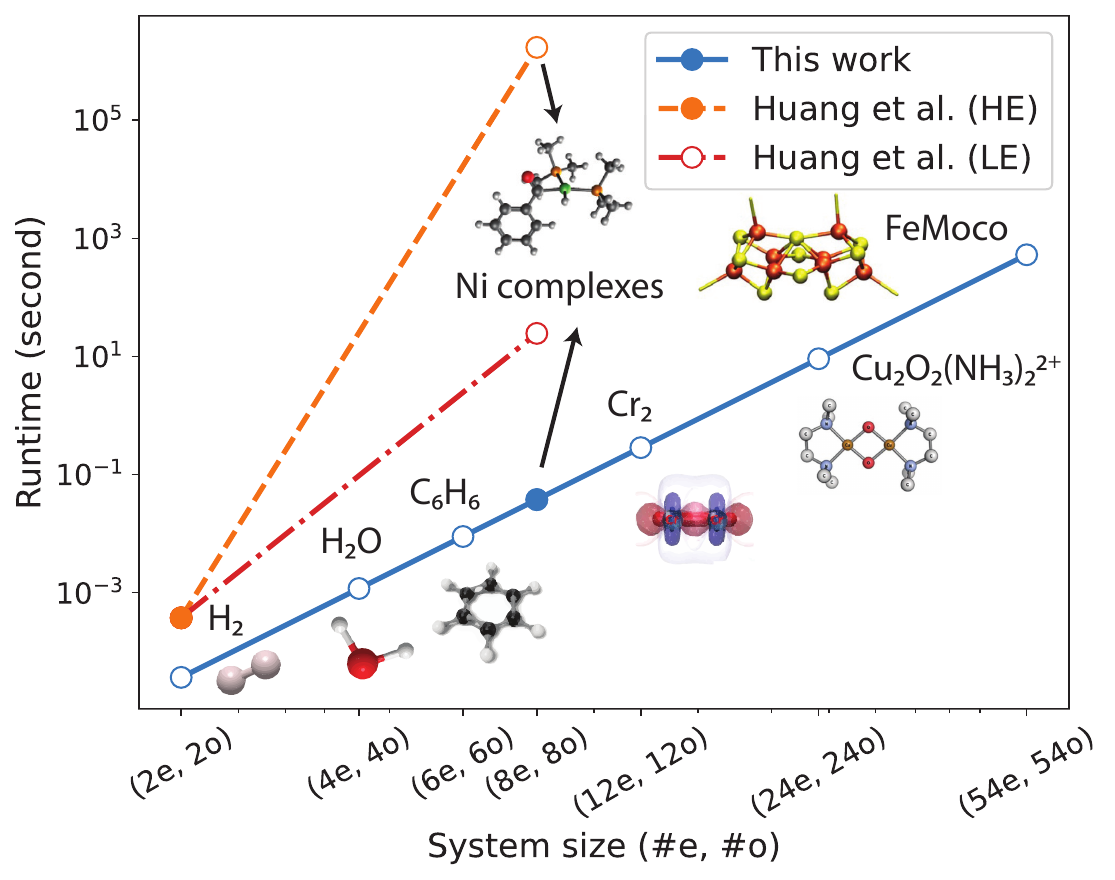}
    \caption{Scaling of classical post-processing runtime in QC-AFQMC, showing the runtime per time step per shadow. The data are from QPU experiments (solid circles), as well as from estimated runtime (hollow circles). Ni complexes is the molecular system studied in this work. The estimated run time for this work is calculated by scaling the data from the QPU experiment with $\mathcal{O}(N_q^5)$. The estimated run time for the work of Huang {\it et al.} is calculated by scaling the data from the QPU experiment with $\mathcal{O}(N_B^4N_q^4)$. The orange dashed line scales to 130 spatial orbitals (higher-end estimate, HE) while the red dash-dot line scales to 8 spatial orbitals (lower-end estimate, LE) at problem size (8e, 8o).}
    \label{fig:qmc_scaling}
\end{figure}

The details are shown in  Table~\ref{tab:huang_comparison2}, where we compare the amount of time needed to compute one local
energy with a single shadow measurement between the two implementations. 
In Huang~{\it et al.}, it takes 1 minute to process 16,000 shadow circuits for each walker (roughly 10 unique measurement outcomes per circuit with 1024 shots, yielding 160,000 shadow measurements in total), so that the time per shadow measurement is $3.75\times 10^{-4}$~sec. Projecting this to a 16-qubit system (an 8-electron, 8-orbital active space) with big-$\mathcal{O}$ scaling as $\mathcal{O}(N_q^8)$, we 
obtain the projected time of 24.58~sec. The $\mathcal{O}(N_q^8)$ scaling arises from two factors: first, there are $\mathcal{O}(N_q^4)$ terms in the Hamiltonian; and second, it costs $\mathcal{O}
(N_q^4)$ to process one shadow measurement for each term. We then further project it to
include VCE, and the big-$\mathcal{O}$ scaling of VCE becomes $\mathcal{O}(N_B^4N_q^4)$, in
which $N_B$ is the total number of orbitals in the full orbital space. With this taken into
account, for a system with 130 spatial orbitals (the same as the systems studied in this work), 
we arrive at the final projection of $t_1=1.71 \times 10^6$~sec per step per shadow. The algorithm implemented in this work solves the problem of the same size but with 179,858 shadows with $t_2=7.5\times 10^{-4}$~sec per step per shadow on an NVIDIA H200 GPU, which translates to $t_2=3.7\times 10^{-2}$~sec of CPU time under the assumption of $50\times$ GPU-over-CPU acceleration. As shown in Fig.~\ref{fig:qmc_scaling}, the ratio between the classical processing duration for Ni-complexes is $t_1/t_2\simeq 2\times 10^9$. Even when adjusted for GPU acceleration, the ratio is still on the order of $10^7$ speedup. This dramatic computational cost reduction is primarily driven by the reduced asymptotic algorithmic cost achieved through automatic differentiation and the use of Cholesky decomposition of Hamiltonian. As shown in Fig.~\ref{fig:qmc_scaling}, the combination of the algorithmic improvement and the usage of GPUs put the post-processing of even larger chemical systems in practical regime.

Finally, we analyze the scalability of QC-AFQMC with the size
of the basis set. For a fixed active space, the number of qubits is independent of the basis set size, and so is the quantum part of time to
solution. However, the situation is different for classical post-processing. As as have discussed in Section~\ref{subsec:improvement},
the cost of computing force bias and local energy scales as $\mathcal{O}(N^{4})$ and $\mathcal{O}(N^{5})$ per shadow measurement. To be more specific, the more 
precise scaling is $\mathcal{O}(N_BN_q^{3})$ and $\mathcal{O}(N_BN_q^{4})$ respectively, in which $N_B$ is the size of the basis 
set and $N_q$ is the number of qubits. Therefore, the cost of post-processing scales linearly with the size of the basis set. 

\section{Conclusions}

We demonstrate the operation of an end-to-end workflow based on the QC-AFQMC method to model the reaction barrier of a step of the nickel catalyzed Suzuki--Miyaura reaction. Operating in a high-performance hybrid quantum-classical compute environment, the workflow includes state preparation on a quantum circuit simulator, sampling of the trial state on the IonQ Forte QPU via the matchgate shadow protocol, and execution of the QC-AFQMC imaginary time propagation on GPU-accelerated cluster on AWS. While previous studies of QC-AFQMC concluded that the method is too computationally demanding to be of practical use, we demonstrate that it in fact could be applied to problems as large as catalytic transition metal complexes. This is achieved through the use of the active space approximation and virtual correlation energy. The accuracy of estimating the reaction barriers of the chemical reactions with QC-AFQMC+VCE based on ideally sampled matchgates is within the uncertainty interval of AFQMC ($\pm4$~kcal/mol) relative to the reference CCSD(T) method. With QPU sampled matchgates, the discrepancy increases to 10~kcal/mol calling for advanced error mitigation techniques.

We show the importance of tuning the system performance of a QPU for the workload of a specific application. By applying performance optimizations to the control software of the QPU, we achieve a $9\times$ speedup in collecting matchgate circuit measurements compared to a baseline without application-specific tuning.
In the QC-AFQMC imaginary time propagation step, we demonstrate the use of analytic differentiation for computing the energy and force bias leading to a reduction in the complexity of the algorithm from $\mathcal{O}(N^{8.5})$ to $\mathcal{O}(N^{5.5})$, which translates into a $656\times$ time-to-solution improvement over a baseline projected from the state of the art implementation of the method by Huang {\it et al.} \cite{Huang2024-hk}.

This work represents a significant step toward practical quantum chemistry simulations on quantum computing devices and offers opportunities for further improvements. Future work should examine the choice of the trial state that could be prepared efficiently and reliably on a quantum computer and their effect on accuracy; novel efficient classical shadow techniques for fermionic systems; and the scaling of cost and accuracy for molecular systems with strong correlation and larger basis sets.

\clearpage

\bibliographystyle{unsrt_abbrv_custom}
\bibliography{Journal_Short_Name, goings_references,AZ_refs,ionq, error_mitigation_refs}

\clearpage
\onecolumngrid
\section{Supplemental Information}

\subsection{Matrix Definitions}
\begin{enumerate}
\item The $\mathbf{W}$ matrix for a $N$ qubit, $\zeta$ electron system is  
\begin{equation}
    \mathbf{W}=\bigoplus_{j=1}^\xi\frac{1}{\sqrt{2}}{\begin{pmatrix}
                                            1 & -i \\
                                            1 & i 
                                            \end{pmatrix}} \smalloplus \bigoplus_{j=\xi+1}^N{\begin{pmatrix}
                                            1 & 0 \\
                                            0 & 1 
                                            \end{pmatrix}}
\end{equation}

\item The covariance matrix is: 
\begin{equation}
    \mathbf{C}_{\ket{b}} = \bigoplus_{j=1}^N \begin{pmatrix} 0 & (-1)^{b_j} \\ -(-1)^{b_j} & 0 \end{pmatrix}
\end{equation}
\item For a $\zeta$-electron Slater determinant (walker), it could be written as
\begin{equation}
    \ket{\phi}=\tilde{a}^\dagger_1\cdots \tilde{a}^\dagger_\zeta\ket{\textbf{0}}, \quad \mathrm{where} \quad \tilde{a}_j=\sum_{k=1}^N{V_{jk}a_k},
\end{equation}
in which one could think that $a_k$ is the annihilation operator in the HF molecular orbital basis, and $\textbf{V}$ is the orbital rotation matrix. The definition of the 
$\textbf{M}_{\phi}$ matrix is
\begin{equation}
    \textbf{M}_{\phi}={\begin{pmatrix}
                R_{11} & \cdots & R_{1n} \\
                \vdots & \ddots & \vdots \\
                R_{n1} & \cdots & R_{nn}
                \end{pmatrix}}
\end{equation}
with blocks 
\begin{equation}
    R_{jk}={\begin{pmatrix}
                \mathrm{Re}(V_{jk}) & -\mathrm{Im}(V_{jk}) \\
                \mathrm{Im}(V_{jk}) & \mathrm{Re}(V_{jk})
                \end{pmatrix}}
\end{equation}
\end{enumerate}

\subsection{Virtual Correlation Energy Derivation}
First, we write our trial wave function as (assuming an unrestricted framework), 
\begin{equation}
    \ket{\Psi_T}=\ket{\Xi^\alpha_c}\otimes\ket{\Xi^\beta_c}\otimes\ket{\Psi_{T,a}}\otimes\ket{0^\alpha_v}\otimes\ket{0^\beta_v},
\end{equation}
in which $\ket{\Psi_{T,a}}$ is the trial state prepared on a quantum computer within the active space with $N_a$ 
electrons, $\ket{\Xi^{\alpha (\beta)}_c}$ is a Slater determinant with all the frozen occupied $\alpha$ ($\beta$) orbitals with $N^{\alpha (\beta)}_c$ electrons, and $\ket{0^{\alpha (\beta)}_v}$ is the vacuum state in the space 
of virtual $\alpha$ ($\beta$) orbitals. The trial wave function could always be written as 
\begin{equation}
    \ket{\Psi_T}=\ket{\Xi^\alpha_c}\otimes\ket{\Xi^\beta_c}\otimes\sum_{i}{c_i\ket{\chi_i^\alpha}\otimes\ket{\chi_i^\beta}}\otimes\ket{0^\alpha_v}\otimes\ket{0^\beta_v},
\end{equation}
where $\ket{\chi_i^{\alpha (\beta)}}$ is the $\alpha$ ($\beta$) component of the $i$-th Slater determinant within the 
active space. 

The overlap between the trial state and a walker defined in the full space is 
\begin{equation}
    \sum_i{c_i\braket{\Xi_c^\alpha\Xi_c^\beta\chi^\alpha_i\chi^\beta_i0^\alpha_v0^\beta_v|\phi}}=\sum_i{c_i\mathrm{det}\left(\begin{pmatrix}
\Xi_c^\alpha & 0 & 0 & 0\\
0 & \Xi_c^\beta & 0 & 0 \\
0 & 0 & \chi^\alpha_i & 0 \\
0 & 0 & 0 & \chi^\beta_i \\
0 & 0 & 0 & 0 \\
0 & 0 & 0 & 0 \\
\end{pmatrix}^\dagger\begin{pmatrix}
\phi^\alpha_c & 0 \\
\phi^\alpha_a & 0 \\
\phi^\alpha_v & 0 \\
0 & \phi^\beta_c  \\
0 & \phi^\beta_a  \\
0 & \phi^\beta_v  \\
\end{pmatrix}\right)}
\end{equation}
which becomes
\begin{equation}
    \sum_i{c_i\braket{\Xi_c^\alpha\Xi_c^\beta\chi^\alpha_i\chi^\beta_i|\phi}}=\sum_i{c_i\mathrm{det}\begin{pmatrix}
\Xi_c^{\alpha\dagger}\phi^\alpha_c & 0 \\
0 & \Xi_c^{\beta\dagger}\phi^\beta_c \\
\chi^{\alpha\dagger}_i\phi^\alpha_a & 0 \\
0 & \chi^{\beta\dagger}_i\phi^\beta_a  \\
\end{pmatrix}}
\end{equation}
where $\phi^{\alpha (\beta)}_c$ and $\phi^{\alpha (\beta)}_a$ are $N^{\alpha(\beta)}_a+N^{\alpha(\beta)}_c$ column 
molecular orbital coefficients of the walker. $\Xi_c^{\alpha(\beta)}$ is diagonal with ones up to the number of $\alpha$($\beta$) core
electrons and zeros elsewhere. As one could see, the virtual degrees of freedom no longer appear. 

To further remove the core degrees of freedom, we perform singular value decompositions (SVD),
\begin{equation}
    \begin{split}
        \Xi_c^{\alpha\dagger}\phi^\alpha_c=U^\alpha_c\Sigma_c^\alpha V^{\alpha\dagger}_c \\
        \Xi_c^{\beta\dagger}\phi^\beta_c=U^\beta\Sigma_c^\beta V^{\beta\dagger}_c, \\
    \end{split}
\end{equation}
where $U^\alpha_c\in\mathbb{C}^{N^\alpha_c\times N^\alpha_c}$, $V^\alpha_c\in\mathbb{C}^{(N^\alpha_c+N^\alpha_a)\times N^\alpha_c}$, $U^\beta_c\in\mathbb{C}^{N^\beta_c\times N^\beta_c}$, $V^\beta_c\in\mathbb{C}^{(N^\beta_c+N^\beta_a)\times N^\beta_c}$. We then define unitary matrices $U$ and $V$ as 
\begin{equation}
    U=\begin{pmatrix}
U^\alpha_c & 0 & 0 & 0 \\
0 & U^\beta_c & 0 & 0 \\
0 & 0 & I & 0 \\
0 & 0 & 0 & I  \\
\end{pmatrix}
\end{equation}

\begin{equation}
    V=\begin{pmatrix}
V^\alpha_c & 0 & V^{\alpha\prime}_c & 0 \\
0 & V^\beta_c & 0 & V^{\beta\prime}_c \\
\end{pmatrix}
\end{equation}
in which $V^{\alpha(\beta)\prime}_c$ are added orthonormal columns to $V^{\alpha(\beta)}_c$, and usually this is obtained
automatically from SVD. 

We can then write the overlap as 
\begin{equation}
    \begin{split}
        \braket{\Xi_c^\alpha\Xi_c^\beta\chi^\alpha_i\chi^\beta_i|\phi}&=\mathrm{det\left(U^\dagger\begin{pmatrix}
\Xi_c^\alpha & 0 & 0 & 0\\
0 & \Xi_c^\beta & 0 & 0 \\
0 & 0 & \chi^\alpha_i & 0 \\
0 & 0 & 0 & \chi^\beta_i \\
\end{pmatrix}^\dagger\begin{pmatrix}
\phi^\alpha_c & 0 \\
0 & \phi^\beta_c  \\
\phi^\alpha_a & 0 \\
0 & \phi^\beta_a  \\
\end{pmatrix}V\right)} / (\mathrm{det}(U^\dagger)\mathrm{det}(V)) \\
&=\mathrm{det}(\Sigma^\alpha_c)\mathrm{det}(\Sigma^\beta_c)\mathrm{det}(\chi^{\alpha\dagger}_i\tilde{\phi}^{\alpha}_a)\mathrm{det}(\chi^{\beta\dagger}_i\tilde{\phi}^{\beta}_a)\mathrm{det}(R^\alpha)\mathrm{det}(R^\beta) / (\mathrm{det}(U^\dagger)\mathrm{det}(V)), \\
    \end{split}
\end{equation}
in which $\tilde{\phi}^{\alpha(\beta)}_a$ is the normalized Slater determinant within the active space, and $\mathrm{det}(R^{\alpha(\beta)})$ is the normalization matrix obtained by performing QR decomposition of the matrix $\phi^{\alpha(\beta)}_aV^{\alpha(\beta)\prime}$. Therefore, computing the overlap between the trial and Slater determinant in the full space only requires the evaluation of the overlap between the trial wave function and a modified
determinant in the active space. 
\begin{equation}
    \braket{\Psi_T|\phi}=\mathrm{det}(\Sigma^\alpha_c)\mathrm{det}(\Sigma^\beta_c)\mathrm{det}(R^\alpha)\mathrm{det}(R^\beta)\braket{\Psi_{T,a}|\tilde{\phi}_a}/(\mathrm{det}(U^\dagger)\mathrm{det}(V)),
\end{equation}
which we already know how to compute using matchgate shadows. 

\subsection{Reduced Density Matrices}
Besides energy, we could also compute reduced density matrices (RDM) with 
matchgate shadows. For a local fermionic observable $\tilde{\gamma}_{S}$, one
could compute its expectation values as 
\begin{equation}
    \label{eqn:local_fermionic_ob_exp}
    \braket{\Psi|\tilde{\gamma}_{S}|\Psi}=\sum_b{{2n \choose |S|}{n \choose |S|/2}^{-1} \mathrm{pf}\left(i(
    \textbf{Q}^\prime \textbf{Q}_p^T\textbf{C}_{\ket{b}}\textbf{Q}_p\textbf{Q}^{\prime T})|_{S}\right)}
\end{equation}
in which $\tilde{\gamma}_{S}$ is a product of rotated Majorana operators
\begin{equation}
    \label{eqn:majorana_op_prod}
    \tilde{\gamma}_{S}=\tilde{\gamma}_{\mu_1}\cdots\tilde{\gamma}_{\mu_{|S|}}, \quad \tilde{\gamma}_\mu=\sum_{\mu=1}^n{Q^\prime_{\mu\nu}\gamma_\nu}
\end{equation}
with rotation matrix $\textbf{Q}^\prime\in O(2n)$, a $2n$ by $2n$ real orthogonal matrix. $\textbf{C}_{\ket{b}}$ and $\textbf{Q}_p$ is the shadow
covariance matrix and orthogonal matrix used to construct the shadow circuits defined in Equation \ref{eqn:b_mat}. The Majorana operators are defined with 
creation and annihilation operators
\begin{equation}
    \label{eqn:majorana_op}
    \gamma_{2j-1}=a_j+a^\dagger_j, \quad \gamma_{2j}=-i(a_j-a^\dagger_j)
\end{equation}
for $j\in [n]=\{1,\cdots,n\}$. 

Now it's obvious that one could write the reduced density matrix operators in terms 
of $\gamma_{S}$. For example, the off-diagonal element of the 1-RDM is
\begin{equation}
    \label{eqn:1rdm_off_diagonal}
    a^\dagger_pa_q(p\neq q)=\frac{1}{4}\left(\gamma_{2p}\gamma_{2p}+i\gamma_{2p}\gamma_{2q+1}-i\gamma_{2p+1}\gamma_{2q}+\gamma_{2p+1}\gamma_{2q+1}\right)
\end{equation}
and the diagonal part is
\begin{equation}
    \label{eqn:1rdm_diagonal}
    a^\dagger_pa_p=\frac{1}{2}\left(1+i\gamma_{2p}\gamma_{2p+1}\right)
\end{equation}

\subsection{Implementation Details}
To launch the execution of the workload, we use the IonQ Hybrid Service framework. This framework allows the definition of a custom workload input, that can be submitted through the IonQ application programming interface (API), by using IonQ's Python based software development kit (SDK). The SDK allows the creation of the custom workload, with the necessary input parameters needed for the QC-AFQMC workflow, initiation of workflow, asynchronous status tracking, and results retrieval upon the completion of the workflow. It also allows tracking the state of the job, either via API or the Console user interface.

On the backend side of the IonQ Cloud, we tailor the execution of the custom workload to a custom built container image that is capable of validating the input and in this particular case start the execution of the workflow through Amazon Braket. This allows the user to choose different backends, e.g. quantum simulator, to test the implementation before executing it on a quantum hardware.
With this approach we are able to represent the problem as an Application that could be exposed to any perspective users in running the same workflow with different combinations of input parameters, such as the type of molecules and QC-AFQMC hyper parameters, while keeping a record of execution and results.

The custom logic and code for this particular workflow is maintained in a GitHub Repository, and has configuration for continuous integration, which allows versioned Docker images to be rebuilt every time when new features are introduced. To ensure reproducibility and manage dependencies for our cloud-based workflow, executed as a Braket hybrid job, we use the ``Bring Your Own Container'' feature of Braket Hybrid Jobs, which allows us to utilize a custom Docker image. This image, containing essential libraries like MPI, PySCF, our in-house QC-AFQMC library, and NVIDIA's CUDA-Q Python packages. By pushing this pre-built image to Amazon ECR and deploying it within Amazon Braket Hybrid Jobs, we eliminate dependency drift and guarantee a consistent, integrated environment for both the GPU-accelerated classical post-processing and CUDA-Q driven quantum components of the QC-AFQMC algorithm.

\vspace{1em}
\noindent
\textbf{Implementation details.} The repository includes:
\begin{itemize}
    \item A top-level \texttt{Dockerfile} that defines how to install \texttt{pyscf}, \texttt{ipie}, \texttt{cudaq}, our inhouse QC-AFQMC library, and all other dependencies.
    \item A \texttt{container/Dockerfile} for Braket-compatible images that extends the default \texttt{amazon-braket-pytorch-jobs} base image, adding MPI, \texttt{cudaq}, and IonQ-specific libraries.
    \item A build script \texttt{container\_build\_and\_push.sh} that tags the container and pushes it to ECR for use within Braket hybrid jobs.
\end{itemize}
Upon job submission, we launch the hybrid jobs on an \texttt{ml.m5.12xlarge} instance via the Braket SDK \cite{braket} and specify IonQ Forte as the desired backend. In particular, we set:
\begin{lstlisting}[language=Python, basicstyle=\ttfamily\small]
device_arn = "arn:aws:braket:us-east-1::device/qpu/ionq/Forte-Enterprise-1"
cudaq.set_target("braket", machine=device_arn)
\end{lstlisting}This instructs CUDA-Q to send quantum circuits to IonQ's hardware through Braket. We wrap the entire AFQMC procedure (geometry-building, matchgate shadow circuits, GPU-accelerated overlap post-processing, etc.) in a single \texttt{@hybrid\_job} function. The result is a ``submit-once'' HPC workflow that handles both classical and quantum parts consistently: the classical AFQMC steps run on the CPU/GPU instance in the same container, while quantum measurement tasks are delegated to IonQ's Forte system through Amazon Braket.

In summary, the container-based QC-AFQMC pipeline provides a straightforward route to run GPU-accelerated classical post-processing and IonQ-based quantum circuits in a single HPC job on AWS Parallel Cluster. We found this approach substantially simplified environment management, ensured reproducible results across HPC executions, and  facilitates scalability to larger systems.

\subsection{Restart capability}

As each step in the AFQMC imaginary time propagation depends on the previous steps, a robust checkpoint-restart mechanism becomes essential. Our implementation addresses this through an HDF5-based framework where each MPI rank manages its assigned walker subset. The framework captures complete snapshots of the simulation state, preserving walker configurations, weights, propagator states, trial wave function parameters, and random number generator states. This state preservation serves multiple purposes. It enables recovery from hardware failures, allows calculations to resume from previous points, and supports analysis of state evolution over time. The system validates simulation states during restart operations by verifying Hamiltonian parameters and simulation variables while maintaining walker distribution across ranks. The HDF5 format's parallel I/O capabilities and efficient handling of large datasets make it suited for production-scale calculations spanning weeks or months.

\subsection{DFT Reaction Energy Profiles}

To assess the impact of molecular truncation on reaction energetics, we computed reaction energy profiles using four density functionals: B3LYP, M06-2X, $\omega$B97X, and PBE0. All calculations were performed with the minimal STO-3G basis set to provide a consistent and computationally efficient framework for comparing the different truncation levels.

Table~\ref{tab:dft_energies} presents the computed electronic energies and relative energies for each truncation level and functional. The energy of structure B is set as the reference (0.00 kcal/mol) within each truncation category. The subsequent figures illustrate these relative energies as reaction energy profiles.

\begin{table*}[h]
    \centering
    \caption{Electronic energies and relative energies for each truncation level and structure, computed using different DFT functionals at the STO-3G level.}
    \label{tab:dft_energies}
    \renewcommand{\arraystretch}{1.2}
    \begin{tabular}{lcrrrrrrrr}
\toprule
Model & Structure &  
$E_{\text{PBE0}}$ &  $E_{\text{B3LYP}}$ &  $E_{\text{M06-2X}}$ &  $E_{\omega\text{B97X}}$ &  
$\Delta E_{\text{PBE0}}$ &  $\Delta E_{\text{B3LYP}}$ &  $\Delta E_{\text{M06-2X}}$ &  $\Delta E_{\omega\text{B97X}}$ \\
\midrule
        original &        BC & -3208.6137 & -3210.1846 & -3210.0978 & -3210.1482 &  61.63 &  63.96 &  58.08 &  68.16 \\
        original &         C & -3208.6846 & -3210.2603 & -3210.1916 & -3210.2301 &  17.17 &  16.49 &  -0.83 &  16.76 \\
        original &         B & -3208.7120 & -3210.2866 & -3210.1903 & -3210.2568 &   0.00 &   0.00 &   0.00 &   0.00 \\
         reduced &        BC & -2742.9552 & -2744.0196 & -2743.9444 & -2743.9880 &  63.20 &  64.35 &  57.23 &  69.08 \\
         reduced &         C & -2743.0313 & -2744.0947 & -2744.0287 & -2744.0688 &  15.49 &  17.21 &   4.37 &  18.39 \\
         reduced &         B & -2743.0559 & -2744.1221 & -2744.0356 & -2744.0981 &   0.00 &   0.00 &   0.00 &   0.00 \\
         minimal &        BC & -2553.7526 & -2554.6277 & -2554.5355 & -2554.6115 &  61.54 &  62.42 &  57.81 &  66.69 \\
         minimal &         C & -2553.8386 & -2554.7046 & -2554.6316 & -2554.6999 &   7.59 &  14.16 &  -2.47 &  11.24 \\
         minimal &         B & -2553.8507 & -2554.7272 & -2554.6276 & -2554.7178 &   0.00 &   0.00 &   0.00 &   0.00 \\
\bottomrule
\end{tabular}
    \begin{flushleft}
    \textbf{Notes:} All energies ($E$) are in Hartree, and relative energies ($\Delta E$) are in kcal/mol. $\Delta E$ values are referenced to structure B within each truncation level. Structures: B (reactant), BC (transition state), C (product).
    \end{flushleft}
\end{table*}

The computed reaction energy profiles are presented in the following figures. Each plot displays the relative energies from Table~\ref{tab:dft_energies}, illustrating the energetic differences between the full, reduced, and minimal models. As before, the energy of complex B is set as the zero-point reference for relative energies. The solid line represents the full 77-atom model, while the dashed and dotted lines represent the 41-atom and 34-atom truncated models, respectively. The close agreement across truncation levels suggests that the reduced models retain essential features of the reaction energetics.

\subsection{Single Orbital Entropy Analysis}

Single-orbital entropy profiles were computed for the original, reduced, and minimal models to assess the robustness of the truncation strategy. The entropy values quantify the degree of entanglement for each molecular orbital, providing insight into the active space selection process.

\begin{figure}[h]
    \centering
    \includegraphics[width=0.8\textwidth]{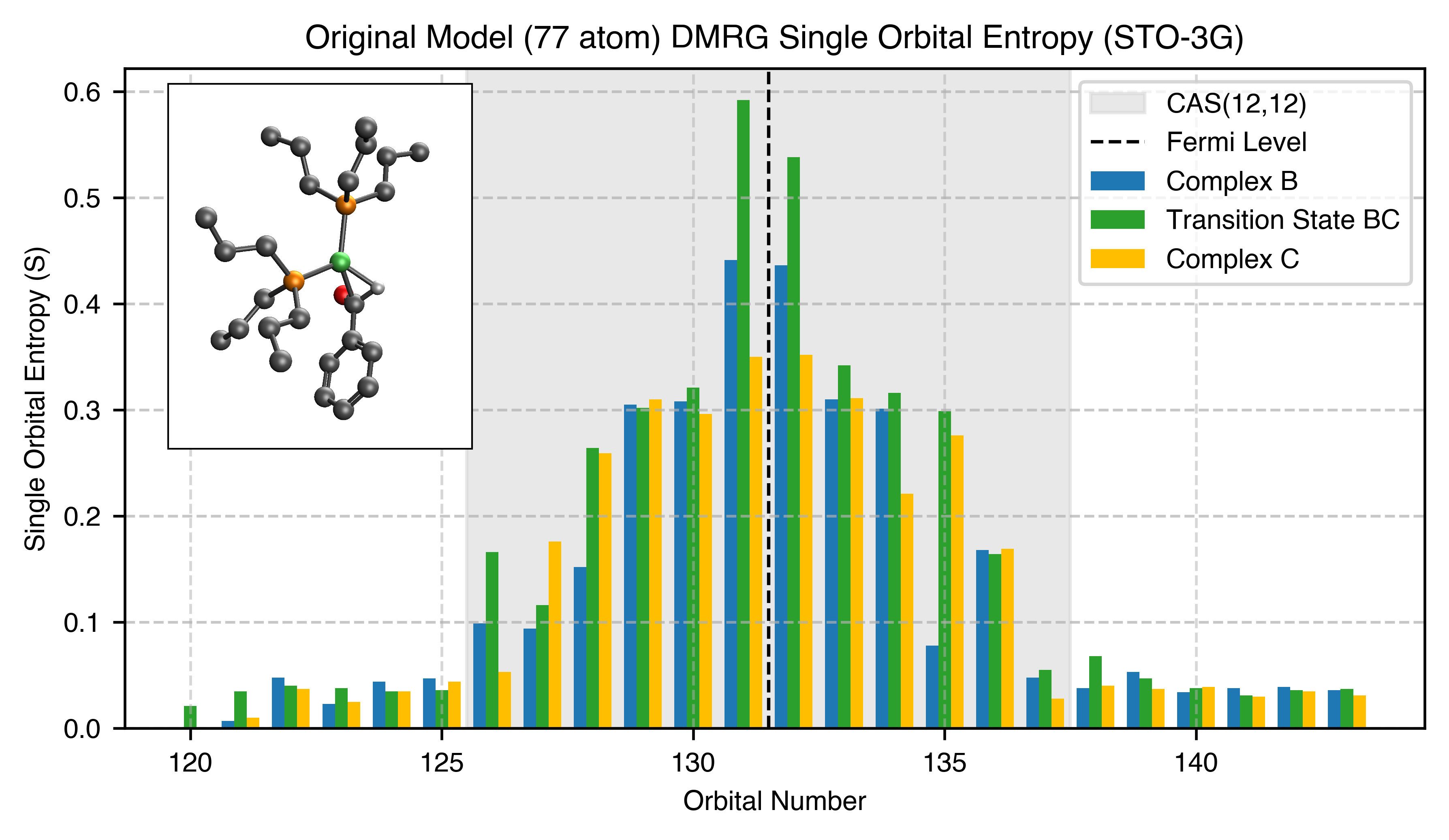}
    \caption{Single Orbital Entropy for the full model (77 atoms). The entanglement profile indicates a well-defined active space, with significant contributions from orbitals near the Fermi level.}
    \label{fig:original_entropy}
\end{figure}

\begin{figure}[h]
    \centering
    \includegraphics[width=0.8\textwidth]{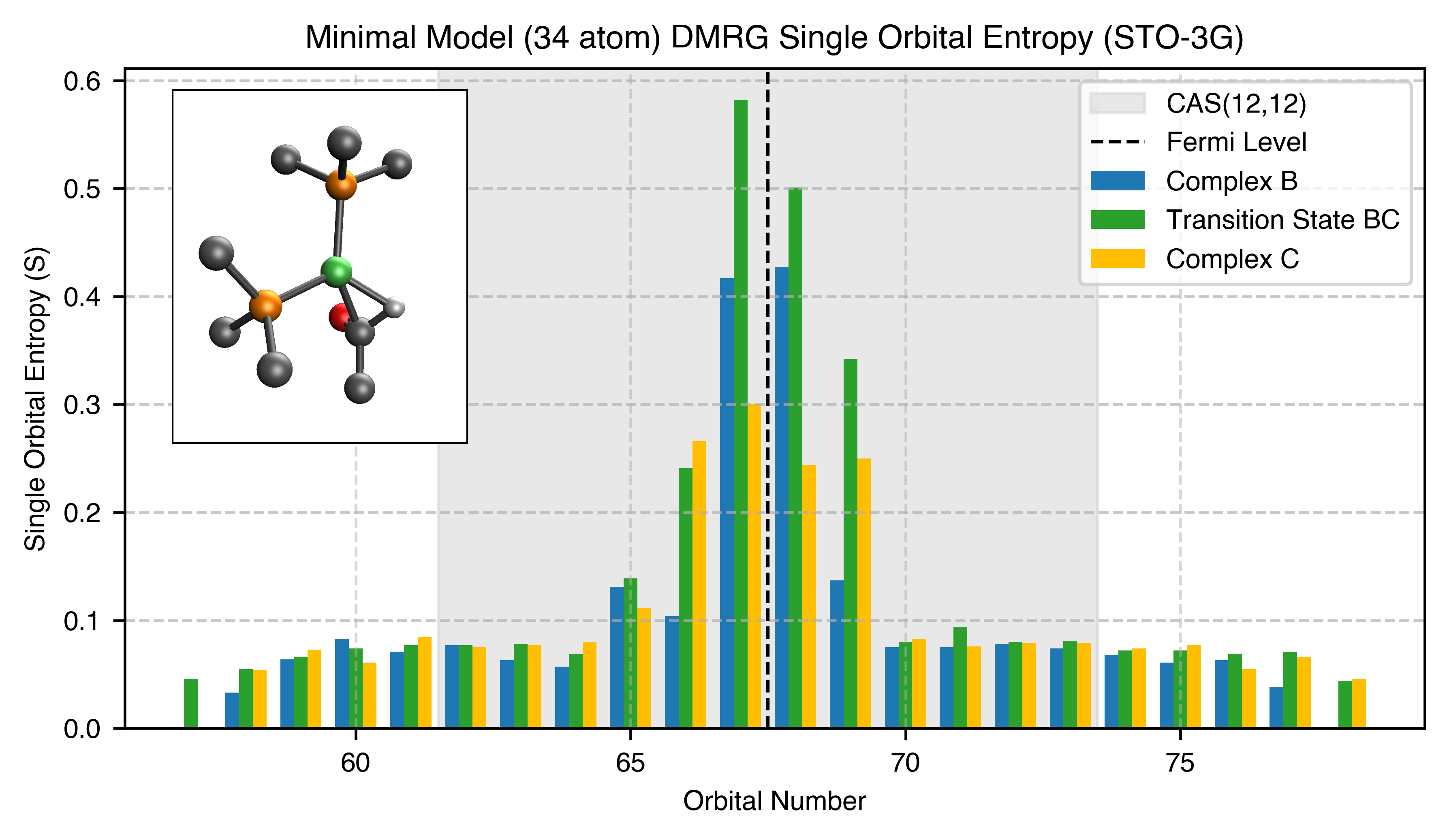}
    \caption{Single Orbital Entropy for the minimal model (34 atoms). Despite the aggressive truncation, key frontier orbitals maintain high entanglement, supporting the model reduction strategy.}
    \label{fig:super_truncated_entropy}
\end{figure}

\clearpage
\subsection{Orbital Entanglement and Mutual Information Networks}
To further analyze the impact of molecular truncation on active space selection, we visualize orbital entanglement using mutual information networks. In these figures, the node intensity represents the single-orbital entropy, indicating the degree of entanglement of an individual orbital. The connection width represents the two-orbital mutual information, derived from the density cumulant matrix, quantifying the correlation strength between orbitals.

\clearpage
\subsection{XYZ Structures}

This section presents the molecular geometries used in this study. The structures are grouped by truncation level, with the atom count reflecting the extent of the simplification:

\begin{itemize}
    \item Full Model (77 atoms): The original, unmodified complex.
    \item Reduced Model (41 atoms): Peripheral aliphatic \ce{-CH2-CH2-CH3} groups pruned while preserving coordination environment.
    \item Minimal Model (34 atoms): Further reduction through simplification of the phenyl framework.
\end{itemize}

Each structure is labeled with its molecular complex designation---\textbf{(B)}, \textbf{(BC)}, \textbf{(C)}---and its corresponding truncation level.

\subsubsection{Full Model (77 atoms, original complex)}

\textbf{Nickel-Complex (B) (77 atoms)}
\begin{lstlisting}[basicstyle=\ttfamily\footnotesize]
77

Ni         0.01087       -0.65034       -0.82192
P         -2.11214       -0.34418       -0.26552
P          1.28203        1.03861       -0.33963
C          1.20132       -1.99109       -1.53283
O         -0.03666       -2.40446       -1.42497
C          2.24029       -2.42578       -0.56130
C          3.59793       -2.18194       -0.81437
C          4.20131       -3.03680        1.36492
C          1.88519       -3.02003        0.65925
C          2.85488       -3.31682        1.61216
H          3.89145       -1.74659       -1.77460
H          4.95994       -3.26846        2.11574
H          0.82860       -3.22764        0.84174
H          2.56065       -3.77256        2.56099
C          2.49398        1.41965       -1.69188
C          0.55710        2.69169        0.08673
C          2.31701        0.64211        1.13840
C         -2.39241       -0.53396        1.56040
C         -2.85360        1.28411       -0.75583
C         -3.17221       -1.64013       -1.04980
C          3.14662        2.79957       -1.77392
H          1.96669        1.20535       -2.63559
H          3.26057        0.63088       -1.60119
C          4.20054        2.87265       -2.87486
H          2.37283        3.56178       -1.96254
H          3.60512        3.06631       -0.80900
C          3.39381        1.62252        1.59431
H          2.76905       -0.33858        0.92618
H          1.59722        0.45366        1.95357
C          4.06936        1.16241        2.88306
H          4.15614        1.72156        0.80459
H          2.96513        2.62955        1.73825
C         -0.19324        2.70656        1.41860
H         -0.13033        2.94971       -0.73711
H          1.34114        3.46536        0.09788
C         -0.99885        3.98304        1.63496
H         -0.86235        1.83350        1.47767
H          0.52690        2.58537        2.24448
C         -4.36400        1.44558       -0.92492
H         -2.35476        1.53214       -1.70805
H         -2.47198        2.02144       -0.03040
C         -4.74076        2.86511       -1.33943
H         -4.72597        0.73416       -1.68475
H         -4.88397        1.18794        0.00983
C         -4.51676       -2.01652       -0.42772
H         -2.51327       -2.52291       -1.09866
H         -3.30102       -1.32443       -2.09993
C         -5.21259       -3.13613       -1.19683
H         -4.36087       -2.34058        0.61457
H         -5.18230       -1.14067       -0.37766
C         -3.59819        0.12945        2.22352
H         -1.46818       -0.17479        2.04124
H         -2.39896       -1.62333        1.73812
C         -3.63942       -0.11903        3.72826
H         -3.56667        1.21603        2.03555
H         -4.53014       -0.23474        1.76458
H          3.76169        2.64412       -3.85885
H          5.00967        2.14664       -2.69754
H         -2.73098        0.26507        4.21861
H         -3.70600       -1.19573        3.95004
H         -4.58606       -4.04100       -1.23069
H         -5.41569       -2.83690       -2.23719
H         -4.25539        3.14312       -2.28829
H         -4.42449        3.59701       -0.57908
H         -1.76636        4.10724        0.85359
H         -0.35102        4.87304        1.60296
H          4.52181        0.16716        2.75050
H          3.34077        1.08750        3.70586
H          4.86094        1.85943        3.19650
H          4.65367        3.87347       -2.93389
H         -5.82726        2.97152       -1.47493
H         -1.51024        3.97454        2.60916
H         -4.50639        0.37329        4.19339
H         -6.17211       -3.40630       -0.73105
C          4.56898       -2.47561        0.14126
H          5.62137       -2.27113       -0.07118
H          1.59406       -1.76651       -2.54558
\end{lstlisting}

\textbf{Nickel-Complex (BC) (77 atoms)}
\begin{lstlisting}[basicstyle=\ttfamily\footnotesize]
77

Ni        -0.20821       -0.95408        0.00558
P         -2.37070       -0.53276       -0.06785
P          1.08758        0.78556       -0.05782
C          1.06458       -2.28458        0.46662
O          0.68222       -2.39432        1.64814
C          2.49962       -2.36556        0.06722
C          2.85880       -2.55212       -1.27126
C          5.19044       -2.30843       -0.69430
C          3.49997       -2.18699        1.02936
C          4.83835       -2.14137        0.64729
H          2.07336       -2.69600       -2.01903
H          6.24111       -2.27787       -0.99194
H          3.20177       -2.07778        2.07438
H          5.61493       -1.98430        1.39973
C          2.50895        0.78837       -1.24987
C          0.22541        2.40863       -0.34766
C          1.82857        0.95422        1.63712
C         -2.88734        0.43138       -1.56959
C         -3.43355       -2.05643       -0.05052
C         -2.94845        0.46074        1.39083
C          3.22492        2.07841       -1.64589
H          2.11156        0.30017       -2.15530
H          3.23538        0.06972       -0.83891
C          4.34123        1.81308       -2.65249
H          2.50449        2.79187       -2.07963
H          3.64752        2.56873       -0.75571
C          3.20963        1.58237        1.81357
H          1.83423       -0.07084        2.04357
H          1.07831        1.49261        2.24231
C          3.68722        1.51561        3.26102
H          3.93412        1.05520        1.17134
H          3.20069        2.63157        1.47653
C          0.81643        3.70654        0.20401
H         -0.78417        2.27182        0.07409
H          0.07648        2.48267       -1.43982
C         -0.06220        4.91598       -0.10293
H          0.94104        3.61544        1.29540
H          1.82438        3.87350       -0.20306
C         -4.89803       -1.99906        0.38680
H         -2.88615       -2.75383        0.60532
H         -3.34526       -2.48716       -1.06390
C         -5.56322       -3.37206        0.34741
H         -4.95749       -1.59830        1.41215
H         -5.46508       -1.30140       -0.24773
C         -4.21643        1.30958        1.30080
H         -2.09632        1.10825        1.65615
H         -3.02093       -0.26415        2.22052
C         -4.51228        2.04369        2.60547
H         -4.10757        2.04631        0.48724
H         -5.07944        0.68220        1.03049
C         -4.32780        0.38366       -2.07862
H         -2.20840        0.08471       -2.36628
H         -2.58279        1.47378       -1.36925
C         -4.52394        1.23460       -3.33015
H         -4.59964       -0.66070       -2.30376
H         -5.02431        0.71943       -1.29538
H          3.94326        1.35995       -3.57426
H          5.08556        1.11460       -2.23834
H         -3.84935        0.91153       -4.13873
H         -4.31161        2.29620       -3.12670
H         -3.67956        2.70789        2.88622
H         -4.66180        1.33319        3.43361
H         -5.03753       -4.08584        1.00074
H         -5.55431       -3.78911       -0.67207
H         -1.06948        4.79370        0.32678
H         -0.18085        5.05701       -1.18895
H          3.74332        0.47184        3.60825
H          2.99811        2.05085        3.93337
H          4.68536        1.96413        3.37743
H          4.86343        2.74073       -2.93100
H         -6.61088       -3.32000        0.67962
H          0.36759        5.84071        0.31028
H         -5.55553        1.16526       -3.70634
H         -5.42031        2.65953        2.52342
C          4.19831       -2.52880       -1.65131
H          4.47123       -2.67342       -2.69905
H          0.31522       -2.60661       -0.39764
\end{lstlisting}

\textbf{Nickel-Complex (C) (77 atoms)}
\begin{lstlisting}[basicstyle=\ttfamily\footnotesize]
77

Ni        -0.09499       -0.70961       -0.88163
P          0.83362        1.14737       -0.08457
P         -2.21681       -0.55889       -0.29118
C         -3.13605        0.95217       -0.87142
C          2.41999        0.79469        0.79744
C          1.25317        2.20721       -1.53618
C         -0.14241        2.21680        1.07659
C         -3.27261       -1.95663       -0.87463
C         -2.46042       -0.58283        1.54476
C         -4.20545        1.55706        0.03972
C         -4.89173        2.76269       -0.59469
C         -4.77175       -1.92790       -0.58688
C         -1.60275       -1.64920        2.22762
C         -1.79088       -1.68032        3.74022
C          2.40555        3.20762       -1.44654
C          2.61943        3.94196       -2.76689
C          0.16783        3.70776        1.20773
C         -0.75782        4.39302        2.20866
C          3.05131        1.83087        1.72525
C          4.35902        1.32247        2.32586
H         -3.57379        0.68046       -1.84732
H         -2.37412        1.71505       -1.10039
H         -4.95870        0.79682        0.29986
H         -3.74585        1.86346        0.99383
H         -5.39994        2.48488       -1.53135
H         -4.16183        3.55157       -0.83622
H         -2.80788       -2.86415       -0.45335
H         -3.08436       -2.02265       -1.95903
C         -5.48112       -3.19130       -1.06626
H         -4.94858       -1.79978        0.49420
H         -5.22564       -1.05010       -1.07637
H         -3.52547       -0.72137        1.79189
H         -2.17907        0.41121        1.92629
H         -1.83347       -2.64219        1.80639
H         -0.54311       -1.46169        1.98469
H          0.31486        2.71529       -1.82141
H          1.46357        1.47721       -2.33672
H          2.22962        3.93929       -0.64243
H          3.33014        2.67041       -1.17992
H         -0.07886        1.72106        2.06204
H         -1.19262        2.10883        0.76309
H          1.21469        3.86007        1.50897
H          0.06010        4.18881        0.22182
H          3.12639        0.49904        0.00354
H          2.23399       -0.13627        1.35850
H          3.23759        2.77142        1.18420
H          2.34849        2.07760        2.53861
H         -2.83594       -1.90272        4.00798
H         -1.53313       -0.70930        4.19227
H          1.72092        4.51090       -3.05473
H          2.83719        3.23190       -3.57933
H          5.10037        1.11509        1.53858
H          4.20107        0.38465        2.88098
H         -0.64347        3.95901        3.21467
H         -1.81409        4.27712        1.91678
H         -5.08179       -4.08544       -0.56268
H         -5.34458       -3.33475       -2.14940
C          1.56619       -1.23037       -1.59752
C          2.42496       -2.19429       -0.79416
O          2.03514       -0.75266       -2.61890
C          1.89374       -2.90355        0.28578
C          4.06956       -3.82561        0.77679
C          3.78588       -2.31996       -1.09047
C          4.60631       -3.12853       -0.30802
H          0.82403       -2.81064        0.49554
H          4.71282       -4.45896        1.39257
H          4.17587       -1.76304       -1.94610
H          5.67011       -3.22016       -0.54079
H          4.79661        2.05732        3.01800
H          3.45822        4.65102       -2.70165
H         -0.54676        5.47001        2.28267
H         -5.64440        3.19628        0.08038
H         -1.15421       -2.44641        4.20685
H         -6.56196       -3.14587       -0.86616
C          2.70980       -3.71689        1.07017
H          2.28588       -4.26964        1.91204
H         -0.34070       -1.99890       -1.57653
\end{lstlisting}

\subsubsection{Reduced Model (41 atoms)}

\textbf{Nickel-Complex (B) (41 atoms)}
\begin{lstlisting}[basicstyle=\ttfamily\footnotesize]
41

Ni         0.01087       -0.65034       -0.82192
P         -2.11214       -0.34418       -0.26552
P          1.28203        1.03861       -0.33963
C          1.20132       -1.99109       -1.53283
O         -0.03665       -2.40446       -1.42497
C          2.24029       -2.42578       -0.56130
C          3.59793       -2.18194       -0.81437
C          4.20131       -3.03680        1.36492
C          1.88519       -3.02003        0.65925
C          2.85488       -3.31682        1.61216
H          3.89145       -1.74659       -1.77460
H          4.95994       -3.26846        2.11574
H          0.82860       -3.22764        0.84174
H          2.56065       -3.77256        2.56099
C          2.49398        1.41965       -1.69188
C          0.55710        2.69169        0.08673
C          2.31701        0.64211        1.13840
C         -2.39241       -0.53396        1.56040
C         -2.85360        1.28411       -0.75583
C         -3.17221       -1.64013       -1.04980
H          1.96669        1.20535       -2.63559
H          3.26057        0.63088       -1.60119
H          2.76905       -0.33858        0.92618
H          1.59722        0.45366        1.95357
H         -0.13033        2.94971       -0.73711
H          1.34114        3.46536        0.09788
H         -2.35476        1.53214       -1.70805
H         -2.47198        2.02144       -0.03040
H         -2.51327       -2.52291       -1.09866
H         -3.30102       -1.32443       -2.09993
H         -1.46818       -0.17479        2.04124
H         -2.39896       -1.62333        1.73812
C          4.56898       -2.47561        0.14126
H          5.62137       -2.27113       -0.07118
H          1.59406       -1.76651       -2.54558
H          2.88434        2.41508       -1.65140
H          0.08173        2.64722        1.04430
H          3.03552        1.40151        1.36634
H         -3.27294       -0.03374        1.90585
H         -3.92220        1.26995       -0.80860
H         -4.10183       -1.80441       -0.54610
\end{lstlisting}

\textbf{Nickel-Complex (BC) (41 atoms)}
\begin{lstlisting}[basicstyle=\ttfamily\footnotesize]
41

Ni        -0.20821       -0.95408        0.00558
P         -2.37070       -0.53276       -0.06785
P          1.08758        0.78556       -0.05782
C          1.06458       -2.28458        0.46662
O          0.68222       -2.39432        1.64814
C          2.49962       -2.36556        0.06722
C          2.85880       -2.55212       -1.27126
C          5.19044       -2.30843       -0.69430
C          3.49997       -2.18699        1.02936
C          4.83835       -2.14137        0.64729
H          2.07336       -2.69600       -2.01903
H          6.24111       -2.27787       -0.99195
H          3.20177       -2.07778        2.07438
H          5.61493       -1.98430        1.39973
C          2.50895        0.78837       -1.24987
C          0.22541        2.40863       -0.34766
C          1.82857        0.95422        1.63712
C         -2.88734        0.43138       -1.56959
C         -3.43355       -2.05643       -0.05052
C         -2.94845        0.46074        1.39083
H          2.11156        0.30017       -2.15530
H          3.23538        0.06972       -0.83891
H          1.83423       -0.07084        2.04357
H          1.07831        1.49261        2.24231
H         -0.78417        2.27182        0.07409
H          0.07648        2.48267       -1.43982
H         -2.88615       -2.75383        0.60532
H         -3.34526       -2.48716       -1.06390
H         -2.09632        1.10825        1.65615
H         -3.02093       -0.26415        2.22052
H         -2.20840        0.08471       -2.36628
H         -2.58279        1.47378       -1.36925
C          4.19831       -2.52880       -1.65131
H          4.47123       -2.67342       -2.69905
H          0.31522       -2.60661       -0.39764
H          2.92268        1.75996       -1.42232
H          0.75415        3.24576        0.05797
H          2.78699        1.42995        1.63332
H         -3.92606        0.33315       -1.80690
H         -4.44707       -1.87799        0.24247
H         -3.85493        0.99866        1.20692
\end{lstlisting}

\textbf{Nickel-Complex (C) (41 atoms)}
\begin{lstlisting}[basicstyle=\ttfamily\footnotesize]
41

Ni        -0.09499       -0.70961       -0.88163
P          0.83362        1.14737       -0.08457
P         -2.21681       -0.55889       -0.29118
C         -3.13605        0.95217       -0.87142
C          2.41999        0.79469        0.79744
C          1.25317        2.20721       -1.53618
C         -0.14241        2.21680        1.07659
C         -3.27261       -1.95663       -0.87463
C         -2.46042       -0.58283        1.54476
H         -3.57379        0.68046       -1.84732
H         -2.37412        1.71505       -1.10039
H         -2.80788       -2.86415       -0.45335
H         -3.08436       -2.02265       -1.95903
H         -3.52547       -0.72137        1.79189
H         -2.17907        0.41121        1.92629
H          0.31486        2.71529       -1.82141
H          1.46357        1.47721       -2.33672
H         -0.07886        1.72106        2.06204
H         -1.19262        2.10883        0.76309
H          3.12639        0.49904        0.00354
H          2.23399       -0.13627        1.35850
C          1.56619       -1.23037       -1.59752
C          2.42496       -2.19429       -0.79416
O          2.03514       -0.75266       -2.61890
C          1.89374       -2.90355        0.28578
C          4.06956       -3.82561        0.77679
C          3.78588       -2.31996       -1.09047
C          4.60631       -3.12853       -0.30802
H          0.82403       -2.81064        0.49554
H          4.71282       -4.45896        1.39257
H          4.17587       -1.76304       -1.94610
H          5.67011       -3.22016       -0.54079
C          2.70980       -3.71689        1.07017
H          2.28588       -4.26964        1.91204
H         -0.34070       -1.99890       -1.57653
H         -3.85827        1.28732       -0.15660
H          2.76064        1.60748        1.40426
H          2.04728        2.90226       -1.35955
H          0.19257        3.23288        1.09329
H         -4.30612       -1.84521       -0.62100
H         -1.87887       -1.37561        1.96687
\end{lstlisting}

\subsubsection{Minimal Model (34 atoms)}

\textbf{Nickel-Complex (B) (34 atoms)}
\begin{lstlisting}[basicstyle=\ttfamily\footnotesize]
34

Ni         0.01087       -0.65034       -0.82192
P         -2.11214       -0.34418       -0.26552
P          1.28203        1.03861       -0.33963
C          1.20132       -1.99109       -1.53283
O         -0.03665       -2.40446       -1.42497
C          2.24029       -2.42578       -0.56130
C          2.49398        1.41965       -1.69188
C          0.55710        2.69169        0.08673
C          2.31701        0.64211        1.13840
C         -2.39241       -0.53396        1.56040
C         -2.85360        1.28411       -0.75583
C         -3.17221       -1.64013       -1.04980
H          1.96669        1.20535       -2.63559
H          3.26057        0.63088       -1.60119
H          2.76905       -0.33858        0.92618
H          1.59722        0.45366        1.95357
H         -0.13033        2.94971       -0.73711
H          1.34114        3.46536        0.09788
H         -2.35476        1.53214       -1.70805
H         -2.47198        2.02144       -0.03040
H         -2.51327       -2.52291       -1.09866
H         -3.30102       -1.32443       -2.09993
H         -1.46818       -0.17479        2.04124
H         -2.39896       -1.62333        1.73812
H          1.59406       -1.76651       -2.54558
H          2.88434        2.41508       -1.65140
H          0.08173        2.64722        1.04430
H          3.03552        1.40151        1.36634
H         -3.27294       -0.03374        1.90585
H         -3.92220        1.26995       -0.80860
H         -4.10183       -1.80441       -0.54610
H          2.60894       -3.39050       -0.84114
H          1.81313       -2.47760        0.41837
H          3.04622       -1.72195       -0.56222
\end{lstlisting}

\textbf{Nickel-Complex (BC) (34 atoms)}
\begin{lstlisting}[basicstyle=\ttfamily\footnotesize]
34

Ni        -0.20821       -0.95408        0.00558
P         -2.37070       -0.53276       -0.06785
P          1.08758        0.78556       -0.05782
C          1.06458       -2.28458        0.46662
O          0.68222       -2.39432        1.64814
C          2.49962       -2.36556        0.06722
C          2.50895        0.78837       -1.24987
C          0.22541        2.40863       -0.34766
C          1.82857        0.95422        1.63712
C         -2.88734        0.43138       -1.56959
C         -3.43355       -2.05643       -0.05052
C         -2.94845        0.46074        1.39083
H          2.11156        0.30017       -2.15530
H          3.23538        0.06972       -0.83891
H          1.83423       -0.07084        2.04357
H          1.07831        1.49261        2.24231
H         -0.78417        2.27182        0.07409
H          0.07648        2.48267       -1.43982
H         -2.88615       -2.75383        0.60532
H         -3.34526       -2.48716       -1.06390
H         -2.09632        1.10825        1.65615
H         -3.02093       -0.26415        2.22052
H         -2.20840        0.08471       -2.36628
H         -2.58279        1.47378       -1.36925
H          0.31522       -2.60661       -0.39764
H          2.92268        1.75996       -1.42232
H          0.75415        3.24576        0.05797
H          2.78699        1.42995        1.63332
H         -3.92606        0.33315       -1.80690
H         -4.44707       -1.87799        0.24247
H         -3.85493        0.99866        1.20692
H          2.58230       -2.24359       -0.99258
H          2.89482       -3.31932        0.34838
H          3.05103       -1.59186        0.55938
\end{lstlisting}

\textbf{Nickel-Complex (C) (34 atoms)}
\begin{lstlisting}[basicstyle=\ttfamily\footnotesize]
34

Ni        -0.09499       -0.70961       -0.88163
P          0.83362        1.14737       -0.08457
P         -2.21681       -0.55889       -0.29118
C         -3.13605        0.95217       -0.87142
C          2.41999        0.79469        0.79744
C          1.25317        2.20721       -1.53618
C         -0.14241        2.21680        1.07659
C         -3.27261       -1.95663       -0.87463
C         -2.46042       -0.58283        1.54476
H         -3.57379        0.68046       -1.84732
H         -2.37412        1.71505       -1.10039
H         -2.80788       -2.86415       -0.45335
H         -3.08436       -2.02265       -1.95903
H         -3.52547       -0.72137        1.79189
H         -2.17907        0.41121        1.92629
H          0.31486        2.71529       -1.82141
H          1.46357        1.47721       -2.33672
H         -0.07886        1.72106        2.06204
H         -1.19262        2.10883        0.76309
H          3.12639        0.49904        0.00354
H          2.23399       -0.13627        1.35850
C          1.56619       -1.23037       -1.59752
C          2.42496       -2.19429       -0.79416
O          2.03514       -0.75266       -2.61890
H         -0.34070       -1.99890       -1.57653
H         -3.85827        1.28732       -0.15660
H          2.76064        1.60748        1.40426
H          2.04728        2.90226       -1.35955
H          0.19257        3.23288        1.09329
H         -4.30612       -1.84521       -0.62100
H         -1.87887       -1.37561        1.96687
H          3.33101       -2.39393       -1.32718
H          1.89011       -3.10891       -0.64477
H          2.65807       -1.75835        0.15479
\end{lstlisting}

\end{document}